\documentclass[a4paper]{ociamthesis}

\usepackage[style=numeric-comp, sorting=none, backend=bibtex, doi=false, isbn=false]{biblatex}

\usepackage{graphicx}
\usepackage{dcolumn}

\usepackage{savesym}
\usepackage{amssymb}   
\usepackage{booktabs}
\usepackage{multirow}
\usepackage{siunitx}
\usepackage{amsmath,amsfonts,amssymb,amsthm,bm,upgreek, mathtools, dsfont}
\usepackage[mathscr]{eucal}
\usepackage{float}
\usepackage{braket}
\usepackage{mathrsfs}
\DeclareMathOperator{\Tr}{Tr}
\usepackage{color,soul}
\usepackage{latexsym}
\usepackage{amsbsy}
\usepackage[utf8]{inputenc}
\usepackage[T1]{fontenc}
\usepackage{mathtools}
\usepackage{physics}
\usepackage[english]{babel}
\usepackage{bbm}

\DeclareUnicodeCharacter{2212}{-}

\theoremstyle{plain}

\newtheorem*{lemma*}{Lemma}

\newtheorem{definition}{Definition}
\newtheorem{property}{Property}
\newtheorem{criterion}{Criterion}
\newtheorem{claim}{Claim}

\makeatletter
\newsavebox{\@brx}
\newcommand{\llangle}[1][]{\savebox{\@brx}{\(\m@th{#1\langle}\)}%
  \mathopen{\copy\@brx\kern-0.5\wd\@brx\usebox{\@brx}}}
\newcommand{\rrangle}[1][]{\savebox{\@brx}{\(\m@th{#1\rangle}\)}%
  \mathclose{\copy\@brx\kern-0.5\wd\@brx\usebox{\@brx}}}
\makeatother

\newcommand*{\bibtitle}{References}

\title{Quantum Correlations in Space-Time: Foundations and Applications}
\author{Tian Zhang}
\college{Somerville College}

\degree{Doctor of Philosophy}
\degreedate{Revised Version, 2020}

\begin{document}

\setlength{\textbaselineskip}{22pt plus2pt}

\setlength{\frontmatterbaselineskip}{17pt plus1pt minus1pt}

\setlength{\baselineskip}{\textbaselineskip}


\setcounter{secnumdepth}{2}
\setcounter{tocdepth}{2}

\begin{abstractseparate}

The absolute/relative debate on the nature of space and time is ongoing for thousands of years. 
Here we attempt to investigate space and time from the information theoretic point of view to understand spatial and temporal correlations under the relative assumption. 
Correlations, as a measure of relationship between two quantities, do not distinguish space and time in classical probability theory; quantum correlations in space are well-studied but temporal correlations are not well understood. 
The thesis investigates quantum correlations in space-time, by treating temporal correlations equally in form as spatial correlations and unifying quantum correlations in space and time. 
In particular, we follow the pseudo-density matrix formalism in which quantum states in spacetime are properly defined by correlations from measurements. 

We first review classical correlations, quantum correlations in space and time, to motivate the pseudo-density matrix formalism in finite dimensions. 
Next we generalise the pseudo-density matrix formulation to continuous variables and general measurements. 
Specifically, we define Gaussian spacetime states by the first two statistical moments, and for general continuous variables spacetime states are defined via the Wigner function representation. 
We also define spacetime quantum states in position measurements and weak measurements for general measurement processes. 
Then we compare the pseudo-density matrix formalism with other spacetime formulations: indefinite causal structures, consistent histories, generalised non-local games, out-of-time-order correlation functions, and path integrals. We argue that in non-relativistic quantum mechanics, different spacetime formulations are closely related and almost equivalent via quantum correlations, except path integrals. Finally, we apply the pseudo-density matrix formulation to time crystals. By defining time crystals as long-range order in time, we analyse continuous and discrete time translation symmetry as well as discuss the existence of time crystals from an algebraic point of view. Finally, we summarise our work and provide the outlook for future directions.

\end{abstractseparate}

\begin{romanpages}

\maketitle

\begin{dedication}

“Space and time are the pure forms thereof; sensation the matter.”\\
\ \\
― Immanuel Kant, \textit{Critique of Pure Reason}
\end{dedication}

\begin{acknowledgements}

I find myself very lucky to have the opportunity to study at Oxford and around the world with so many talented mentors and friends, from whom I have learnt so much. I am very grateful for all the help and support they offer and would express my sincere gratitude to all who have made my DPhil days special.

First of all, I would like to thank my supervisor, Prof. Vlatko Vedral, for inviting me to his group, for insightful discussions during these years, and for his enthusiasm for physics which always inspires me. 
I would also like to thank Dr. Tristan Farrow, who is always so kind and thoughtful, like a big elder brother, not only teaches me how to cope with different academic situations, but also cares about my personal development. 
I am grateful to Prof. Oscar Dahlsten, who has invited me twice to SUSTech in Shenzhen, China, collaborated with me, especially for his patience on teaching me how to write my first paper, as well as continually giving me lots of guidance and feedback. 
I also thank Dr. Felix Tennie, who has been so patient and helpful to revise my first-year transfer report and all the suggestions on academic writing. 
I would like to thank Dr. Chiara Marletto, who collaborated with me on my first project on time crystals, for being so helpful and considerate to give me suggestions on research and writing. 
I would love to thank the rest of the group, including Anupama Unnikrishnan, Nana Liu, Christian Schilling, Davide Girolami, Reevu Maity, Pieter Bogaert, Thomas Elliott, Benjamin Yadin, Aditya Iyer, Jinzhao Sun, Sam Kuypers, David Felce, and Anicet Tibau Vidal. 
I would also like to thank the quantum group in the department of computer science, especially Prof. Giulio Chiribella and Prof. Jonathan Barrett for insightful discussion, and Prof. Bob Coecke and lots of others for organising interesting talks, conferences and Wolfson foundation discussions, from which I benefited a lot. 


Then I want to thank Perimeter Institute for Theoretical Physics, in particular for the visiting graduate fellow program.
 I would love to thank my local host and advisor, Prof. Lucien Hardy, for inviting me to Perimeter, for regular meetings with me to answer my questions and check my progress, for always being so encouraging and telling me to think of big problems. 
 It is a great pleasure to meet Prof. Lee Smolin at Perimeter and hold weekly meetings with him. I learned so many fascinating ideas and deep thoughts from Lee and gradually started to build on my own research taste. 
 Thanks to Prof. Rafael Sorkin as well for being so patient and helpful and explaining to me on quantum measure theory, irreversibility, nonunitary, and lots of sightful conversation. 
 And thanks to Beni Yoshida, for introducing the wonderful world of black hole information paradox to me and collaborating with me on the black hole final state projection proposal. 
 I would also thank Guifre Vidal, for his guidance on organising the session of black hole information paradox in the quantum information workshop in Benasque, Spain. 
 I would also love to thank the whole quantum foundation and quantum information group, Daniel Gottesman, Robert Spekkens, David Schmid, Tobias Fritz, Denis Rosset, Thomas Galley, Nitica Sakharwade, Zi-Wen Liu, Nick Hunter-Jones, and lots of others that I am sorry I cannot list for all, for having lunches and discussions together and always being so friendly and helpful. 
 I had a brilliant time at Perimeter and I am so grateful for everyone there.

I thank Prof. Xi Dong, for hosting my visit at Santa Barbara and discussing possible projects on holographic min- and max-entropy. 
I also want to thank the organisers and participants of Boulder Summer School 2018, for organising such a great quantum information summer school, where I learned so much and broadened my understanding of quantum information science. 
Thanks to Felix Leditzky, Graeme Smith, Mark M. Wilde for the opportunity to present my work at Rocky Mountain Summit on Quantum Information. 
Thanks to Hilary Carteret and John Donohue for inviting me to present my work at Institute of Quantum Computing, University of Waterloo.
And thanks to Prof. Otfried Guehne for inviting me to visit and work with his group at University of Siegen.

I am very grateful to Prof. Renato Renner at ETH Zurich and Prof. Simon Benjamin at Oxford Materials for being my examiners and all the discussion and suggestions for my thesis. 



I would thank my valuable friends at Oxford, at Perimeter Institute, at Boulder Summer School, at different conferences, from my undergraduate, Peking University and even long time before. I wish I could list all their names here but I am afraid there are too many of them and I am more afraid I may miss some of their names. 

Thank you so much, my dearest mum and dad. Thanks to all my family members for their love and understanding.  

I have had very difficult time during my DPhil. I am so grateful that I have received so much love and support to get through all the hard time. 
I would express my gratitude, again to so many people along the way for their companion. 
And thank you, my reader, for taking your time to have a look at this thesis.

\end{acknowledgements}

\begin{abstract}
	
\end{abstract}

\dominitoc 

\flushbottom

\tableofcontents





\end{romanpages}

\flushbottom
%

\chapter{\label{ch:1-intro}Introduction} 

\minitoc 

\clearpage

What is time?

The intrinsic motivation for all the work in the thesis is to get a little bit closer to this question.


In general, there are three schools that hold different views on time. As Page and Wootters argue in their famous ``evolution without evolution'' paper~\cite{page1983evolution}, all the observables which commute with the Hamiltonian are stationary and the dynamics of a system we observe can be fully described by stationary observable dependent upon internal clock readings. Barbour~\cite{barbour2001end} also believes in the timeless universe where time does not exist and is merely an illusion. They claim that in general relativity, especially in the equivalent Arnowitt-Deser-Misner (ADM) formalism~\cite{arnowitt1959dynamical}, the dynamics is embedded in three-dimensional Riemannian spaces rather than the four-dimensional spacetime since one dimension can be arbitrarily chosen. Not to mention quantum cosmology~\cite{hartle1983wave}, where quantum mechanics is applied to the whole universe, the Wheeler-Dewitt equation~\cite{dewitt1967quantum} 
serves as a stationary Schr\"{o}dinger equation for the wave function of the universe. 

However, Smolin and his colleagues~\cite{smolin2013time, cortes2014quantum} hold the opposite point of view; that is, time is fundamental in nature. They claim that in the Newtonian paradigm~\cite{smolin2015temporal}, questions such as why the laws and why these initial conditions remain unanswered. They believe that the reality of time is important in selecting the fundamental laws of physics and construct an ultimate theory of the whole universe instead of part of the universe. 

Nevertheless, we have no evidence to judge the above two views on time, whether time does not exist or time is fundamental so far. Instead in this thesis, we would take a practical point of view from the lesson of relativity: time may be treated as an equal footing as space. 
Both special relativity and general relativity treat time as part of spacetime and gain beautiful results which have already been verified. What's more, treating time operationally equal as space, also provides one possible method to study time following the methods for investigating space. 

More specifically, we investigate time from the quantum information perspective in terms of temporal correlations, as an analogue of spatial correlations. Spatial correlations like entanglement, nonlocality, steering, and discord, are well-studied in quantum information. 
We know that physicists have been working for decades in search for a way to quantise space-time and trying to build a theory for quantum gravity. That is not the goal for this thesis. 
Here, we focus on the quantum information side of spacetime; more precisely, our topic is restricted to quantum correlations in non-relativistic space-time. 

We start from a particular kind of space-time formulation called pseudo-density matrix formalism which treats temporal correlations as spatial correlations, further generalise this formulation to continuous variables and general measurement processes, compare it with other space-time formulations via quantum correlations and argue that these non-relativistic space-time formulations are very much related, and apply the pseudo-density matrix formalism to time crystals to show its practical power. 

The thesis proceeds as follows. In Chapter 2, we introduce quantum correlations in space-time. We first introduce classical correlations in probability theory and statistical mechanics. After introducing the basics for quantum mechanics, we review quantum correlations in space. In bipartite quantum correlations, we discuss correlation and entanglement, the difference among Bell nonlocality, steering and entanglement, other quantum correlations as quantum discord, and formulate the hierarchy of quantum correlations in space based on operator algebra. We briefly mention multipartite quantum correlations. Then we move on to quantum correlations in time. From the correlations in field theory, we explore a further possibility for temporal correlations and propose a unified approach for quantum correlations in space and time to motivate pseudo-density matrices. Finally we formally introduce the pseudo-density matrix formalism. 

In Chapter 3, we fully generalise the pseudo-density matrix formalism to continuous variables and general measurement processes. Pseudo-density matrix formalism is based on building measurement correlations; the key for generalisation is to choose the right measurement operators. For the Gaussian case, we simply extend the correlations to the temporal domain by quadratures measurements and compare the spatial vs temporal Gaussian states. For general continuous variables, we use the Wigner function representation and its one-to-one correspondence with the density matrix formalism to define spacetime states, and compare the properties of spacetime Wigner functions with the uniquely determined properties of normal Wigner functions. We further generalise the formalism for general measurement processes like position measurements and weak measurements. We also give an experimental proposal for tomography in the Gaussian case. Before coming to the end, we compare spacetime states in the generalised pseudo-density matrix formalism and make further comments. 

In Chapter 4, we compare spatial-temporal correlations in pseudo-density matrix formalism with correlations in other spacetime formulations. In particular, we analyse indefinite causal structures, consistent histories, generalised non-local games, out-of-time-order correlation functions, and path integrals. We aim to argue, in non-relativistic quantum mechanics, spacetime formulations are closely related via quantum correlations. We also take lessons from these spacetime formulations and further develop the pseudo-density matrix formalism. In the section of out-of-time-order correlation functions, we discuss their possible application in the black hole final state projection proposal, as one of possible explanations for black hole information paradox. The path integral approach gives a different representation of quantum correlations and suggests interesting properties for quantum measure and relativistic quantum information. 

In Chapter 5, we use time crystals as an illustration of temporal correlations, or more specifically, long-range order in time. We first review spontaneous symmetry breaking, time translation symmetry breaking and different mathematical definitions for time crystals. After formally introducing long-range order, we define time crystals as long-range order in time in the pseudo-density matrix formulation. To illustrate what time crystals are, we consider continuous time translation symmetry in terms of general decoherent processes and a generalised version of Mermin-Wagner theorem, and discuss discrete time translation symmetry via a stabilisation protocol of quantum computation, phase flip codes of quantum error correction and Floquet many-body localisation. We also use an algebraic point of view to analyse the existence of time crystals. 

Chapter 6 is for the conclusion and outlook.

\chapter{\label{ch:2-correlation}Quantum correlations in space-time}

\clearpage

\minitoc

\clearpage

\section{Classical correlations}
In this section we introduce classical correlations in probability theory and statistical mechanics. In the classical case, it is not necessary to distinguish spatial or temporal correlations; that is, classical correlations are defined whatever the spatio-temporal structures are. 

\subsection{Correlations in probability theory}
Now we introduce correlations defined in probability theory based on Ref.~\cite{sheldon2002first}.
For a discrete random variable $X$ with the probability mass function $p(x) = P\{X=x\}$, the expectation value of $X$ is defined as
$E[X] = \sum_{x: p(x)>0} xp(x)$.
For a continuous random variable $X$ with the probability density function $f(x)$ such that $P\{a\leq X \leq b\} = \int_a^b f(x) \textrm{d}x$, the expectation value of $X$ is defined as
$E[X] = \int_{-\infty}^{\infty} xf(x)\textrm{d}x$.
The variance of $X$ is defined as 
$\text{Var}(X) = E[(X-E[X])^2]$.
This definition is equivalent to 
$\text{Var}(X) = E[X^2] - (E[X])^2$.

For two random variables $X$ and $Y$, the covariance is defined as 
$\text{Cov}(X, Y) = E[(X-E[X])(Y-E[Y])]$.
It is easy to see that
$\text{Cov}(X, Y) = E[XY]-E[X]E[Y]$.
Then we define the correlation of $X$ and $Y$ as
\begin{equation}\label{classicalcorr}
\text{Corr}(X, Y) = \frac{\text{Cov}(X,Y)}{\sqrt{\text{Var}(X)\text{Var}(Y)}}
\end{equation}
It is also referred to the Pearson product-moment correlation coefficient or the bivariate correlation, as a measure for the linear correlation between $X$ and $Y$. 




\subsection{Correlations in statistical mechanics}

In statistical mechanics~\cite{sethna2006statistical}, the equilibrium correlation function for two random variables $S_1$ at position $\mathbf{x}$ and time $t$ and $S_2$ at position $\mathbf{x}+\mathbf{r}$ and time $t+\tau$ is defined as
\begin{equation}
C(\mathbf{r}, \tau) = \langle S_1(\mathbf{x}, t) S_2(\mathbf{x}+\mathbf{r}, t+\tau) \rangle - \langle S_1(\mathbf{x}, t) \rangle  \langle S_2(\mathbf{x}+\mathbf{r}, t+\tau) \rangle,
\end{equation}
where $\langle O \rangle$ is the thermal average of the random variable $O$; it is usually averaged over the whole phase space of the system. That is, 
\begin{equation}
\langle O \rangle = \frac{\int O e^{-\beta H(q_1, \dots, q_m, p_1, \dots, p_n)}\textrm{d}\tau}{\int e^{-\beta H(q_1, \dots, q_m, p_1, \dots, p_n)}\textrm{d}\tau},
\end{equation}
where $\beta = 1/k_BT$, $k_B$ is Boltzmann constant and $T$ is the temperature, $H$ is the Hamiltonian of the classical system in terms of coordinates $q_i$ and their conjugate generalised momenta $p_i$, and $\textrm{d}\tau$ is the volume element of the classical phase space. 
In particular, the equal-time spin-spin correlation function for two Ising spins is given as 
\begin{equation}
C_t(\mathbf{r}) = \langle S(\mathbf{x}, t) S(\mathbf{x}+\mathbf{r}, t) \rangle - \langle S(\mathbf{x}, t) \rangle  \langle S(\mathbf{x}+\mathbf{r}, t) \rangle.
\end{equation}
It is used as a measure for spatial coherence for how much information a spin can influence its distant neighbours. Taking the limit of $\mathbf{r}$ to infinity, we obtain the long-range order for which correlations remain non-zero even in the long distance.

\section{Quantum correlations in space}

In this section, we introduce quantum correlations in space. First we review briefly on basics of quantum mechanics. Then we introduce bipartite quantum correlations, in terms of entanglement, steering, nonlocality and discord. We also list the hierarchy of spatial quantum correlations in terms of operator algebra. Finally we mention multipartite quantum correlations in brief. 

\subsection{Basics for quantum mechanics}
In this subsection we briefly review the axioms of quantum mechanics and introduce the concept of quantum states.

\subsubsection{Axioms of quantum mechanics}
We introduce the five axioms of quantum mechanics ~\cite{weinberg1995quantum, nielsen2002quantum, preskill1998lecture}. 


(1) The state in an isolated physical system is represented by a vector, for example, $\ket{\psi}$, in the Hilbert space $\mathcal{H}$ which is a complex vector space with an inner product. 
A system is completely described by normalised state vectors in the Hilbert space.


(2) An observable is represented by an Hermitian operator with $A^{\dag} = A$. 

(3) Suppose the system is measured by a collection of measurement operators $\{M_m\}$ with measurement outcomes $\{m\}$. With the initial state $\ket{\psi}$, after measurements the result $m$ comes with the probability 
\begin{equation}
p(m) = \bra{\psi}M_m^{\dag}M_m\ket{\psi}
\end{equation}
and the state becomes
\begin{equation}
\frac{M_m\ket{\psi}}{\sqrt{\bra{\psi}M_m^{\dag}M_m\ket{\psi}}}
\end{equation}
The measurement operators satisfy $\sum_m M_m^{\dag}M_m = \mathbbm{1}$, then the probabilities sum to 1.


According to Wigner's theorem, for any transformation $\ket{\psi} \rightarrow \ket{\psi'}$ in which the probabilities for a complete set of states collapsing into another complete set $|\braket{\psi}{\psi_n}|^2 = |\braket{\psi'}{\psi'_n}|^2$ hold the same, we may define an operator $U$ such that $\ket{\psi'} = U\ket{\psi}$. Then $U$ is either unitary and linear or else anti-unitary and anti-linear. 
Thus, we have 

(4) A closed quantum system evolves under unitary transformation. That is, the state of the system at two times $t_1$ and $t_2$ are related by a unitary operator $U$ defined by $U^{\dag}U = UU^{\dag} = \mathbbm{1}$ such that
\begin{equation}
\ket{\psi(t_2)} = U \ket{\psi(t_1)}.
\end{equation} 

It is equivalent to 

(4') A closed quantum system evolves under Schr\"odinger equation:
\begin{equation}
i\hbar \frac{d\ket{\psi}}{dt} = H \ket{\psi}
\end{equation}
$H$ is the Hamiltonian of the quantum system.

In addition, we have another postulate for composite quantum systems.

(5) The Hilbert space of the composite system $AB$ is the tensor product $\mathcal{H_A}\otimes\mathcal{H_B}$ of the Hilbert spaces $\mathcal{H_A}$ and $\mathcal{H_B}$ for systems $A$ and $B$. That is, if the system $A$ is in the state $\ket{\psi}_A$ and the system $B$ is in the state $\ket{\phi}_B$, then the composite system $AB$ is in the state $\ket{\psi}_A \otimes \ket{\phi}_B$.

\subsubsection{Quantum states}
Here we define quantum states for discrete finite systems, and leave continuous variables to next chapter. 

The state vector is defined as before in terms of a normalised vector in the Hilbert space: $\ket{\psi} \in \mathcal{H}$. 
A pure state is then given by $\pi = \ket{\psi}\bra{\psi} \in \mathcal{P}$. 
If a quantum system is in the state $\ket{\psi_i}$ with the probability $p_i$, we call the set $\{p_i, \ket{\psi_i}\}$ as an ensemble of pure states. An arbitrary quantum state is represented by a density matrix defined as $\rho = \sum_i p_i\ket{\psi_i}\bra{\psi_i} \in \mathcal{D}$~\cite{nielsen2002quantum}. On the one hand, the set of all possible states $\mathcal{D}$ is a convex set, that is, $\mathcal{D} = \text{Conv} \mathcal{P}$; on the other hand, the extreme points in the state space are pure states, i.e., $\mathcal{P} = \text{Extr} \mathcal{D}$~\cite{ohya2004quantum}. Note that the convex hull $\text{Conv}\mathcal{P}$ of the set $\mathcal{P}$ in the complex state space is defined to be the intersection of all convex sets in the state space that contain $\mathcal{P}$.  An extreme point $x$ of a convex set $\mathcal{D}$ is a point such that for $y, z \in \mathcal{D}$, $0 < \lambda < 1$, $x = \lambda y + (1-\lambda) z$ implies that $x=y=z$. ~\cite{grunbaum1967convex}

A simple criterion to check whether the state $\rho$ is pure or mixed is that $\Tr \rho^2 = 1$ for pure states and $\Tr \rho^2 < 1$ for mixed states~\cite{nielsen2002quantum}. 
Another measure of mixedness for quantum states is given by the von Neumann entropy $S(\rho) = - \Tr \rho \log \rho$~\cite{von2018mathematical}. It is non-negative and vanishes if and only if $\rho$ is a pure state. The von Neumann entropy is concave, subadditive and strongly subadditive. According to Schumacher's quantum noiseless channel coding theorem~\cite{schumacher1995quantum}, it is the amount of quantum information as the minimum compression scheme of rate. 

The distinguishability of states~\cite{nielsen2002quantum} is measured by the quantum relative entropy $D(\rho || \sigma) = \Tr \rho(\log\rho - \log\sigma)$ based on quantum Stein's lemma. The quantum relative entropy is jointly convex, non-negative, and vanishes if and only if $\rho = \sigma$. Other distance measures for quantum states include the trace distance $D(\rho, \sigma) = \frac{1}{2}\Tr|\rho - \sigma|$ and the fidelity $F(\rho, \sigma) = \Tr \sqrt{\rho^{1/2}\sigma\rho^{1/2}}$ or $F(\rho, \sigma) = (\Tr \sqrt{\rho^{1/2}\sigma\rho^{1/2}})^2$.

\subsection{Bipartite quantum correlations}
In this subsection, we focus on bipartite quantum correlations. First we introduce quantum correlation measures based on the distance of the states and compare correlation with entanglement. Then we compare three types of quantum correlations: entanglement, steering and Bell nonlocality. After introducing other measures of quantum correlations such as discord, we use the operator algebraic language to present the hierarchy of quantum correlations in space. 

\subsubsection{Correlation and entanglement}
As an analog of classical correlations in probability theory, the correlation for the quantum state itself is defined for $\Gamma = \rho - \rho_1 \otimes \rho_2$ where $\rho_i$ is the reduced state for the subsystem $i (i = 1, 2)$. The covariance for two observables $A$ and $B$ on the two subsystems separately is then given by
\begin{equation}
\text{Cov}(A, B) = \Tr \Gamma A \otimes B
\end{equation}
Recall that Eqn.~\eqref{classicalcorr} $\text{Corr}(X, Y) = \frac{\text{Cov}(X,Y)}{\sqrt{\text{Var}(X)\text{Var}(Y)}}$. Then we say that the state is uncorrelated, if and only if $\text{Corr}(A , B) = 0$ for all observables $A, B$ for two subsystems. This condition is equivalent to $\langle AB \rangle = \langle A \rangle \langle B \rangle$, as well as $\rho = \rho_1\otimes\rho_2$.

For a pure state, if the state is correlated, we call it entanglement. 
For a mixed state, the state is uncorrelated if and only if $\rho = \rho_1 \otimes \rho_2 \in \mathcal{D}_{unc}$, otherwise we call it correlated. 
At the same time, the state is separable for a possible decomposition $\rho = \sum_i p_i \pi_{1, i} \otimes \pi_{2, i} \in \mathcal{D}_{sep}$; otherwise it is entangled. 
Note that 
$\mathcal{D}_{sep} 
= \text{Conv} \mathcal{D}_{unc}$. 
The correlation measure can be given by the distinguishability as the relative entropy
\begin{equation}
C(\rho) = \min_{\sigma \in \mathcal{D}_{unc}}D(\rho || \sigma) = D(\rho || \rho_1 \otimes \rho_2) = S(\rho_1) + S(\rho_2) - S(\rho) = I(\rho);
\end{equation}
it is equal to the mutual information of the state. 
Here we only discuss whether the state is correlated or not; for general quantum correlations, we will introduce entanglement, steering, Bell nonlocality and discord later. 

\subsubsection{Entanglement, steering and Bell nonlocality}
Here we compare three types of quantum correlations: entanglement, steering and Bell nonlocality~\cite{uola2019quantum}. 

Bell nonlocality is characterised by the violation of Bell inequalities~\cite{bell1964einstein, brunner2014bell}. In a typical Bell experiment, two spatially separated systems are measured by two distant observers, say Alice and Bob, respectively. Alice may select her measurement from several possible ones and denote her choice of measurement by $x$, and gain the outcome $a$ after the measurement. Bob makes the measurement denoted by $y$ and gains the outcome $b$. If there exists a local hidden variable model, then the probability to obtain the results $a$ and $b$ under the measurements $x$ and $y$ can be written as
\begin{equation}\label{lhv}
p(a, b|x, y) = \int \textrm{d}\lambda p(\lambda) p(a|x, \lambda) p(b|y, \lambda),
\end{equation}
where the hidden variable $\lambda$ gives the probability function $p(\lambda)$, Alice and Bob yield the outcome under their local probability distributions with the parameter $\lambda$. Given the measurements $x, y$ and the outcomes $a, b$, the probabilities $p(a, b|x, y)$ in Eqn.~\eqref{lhv} satisfy certain linear inequalities which are referred to Bell inequalities. For some experiments, for example with a pair of entangled qubits, the local hidden variable model cannot exist and Bell inequalities are violated. 

Entanglement is defined as before when a bipartite state cannot be written in terms of a convex combination of the tensor product of pure states
\begin{equation}
\rho_{AB} = \sum_i p_i \rho_i^A \otimes \rho_i^B,
\end{equation}
otherwise the state is separable. 
General measurements are represented by positive operator-valued measures (POVMs). A set of POVMs $\{E_{a|x}\}$ satisfying $E_{a|x} > 0$, $E_{a|x}^{\dag} = E_{a|x}$, and $\sum_a E_{a|x} = \mathbbm{1}$ give the probability of gaining the result $a$ in the state $\rho$ as $p(a) = \Tr(\rho E_{a|x})$. 
For a separable state, the probability for the measurements $E_{a|x}$ and $E_{b|y}$ is given as
\begin{equation}
p(a, b | x, y) = \sum_i p_i \Tr(E_{a|x} \rho_i^A) \Tr(E_{b|y} \rho_i^B). 
\end{equation}
It is easy to see that it belongs to the local hidden variable models and separable states are a convex subset of the local hidden variable states. 

Quantum steering in a sense lies in-between of entanglement and Bell nonlocality, where Alice is described by a classical hidden variable and Bob makes a quantum mechanical measurement. That is, the probability is given by
\begin{equation}\label{steering}
p(a, b| x, y) = \int \textrm{d}\lambda p(\lambda) p(a|x, \lambda) \Tr(E_{b|y}\sigma_{\lambda}^B). 
\end{equation}
In the steering scenario, Alice and Bob share a bipartite quantum state $\rho_{AB}$. For each measurement $x$ and the corresponding outcome $a$ in Alice's lab, Bob has the conditional state $\rho_{a|x}$ such that $\rho_B = \sum_a \rho_{a|x}$ is independent of Alice's choice for the measurement $x$. 
The state $\rho_{AB}$ is said to be unsteerable or have a local hidden state model if there is a representation from some parameter $\lambda$ that
\begin{equation}
\rho_{a|x} = \int \textrm{d}\lambda p(\lambda) p(a|x, \lambda) \sigma_{\lambda};
\end{equation}
otherwise the state is steerable. In Eqn.~\eqref{steering}, the probability can be rewritten as 
\begin{equation}
p(a, b | x, y) = \Tr(E_{b|y}\rho_{a|x}), \quad \rho_{a|x} = \int \textrm{d}\lambda p(\lambda) p(a|x, \lambda) \sigma_{\lambda}^B;
\end{equation}
thus, the local hidden state model exists. 

We can summarise that, the states that have a local hidden variable model and do not violate Bell inequalities form the convex set of LHV states; the states that have a local hidden state model and are unsteerable form the convex subset of LHV states, denoted by LHS states; the separable states  form the convex subset of LHS states. 

\subsubsection{Discord and related measures}

Entanglement is crucial in distinguishing quantum correlations from classical ones; however, it cannot represent for all non-classical correlations, and even separable states contain correlations which are not fully classical~\cite{modi2012classical}. One of these non-classical correlation measures is quantum discord~\cite{henderson2001classical, ollivier2001quantum}. Suppose a POVM measurement $E_a$ is made on the subsystem $A$ of the initial state $\rho_{AB}$. For the outcome $a$, Alice observes it with the probability $p(a) = \Tr(E_a \rho_{AB})$ and Bob gains the conditional state $\rho_{B|a} = \Tr_A(E_a \rho_{AB}) / p(a)$. The conditional entropy then has a classical-quantum version of definition as $S(B|\{E_a\}) = \sum_a p_a S(\rho_{B|a})$. The quantum discord is defined as
\begin{equation}
J(B|A) = \max_{\{E_a\}} S(B) - S(B| \{E_a\}).
\end{equation}
It is non-symmetric, non-negative, invariant under local unitary transformations, and vanishes if and only if the state is classical quantum. 
Other measures of quantum correlations include quantum deficit~\cite{horodecki2003local}, distillable common randomness~\cite{devetak2004distilling}, measurement-induced disturbance~\cite{luo2008using}, symmetric discord~\cite{wu2009correlations}, relative entropy of discord and dissonance~\cite{modi2010unified}, and so on. 

\subsubsection{Hierarchy of spatial correlations}

Now we introduce the hierarchy of quantum correlations based on Ref.~\cite{slofstra2019set}:
\begin{equation}
C_c \subseteq C_{q} \subseteq C_{qs} \subseteq C_{qa} \subseteq C_{qc}.
\end{equation}
Here all the sets are convex, and $C_c$, $C_{qa}$ are closed. 
Consider a two-player non-local game $\mathcal{G}$ with finite input sets $\mathcal{I}_A$, $\mathcal{I}_B$, finite output sets $\mathcal{O}_A$, $\mathcal{O}_B$ and a function $V: \mathcal{O}_A \times \mathcal{O}_B \times \mathcal{I}_A \times \mathcal{I}_B \rightarrow \{0, 1\}$. Suppose the two players, Alice and Bob, after given the inputs $x \in \mathcal{I}_A$ and $y \in \mathcal{I}_B$ respectively, cannot communicate with each other, and return outputs $a \in \mathcal{O}_A$ and $b \in \mathcal{O}_B$ respectively. The players win if $V(a, b|x, y) = 1$, or lose if $V(a, b|x, y) = 0$. The probabilities $p(a, b|x, y)$ that Alice and Bob return output $a \in \mathcal{O}_A$ and $b \in \mathcal{O}_B$ given inputs $x \in \mathcal{I}_A$ and $y \in \mathcal{I}_B$ form a collection $\{ p(a, b|x, y) \} \subset \mathbb{R}^{\mathcal{O}_A \times \mathcal{O}_B \times \mathcal{I}_A \times \mathcal{I}_B}$ called a correlation matrix. 

A correlation matrix $\{ p(a, b|x, y) \}$ is said to be classical under classical strategies with classical shared randomness. Specifically, 
\begin{equation}
p(a, b| x, y) = \sum_{i=1}^k \lambda_i p_i(a | x) q_i(b | y) \ \text{for all} (a, b, x, y) \in \mathcal{O}_A \times \mathcal{O}_B \times \mathcal{I}_A \times \mathcal{I}_B,
\end{equation}
for a probability distribution $\{\lambda_i\}$ on $\{1, \dots, k\}$, probability distributions $\{p_i(a | x) \}$ on $\mathcal{O}_A$ for each $1 \leq i \leq k$ and $x \in \mathcal{I}_A$, and probability distributions $\{q_i(b | y) \}$ on $\mathcal{O}_B$ for each $1 \leq i \leq k$ and $y \in \mathcal{I}_B$. Then the set of classical correlation matrices is denoted by $C_c(\mathcal{O}_A, \mathcal{O}_B, \mathcal{I}_A, \mathcal{I}_B)$ or $C_c$. 

A quantum correlation matrix is constructed under
\begin{equation}
p(a, b| x, y) = \bra{\psi} M_a^x \otimes N_b^y \ket{\psi} \text{for all} (a, b, x, y) \in \mathcal{O}_A \times \mathcal{O}_B \times \mathcal{I}_A \times \mathcal{I}_B
\end{equation}
for a quantum state $\ket{\psi}$ on the finite-dimensional Hilbert spaces $\mathcal{H} = \mathcal{H}_A \otimes \mathcal{H}_B$, projective measurements $\{ M_a^x \}_{a \in \mathcal{O}_A}$ on $\mathcal{H}_A$ for every $x \in \mathcal{I}_A$, and projective measurements $\{ N_b^y \}_{b \in \mathcal{O}_B}$ on $\mathcal{H}_B$ for every $y \in \mathcal{I}_B$. Then the set of quantum correlation matrices is denoted by $C_q(\mathcal{O}_A, \mathcal{O}_B, \mathcal{I}_A, \mathcal{I}_B)$ or $C_q$. 

If we allow Hilbert spaces $\mathcal{H}_A $ and $\mathcal{H}_B$ to be infinite-dimensional, we have another set of correlation matrices denoted by $C_{qs}$.
If we take finite-dimensional correlations to the limit, then the closure of $C_q$ constitutes a new set of correlation matrices denoted by $C_{qa}$. It is known that $C_{qs} \subseteq C_{qa}$ and $C_{qa}$ is also the closure of $C_{qs}$~\cite{scholz2008tsirelson}. 

It is easy to see that $C_c \subseteq C_{q} \subseteq C_{qs} \subseteq C_{qa}$. 
Bell's theorem~\cite{bell1964einstein} states that $C_c \neq C_q$. 
Slofstra~\cite{slofstra2019set} suggests that $C_q$ and $C_{qs}$ are not closed; that is, $C_q \neq C_{qa}$ and $C_{qs} \neq C_{qa}$. 

We can even drop the restriction on tensor product structures and define correlation matrices in terms of commuting operators. Then 
\begin{equation}
p(a, b| x, y) = \bra{\psi} M_a^x N_b^y \ket{\psi} \text{for all} (a, b, x, y) \in \mathcal{O}_A \times \mathcal{O}_B \times \mathcal{I}_A \times \mathcal{I}_B
\end{equation}
for $M_a^x N_b^y = N_b^y M_a^x$ with $\{ M_a^x \}_{a \in \mathcal{O}_A}$ on $\mathcal{H}$ for every $x \in \mathcal{I}_A$, and projective measurements $\{ N_b^y \}_{b \in \mathcal{O}_B}$ on $\mathcal{H}$ for every $y \in \mathcal{I}_B$. This set of correlation matrices is denoted by $C_{qc}$. To determine whether $C_{qc}$ is equal to $C_q$, $C_{qs}$ or $C_{qa}$ is known as Tsirelson's problem~\cite{tsirelson2006bell, dykema2016synchronous}. It is proven that $C_{qs} \neq C_{qc}$~\cite{slofstra2019tsirelson}. A recent result further solves the problem and concludes that $C_{qa} \neq C_{qc}$~\cite{ji2020mip}.

\subsection{Multipartite quantum correlations}

As a direct generalisation of bipartite separability, full separability~\cite{horodecki2009quantum} is defined as $n$-separability of $n$ systems $A_1 \dots A_n$: $\rho_{A_1 \dots A_n} = \sum_{i=1}^k p_i \rho_{A_1}^i \otimes \cdots \otimes \rho_{A_n}^i$ where $k \leq \text{dim} \mathcal{H}_{A_1 \dots A_n}^2$ is known as the Caratheodory bound. 
Multipartite quantum correlations are also defined in terms of subsystems and partitions. 
Consider a quantum state $\ket{\psi}$ of $n$ subsystems. If it can be written as the tensor product of $m$ disjoint subsets $\ket{\psi} = \bigotimes_{i=1}^m \ket{\psi_i}$, then it is said to be $m$-separable $(2 \leq m \leq n)$. $\ket{\psi}$ is said to be $k$-producible if the largest subset for $\ket{\psi_i}$ has at most $k$ subsystems. For a mixed state $\rho$, it is $m$-separable or $k$-producible if it has a decomposition of $m$-separable or $k$-producible pure states~\cite{lu2018entanglement}. 
In particular, the state $\rho_{A_1 \dots A_m}$ is semiseparable if and only if it is separable under all $1 - (m-1)$ partitions: $\{I_1 = \{k\}, I_2 = \{1, \dots, k-1, k+1, \dots, m\}\}, 1 \leq k \leq m$. 
Multipartite quantum correlations have much more rich structures and a full characterisation for multipartite quantum correlations remains as an important open problem. 


\section{Quantum correlations in time}
In this section we introduce quantum correlations defined in quantum field theory and explore further possibilities for correlations which are defined in an even-handed manner for space and time. We also aim towards a unified approach for quantum correlations in space and time. 

\subsection{Correlations in quantum field theory}
In quantum field theory~\cite{peskin2018introduction}, the $n$-point correlation function for the field operator $\phi(x)$ is usually defined in the ground state $\ket{\Omega}$ as 
\begin{equation}
C_n(x_1, x_2, \dots, x_n) = \bra{\Omega} \mathcal{T} \phi(x_1) \phi(x_2) \cdots \phi(x_n) \ket{\Omega}
\end{equation}
where $\mathcal{T}O_1(t_1) O_2(t_2) = \theta(t_1 -t_2)O_1(t_1) O_2(t_2) + \theta(t_2 -t_1)O_2(t_2) O_1(t_1)$ is the time-ordering operator. For example, consider a perturbation for interacting fields with the Hamiltonian divided by $H = H_0 + H_{int}$. With the unitary operator $U(t, t_0) = e^{iH_0(t-t_0)}e^{-iH(t-t_0)}$, the Schr\"odinger equation is written equivalently as $i\frac{\partial }{\partial t}U(t, t_0) = H_I(t) U(t, t_0)$ where $H_I(t) = e^{iH_0(t-t_0)}H_{int} e^{-iH_0(t-t_0)}$. Then the two-point correlation function is given as 
\begin{equation}
\bra{\Omega} \mathcal{T}\{ \phi(x) \phi(y)\} \ket{\Omega} = \lim_{T\rightarrow \infty(1-i\epsilon)}\frac{\bra{0}\mathcal{T}\{\phi_I(x)\phi_I(y)\exp[-i\int_{-T}^T \mathrm{d}t H_I(t)]\}\ket{0}} {\bra{0}\mathcal{T}\{ \exp[-i\int_{-T}^T \mathrm{d}t H_I(t)] \}\ket{0} },
\end{equation}
where $\phi_I(x)$ is defined through $\phi(x) = U^{\dag}(t, t_0) \phi_I(x) U(t, t_0)$.

\subsection{Further possibility for temporal correlations}
As we can see from statistical mechanics and field theory, correlations are defined in terms of thermal states or ground states for the background on spacetime. Here we are thinking of possibilities of generalising temporal correlations beyond thermal states or ground states. 

One possibility comes from autocorrelation functions. In statistics, the autocorrelation of a real or complex random process $\{X(t)\}$ is defined as the expectation value of the product of the values at two different times~\cite{park2018fundamentals}:
\begin{equation}
r_{XX}(t_1, t_2) = E[X(t_1)X^*(t_2)]. 
\end{equation}
Here the complex conjugate guarantees the product to be the square of the magnitude of the second momentum for $X(t)$ when $t_1 = t_2$. It is possible to take the expectation values of the product of measurement results for observables at different times to gain the temporal correlations. 

Another choice may be to define quantum states in time. Quantum states are defined across the whole of space but at one instant of time. We associate a Hilbert space for each spatially separated system and assign the tensor product structure; it is possible to associate a Hilbert space for each time and define quantum states in time. Then we may adopt the usual rule for calculating spatial correlations to analyse temporal correlations. 

\subsection{Towards a unified approach for quantum correlations in space and time}
In the previous subsection, we have discussed further possibilities of temporal correlations; here we are looking for a unified approach for quantum correlations in space and time. Following the discussion on quantum states in time, we may define quantum states across spacetime. We assume that the tensor product structure should work for Hilbert spaces at different times. Based on the Hilbert spaces across spacetime, we may define spacetime quantum states and unify temporal correlations and spatial correlations in the spacetime framework. This proposal has already been achieved in the pseudo-density matrix formalism as we are about to introduce in the next section. 

\section{Pseudo-density matrix formalism}

In this section, we introduce the pseudo-density matrix formalism~\cite{fitzsimons2015quantum, zhao2018geometry, pisarczyk2018causal, zhang2019constructing, zhang2019pseudo, zhang2019long} as a unified approach for quantum correlations in space and time. We review the definition of pseudo-density matrices for finite dimensions, present their properties, take bipartite correlations as an example to illustrate how the formalism unifies correlations in space-time. 

\subsection{Definition and properties}

The pseudo-density matrix formulation is a finite-dimensional quantum-mechanical formalism which aims to treat space and time on an equal footing. 
In general, this formulation defines an event via making a measurement in space-time and is built upon correlations from measurement results; thus, it treats temporal correlations just as spatial correlations and unifies spatio-temporal correlations. 
As a price to pay, pseudo-density matrices may not be positive semi-definite. 

An $n$-qubit density matrix can be expanded by Pauli operators in terms of Pauli correlations which are the expectation values of these Pauli operators. In spacetime, instead of considering $n$ qubits, let us pick up $n$ events, where a single-qubit Pauli operator is measured for each. Then, the pseudo-density matrix is defined as 
\begin{equation}\label{pdm}
\hat{R} \equiv \frac{1}{2^n} \sum_{i_1=0}^{3}...\sum_{i_n=0}^{3} \langle \{\sigma_{i_j}\}_{j=1}^{n} \rangle \bigotimes_{j=1}^n \sigma_{i_j},
\end{equation}
where $\langle \{\sigma_{i_j}\}_{j=1}^{n} \rangle$ is the expectation value of the product of these measurement results for a particular choice of events with operators $\{\sigma_{i_j}\}_{j=1}^{n}$.

Similar to a density matrix, it is Hermitian and unit-trace, but not positive semi-definite as we mentioned before. If the measurements are space-like separated or local systems evolve independently, the pseudo-density matrix will reduce to a standard density matrix. Otherwise, for example if measurements are made in time, the pseudo-density matrix may have a negative eigenvalue. For example, we take a single qubit in the state $\ket{0}$ at the initial time and assume the identity evolution between two times. The correlations are 1 for $\langle \{I, I\} \rangle$, $\langle \{X, X\} \rangle$, $\langle \{Y, Y\} \rangle$, $\langle \{Z, Z\} \rangle$, $\langle \{Z, I\} \rangle$, and $\langle \{I, Z\} \rangle$ while all others are given as 0. Then we construct the pseudo-density matrix for two times as
\begin{equation}
R =
\begin{bmatrix}
1 & 0 & 0  & 0\\
0 & 0 & \frac{1}{2} & 0\\
0 & \frac{1}{2} & 0 & 0\\
0 & 0 & 0 & 0
\end{bmatrix},
\end{equation}
with eigenvalues $\{-\frac{1}{2}, 0, \frac{1}{2}, 1\}$. 
Thus it is not positive semi-definite and encodes temporal correlations as a spacetime density matrix. 

Furthermore, the single-time marginal of the pseudo-density matrix is given as the density matrix at that particular time under the partial trace. For any set of operators $O_i$ with eigenvalues $\pm 1$, the expectation values of the measurement outcomes is given as
\begin{equation}\label{pmone}
\langle \{ O_i \}_{i=1}^m \rangle = \Tr \left[ \left( \bigotimes_{i=1}^m O_i \right) R \right].
\end{equation}
Here $O_i$ may be an operator measured on several qubits at the same time. This suggests that any complete basis of operators with eigenvalues $\pm 1$ has the proper operational meaning for correlations of operators, and thus serves as a good alternative basis for pseudo-density matrices. Note that pseudo-density matrices are defined in an operational manner via the measurements of correlations; a strict mathematical characterisation does not exist yet. A full investigation on the all possible basis choices remains an open problem. In the following chapters, we will present several generalisations of pseudo-density matrices with different measurement basis. 

To understand causal relationships, a measure for causal correlations called causality monotone is proposed, similar to the entanglement monotone. This causality monotone $f(R)$ is defined when it satisfies the following criteria: 

(1) $f(R) \geq 0$. In particular, $f(R) = 0$ if $R$ is positive semi-definite; $f(R) = 1$ for a single-qubit closed system at two times. 

(2) $f(R)$ is invariant under local unitary operations (thus under a local change of basis). 

(3) $f(R)$ is non-increasing under local operations. 

(4) $\sum_i p_i f(R_i) \geq f(\sum_i p_i R_i)$. 

\subsection{Characterisation of bipartite correlations in space-time}

\begin{figure}[h]
	\centering
	\includegraphics[scale=0.38]{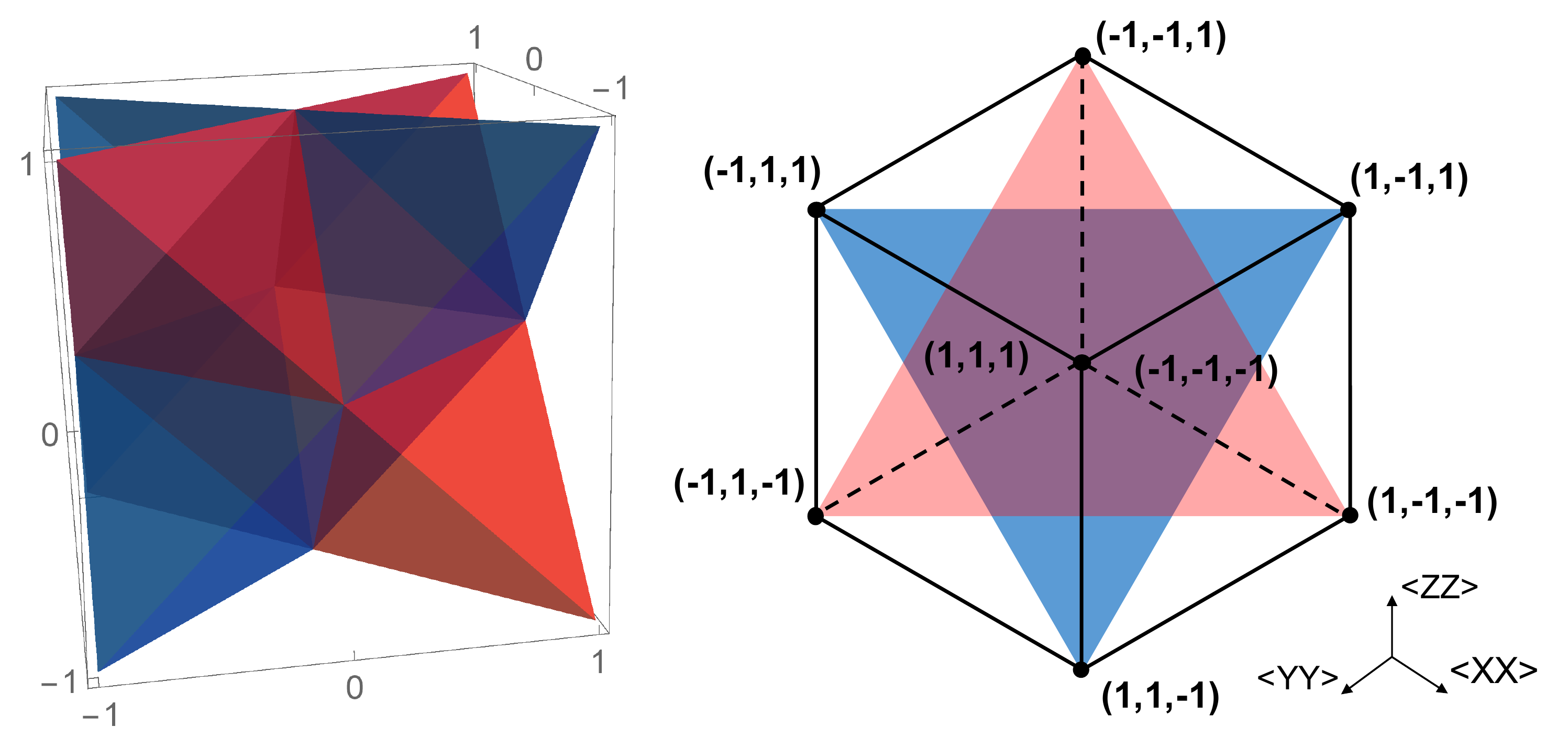}
	\caption{Geometrical representation for bipartite correlations in space and time. The left figure represents the spatial and temporal correlations in the blue and red tetrahedrons, respectively, in 3D modelling of the correlation space $\{\langle \sigma_i\sigma_i \rangle\}_{i=1}^3$. The right figure views the correlation from the $(-1, -1, -1) - (1,1,1)$ direction. The intersection of the spatial and temporal correlations is given by the purple octahedron representing separable states. Thanks to Zhikuan Zhao for providing his original figure in Ref.~\cite{zhao2018geometry}.}
	\label{fig: corrst}
\end{figure}

In this subsection we introduce the work on the characterisation for bipartite correlations in space-time~\cite{zhao2018geometry}. Here the spatial correlations are given by all possible two-qubit density matrices, and compared with temporal correlations in a single-qubit pseudo-density matrix at two times under the unitary evolution. 
There is a reflection between spatial correlations and temporal correlations in the $\langle XX \rangle - \langle ZZ \rangle$ plane of the correlation space $\{\langle \sigma_i\sigma_i \rangle\}_{i=1}^3$. 
The spatial correlations given in terms of Pauli measurements are characterised in Ref.~\cite{horodecki1996information}. The two-point correlations $t_{mn} = \Tr(\rho \sigma_m \otimes \sigma_n)$ form a real matrix $T$. Up to a unitary rotation, $t_{mn}$ is full characterised by its diagonal terms $t_{11}$, $t_{22}$, $t_{33}$. For any two-qubit density matrix $\rho$, the $T$ matrix belongs to the tetrahedron $\mathcal{T}_s$ with vertices $\mathbf{t} = (t_{11}, t_{22}, t_{33})$ given as $(-1, -1, -1)$, $(-1, 1, 1)$, $(1, -1, 1)$, $(1, 1, -1)$. These four vertices correspond to four Bell states.
Now we consider its temporal analog, a pseudo-density matrix for a single qubit at two times under the unitary evolution. This $T$ matrix is represented in another tetrahedron $\mathcal{T}_t$ with vertices $\mathbf{t} = (t_{11}, t_{22}, t_{33})$ given as $(1, 1, 1)$, $(-1, -1, 1)$, $(-1, 1, -1)$, $(1, -1, -1)$. 
Fig.~\ref{fig: corrst} illustrates these relations. On the left, blue and red tetrahedrons $\mathcal{T}_s$ and $\mathcal{T}_t$ show all possible bipartite spatial and temporal correlations. The right figure view these correlations from the $(-1, -1, -1) - (1,1,1)$ direction. It is easy to see that the intersection of the spatial and temporal correlations is given by the purple octahedron representing separable states.
Similarly, temporal correlations of the single-qubit initial state $\frac{I}{2}$ under arbitrary CPTP maps can also be mapped back to bipartite spatial correlations under the partial transpose and given as Fig.~\ref{fig: corrst}. We will see the importance of partial transposition in the continuous-variable generalisation as well. In general, for all possible quantum channel evolution, the set of temporal correlations strictly contains $\mathcal{T}_s$ and is convex on each edge; that is, in the bipartite case the set of all possible temporal correlations is  larger than the set of all possible spatial correlations (entanglement).

\chapter{\label{ch:3-cv}Generalisation of pseudo-density matrix formulation}

\clearpage

\minitoc

\clearpage

\section{Introduction}

In this chapter we follow the paradigm of the pseudo-density matrix~\cite{fitzsimons2015quantum}, which is understood as a particular spacetime state. 
The pseudo-density matrix uses only a single Hilbert space for each spacetime event defined in terms of making measurements in spacetime; 
as a price to pay, it may not be positive semi-definite. 
We take the view from Wigner that ``the function of quantum mechanics is to give statistical correlations between the outcomes of successive observations~\cite{wigner1973epistemological},'' 
and then construct the spacetime states in continuous variables from the observation of measurements of modes and generalise the pseudo-density matrix formulation.
We give six possible definitions for spacetime density matrices in continuous variables or spacetime Wigner functions built upon measurement correlations. 
The choice of measurements to make is a major issue here. They should form a complete basis to extract full information of states in spacetime. 
One natural choice is the quadratures, which turn out to be efficient in analysing Gaussian states. Analogous to the Pauli operators as the basis for a multi-qubit system, another option in continuous variables would be the displacement operators; however, they are anti-Hermitian. Instead, we apply their Fourier transform $T(\alpha)$, twice of displaced parity operators, to the representation of general Wigner functions. We also initialise the discussion of defining spacetime states from position measurements and weak measurements based on previous work on successive measurements~\cite{caves1986quantum1, caves1987quantum2, caves1987quantum3, barchielli1982model}, motivated by linking the pseudo-density matrix formalism to the path integral formalism. We further show that these definitions for continuous variables satisfy natural desiderata, such as those listed in Ref.~\cite{horsman2017can} for quantum joint states over time, as well as additional criteria for spacetime states. An experimental proposal for tomography is presented as well to show how these definitions are operationally meaningful. 

This chapter is based on Ref.~\cite{zhang2019constructing}. It proceeds as follows. First we define spacetime Gaussian states via the characterisation of the first two statistical moments and show that the temporal statistics are different but related to the spatial statistics. Next we define the spacetime Wigner function representation and the corresponding spacetime density matrix, and desirable properties are satisfied analogous to the spatial case. We further discuss the possibility of defining spacetime states via position measurements and weak measurements. A tomographical scheme is suggested for experiments. Then we comment on the pseudo-density matrix paradigm in terms of its properties and basic assumptions, and show its relation with the Choi-Jamiołkowski isomorphism and the path integral formalism. We also set up desirable properties for spacetime quantum states and check whether all the above definitions satisfy them or not. 

\section{Gaussian generalisation of pseudo-density matrix}

In this section we review Gaussian representation in continuous variables, define spacetime Gaussian states motivated from the pseudo-density matrix formalism, analyse simple examples and the differences and similarities of spatial and temporal Gaussian states. 

\subsection{Preliminaries}
Gaussian states are continuous-variable states with a representation in terms of Gaussian functions~\cite{weedbrook2012gaussian, wang2007quantum, adesso2014continuous}. 
The first two statistical moments of the quantum states, the mean value and the covariance matrix, fully characterise Gaussian states, just as normal Gaussian functions in statistics. 
The mean value $\bm{d}$, is defined as the expectation value of the $N$-mode quadrature field operators $\{\hat{q}_k, \hat{p}_k\}_{k=1}^N$ arranged in $\bm{\hat{x}} = (\hat{q}_1, \hat{p}_1, \cdots, \hat{q}_N, \hat{p}_N)^T$, that is, 
\begin{equation}
d_j= \langle \hat{x}_j \rangle _{\rho} \equiv \Tr (\hat{x}_j \hat{\rho}),
\end{equation}
for the Gaussian state $\hat{\rho}$.
The elements in the covariance matrix $\bm{\sigma}$ are defined as 
\begin{equation}
\sigma_{ij} = \langle \hat{x}_i \hat{x}_j + \hat{x}_j \hat{x}_i \rangle_{\rho} - 2 \langle \hat{x}_i \rangle_{\rho}\langle \hat{x}_j \rangle_{\rho}.
\end{equation}
The covariance matrix $\bm{\sigma}$ is real and symmetric, and satisfies the uncertainty principle~\cite{simon1994quantum} as (note that in this thesis we set $\hbar = 1$)
\begin{equation}\label{up}
\bm{\sigma} + i \bm{\Omega} \geq 0,
\end{equation}
in which the elements of $\bm{\Omega}$ is given by commutation relations as
\begin{equation}
[\hat{x}_i, \hat{x}_j] = i\hbar \Omega_{ij}, 
\end{equation}
thus $\bm{\Omega}$ is the $2N \times 2N$ matrix 
\begin{equation}
\bm{\Omega} \equiv \bigoplus_{k=1}^N \bm{\omega} = 
\begin{bmatrix}
\bm{\omega} & \  &\ \\
\ & \ddots & \ \\
\ & \ & \bm{\omega}
\end{bmatrix}
\ \ \text{and} \ \ 
\bm{\omega} = \begin{bmatrix}
0 & 1 \\
-1 & 0
\end{bmatrix}.
\end{equation}
This condition also implies the positive definiteness of $\bm{\sigma}$, i.e., $\bm{\sigma} > 0$.
Then we introduce the Wigner representation for Gaussian states. The Wigner function originally introduced in Ref.~\cite{wigner1932quantum} is a quasi-probability distribution in the phase space and the characteristic function can be given via the Fourier transform of the Wigner function.
By definition, the Wigner representation of a Gaussian state is Gaussian, that is, the characteristic function and the Wigner function~\cite{adesso2014continuous} are given by
\begin{align}
\chi(\bm{\xi}) & = \exp [ -\frac{1}{4} \bm{\xi}^T ( \bm{\Omega} \bm{\sigma} \bm{\Omega}^T ) \bm{\xi} - i(\bm{\Omega} \bm{d})^T \bm{\xi} ], \label{characteristic}\\
W(\bm{x}) & = \frac{ \exp [ -(\bm{x} - \bm{d})^T \bm{\sigma}^{-1} (\bm{x} - \bm{d}) ]}{\pi^N \sqrt{\det \bm{\sigma}}}, \label{wigner}
\end{align}
where $\bm{\xi}, \bm{x} \in \mathbb{R}^{2N}$.

Typical examples of Gaussian states include vacuum states, thermal states and two-mode squeezed states. A one-mode vacuum state $\ket{0}$ has zero mean values and the covariance matrix as the $2 \times 2$ identity matrix $I$. A one-mode thermal state with the mean number of photons $\bar{n}$~\cite{weedbrook2012gaussian} or the inverse temperature $\beta$~\cite{wang2007quantum} is defined equivalently as 
\begin{equation}\label{eqn: thermal}
\hat{\rho}^{th}(\bar{n}) = \sum_{n=0}^{+\infty} \frac{\bar{n}^n}{(\bar{n}+1)^{n+1}} \ket{n}\bra{n},
\end{equation}
or 
\begin{equation}
\hat{\rho}^{th}(\beta) = (1 - e^{-\beta})\exp(-\beta \hat{a}^{\dag}\hat{a}),
\end{equation}
where $\hat{a}, \hat{a}^{\dag}$ are annihilation and creation operators. Note that $\beta = -\ln \frac{\bar{n}}{1+\bar{n}}$. 
The thermal state has zero mean values and the covariance matrix proportional to the identity as $(2\bar{n} + 1)I$ or $\frac{1+e^{-\beta}}{1-e^{-\beta}} I$, respectively to the above two definitions. 
A two-mode squeezed state~\cite{wang2007quantum} is generated from the vacuum state $\ket{0}$ by acting with a two-mode squeezing operator which is defined as 
\begin{equation}
\hat{S}_2(\xi) = \exp[\xi\hat{a}^{\dag}\hat{b}^{\dag} - \xi^*\hat{a}\hat{b}],
\end{equation}
where $\hat{a}^{\dag}$ and $\hat{b}^{\dag}$ ($\hat{a}$ and $\hat{b}$) are creation (annihilation) operators of the two modes, $\xi$ is a complex number where $r=|\xi|$ and $\xi = r e^{i\psi}$.
Then the two-mode squeezed vacuum state is given as $\hat{S}_2(\xi) \ket{00}$. 
From here we omit the phase $\psi$ for simplicity. A two-mode squeezed state with a real squeezed parameter $r$, known as the Einstein-Podolsky-Rosen (EPR) state $\hat{\rho}^{epr}(r) = \hat{S}_2(r)\ket{00}\bra{00}\hat{S}_2^{\dag}(r)$, has zero mean values and the covariance matrix as
\begin{equation}
\bm{\sigma}_{tmss} = 
\begin{bmatrix}
\cosh 2r & 0 & \sinh 2r  & 0\\
0 & \cosh 2r & 0 & -\sinh 2r \\
\sinh 2r  & 0 & \cosh 2r & 0\\
0 & -\sinh 2r  & 0 & \cosh 2r
\end{bmatrix}.
\end{equation}
Taking the partial trace of the two-mode squeezed state, we get the one-mode thermal state: 
$\Tr_b [\hat{\rho}^{epr}(r)] = \hat{\rho}_a^{th}(\bar{n}) = \hat{\rho}_a^{th}(\beta),$
where $\bar{n} = \sinh^2 r$ or $\beta = -\ln \tanh^2 r$~\cite{wang2007quantum}.

\subsection{Spacetime Gaussian states}
Instead of Gaussian states at a specific time as given before, now we define Gaussian states in spacetime. 
Suppose that we are given data associated with single-mode measurements labelled by some index $k=1 , \dots, N$. 
We will use the same recipe, given the data, to create the spacetime state, whether these measurements are made on the same mode at different times or whether they are made on separate modes, or more generally on both different modes and different times. 
This follows the pseudo-density matrix paradigm, in which one wishes to use the same quantum density matrix formalism for all the cases. 

Assume that we are given enough data to characterise a Gaussian state fully, i.e., the mean value and the covariance matrix. 
The expectation values of all quadratures are defined as before. 
The correlation $\langle \{\hat{x}_i, \hat{x}_j\}\rangle$ of two quadratures $\hat{x}_i$ and $\hat{x}_j$ for two events is defined to be the expectation value for the product of measurement results on these quadratures. 
Particularly for measurements or events at the same time, this correlation is defined via a symmetric ordering of two quadrature operators. 
Then the covariance is defined to be related to this correlation and corresponding mean values as the spatial covariance. 


\begin{definition}
We define the Gaussian spacetime state in terms of measurement statistics as being 
(i) a vector $\bm{d}$ of 2N mean values, with j-th entry
\begin{equation}\label{defmv}
d_j= \langle \hat{x}_j \rangle _{\rho}= \Tr (\hat{x}_j \rho).
\end{equation} and (ii) a covariance matrix $\bm{\sigma}$ with entries as
\begin{equation}\label{defcm}
\sigma_{ij} = 2 \langle\{\hat{x}_i, \hat{x}_j\}\rangle_{\rho} - 2 \langle \hat{x}_i \rangle_{\rho} \langle \hat{x}_j \rangle_{\rho}
\end{equation}
where $\langle \{\hat{x}_i, \hat{x}_j\}\rangle_{\rho}$ is the expectation value for the product of measurement results; specifically $\{\hat{x}_i, \hat{x}_j\} = \frac{1}{2} (\hat{x}_i \hat{x}_j + \hat{x}_j \hat{x}_i )$ for measurements at the same time. 
To get the reduced state associated with the mode $k$ one picks out the entries in the $\bm{d}$ and $\bm{\sigma}$ associated with the mode $k$ to create the corresponding Gaussian state of that mode.  
\end{definition}

According to the above definition of reduced states, it is easy to see that the single time marginal is identical to the spatial Gaussian state at that particular time. This is because the mean values and covariances at one time in the spacetime case are defined as the same as them in the spatial case. 

\subsection{Example: vacuum state at two times}

For a simple example, we take a vacuum state at two times with the identity evolution in between. A vacuum state is $\ket{0}$ at the initial time $t_1$ and under the identity evolution it remains $\ket{0}$ at a later time $t_2$. 

Remember that a one-mode vacuum state $\ket{0}$ is a Gaussian state with zero means and the covariance matrix as the identity as stated before. That is, 
at a single time $t_1$ or $t_2$, 
\begin{align}
\langle \hat{q}_1 \rangle = \langle \hat{p}_1 \rangle = \langle \hat{q}_2 \rangle = \langle \hat{p}_2 \rangle= 0; \\
\langle \hat{q}_1\hat{q}_1 \rangle = \langle \hat{p}_1\hat{p}_1 \rangle = \langle \hat{q}_2\hat{q}_2 \rangle = \langle \hat{p}_2\hat{p}_2 \rangle = \frac{1}{2}, \nonumber \\
\langle \hat{q}_1\hat{p}_1 + \hat{p}_1\hat{q}_1 \rangle = \langle \hat{q}_2\hat{p}_2 + \hat{p}_2\hat{q}_2 \rangle = 0.
\end{align}

For measurements at both time $t_1$ and time $t_2$,
\begin{align}
&\langle \{ \hat{q}_1, \hat{q}_2 \} \rangle = \langle \{\hat{q}_2, \hat{q}_1\} \rangle =  \iint \textrm{d}q_1 \textrm{d}q_2 q_1 q_2\Tr( \ket{q_1}\bra{q_1}\ket{0}\bra{0}) \Tr(\ket{q_2}\bra{q_2}\ket{q_1}\bra{q_1}) 
=  \langle \hat{q}_1\hat{q}_1 \rangle = \frac{1}{2}, \nonumber \\
&\langle \{ \hat{q}_1, \hat{p}_2 \} \rangle = \langle \{\hat{p}_2, \hat{q}_1\} \rangle = \iint \textrm{d}q_1 \textrm{d}p_2 q_1p_2\Tr( \ket{q_1}\bra{q_1}\ket{0}\bra{0}) \Tr(\ket{p_2}\bra{p_2}\ket{q_1}\bra{q_1}) = 0, \nonumber \\
&\langle \{ \hat{p}_1, \hat{p}_2 \} \rangle = \langle \{\hat{p}_2, \hat{p}_1\} \rangle = \iint \textrm{d}p_1 \textrm{d}p_2 p_1 p_2\Tr( \ket{p_1}\bra{p_1}\ket{0}\bra{0}) \Tr(\ket{p_2}\bra{p_2}\ket{p_1}\bra{p_1}) = \langle \hat{p}_1\hat{p}_1 \rangle = \frac{1}{2}, \nonumber \\
&\langle \{ \hat{p}_1, \hat{q}_2 \} \rangle = \langle \{\hat{q}_2, \hat{p}_1\} \rangle = \iint \textrm{d}p_1 \textrm{d}q_2 p_1 q_2\Tr( \ket{p_1}\bra{p_1}\ket{0}\bra{0}) \Tr(\ket{q_2}\bra{q_2}\ket{p_1}\bra{p_1}) = 0.
\end{align}

According to the definition given in Eqn.~(\ref{defmv}, \ref{defcm}), the mean values are 0 and the covariance matrix in time is 
\begin{equation}
\bm{\sigma}_{vs} = 
\begin{bmatrix}
1 & 0 & 1 & 0\\
0 & 1 & 0 & 1\\
1& 0 & 1 & 0\\
0 & 1 & 0 & 1
\end{bmatrix}.
\end{equation}

Note that $\bm{\sigma}_{vs}$ is not positive definite and violates the uncertainty principle of Eqn.~(\ref{up}). Thus it is an invalid \emph{spatial} covariance matrix. This illustrates how the covariance statistics for spatial and temporal matrices are different, just as bipartite Pauli correlations in spatial and temporal case are different~\cite{horodecki1996information, zhao2018geometry}, which makes the study of temporal statistics particularly interesting. 

Since the determinant of the covariance matrix is 0, it is impossible to get the inverse of the covariance matrix directly to obtain the temporal Wigner function from Eqn.~(\ref{wigner}). 
From the mean values and the covariance matrix, we gain the temporal characteristic function from Eqn.~(\ref{characteristic}) as
\begin{equation}
\chi (q_1, p_1, q_2, p_2) = \exp(- p_1^2 - 2p_1p_2 - p_2^2 - q_1^2 - 2q_1q_2 - q_2^2),
\end{equation}
Via the Fourier transform, the temporal Wigner function is given as 
\begin{equation}
\mathcal{W}(q_1, p_1, q_2, p_2) = \frac{1}{4\pi} \exp(-p_1^2/4 - q_1^2/4) \delta(- p_1 + p_2) \delta(- q_1 + q_2),
\end{equation}
It is easy to check that the temporal Wigner function is normalised to 1: 
\begin{equation}
\iiiint \mathcal{W}(q_1, p_1, q_2, p_2) \mathrm{d}q_1\mathrm{d}p_1\mathrm{d}q_2\mathrm{d}p_2 = 1.
\end{equation}
However, if we consider the condition that the Wigner function of a pure state is bounded by $\pm \frac{2}{h}$, then this temporal Wigner function is invalid. This may be taken as the temporal signature of the Wigner function.
%

\subsection{Spatial vs temporal Gaussian states}
Now compare spatial Gaussian states and temporal Gaussian states via a simple two-mode example. 
In general, there is not much meaning to comparing an arbitrary spatial state with an arbitrary temporal state.
We need to pick up the spatial state carefully and figure out its temporal analog. 
Remember in the preliminaries we mentioned that taking the partial transpose of a two-mode squeezed state (or to say, the EPR state), we gain a one-mode thermal state. Hence, the temporal analog of the two-mode squeezed state will be the one-mode thermal state at two times. 
Take the one-mode thermal state as the initial state at $t_A$ and further assume that the evolution between $t_A$ and $t_B$ corresponds to the identity operator. 
The mean values are zero. 
The covariance matrix in time becomes
\begin{equation}
\bm{\sigma}_{omts} = 
\begin{bmatrix}
\cosh 2r & 0 & \cosh 2r  & 0\\
0 & \cosh 2r & 0 & \cosh 2r \\
\cosh 2r  & 0 & \cosh 2r & 0\\
0 & \cosh 2r  & 0 & \cosh 2r
\end{bmatrix}.
\end{equation}
Note that again $\bm{\sigma}_{omts}$ is not positive definite and violates the uncertainty principle. 

Compare $\bm{\sigma}_{omts}$ with its spatial analog, the covariance matrix of the two-mode squeezed state $\bm{\sigma}_{tmss}$. 
Under the high temperature approximation as $\beta \rightarrow 0$, $\tanh r \approx 1$ and $\sinh 2r \approx \cosh 2r$. Since $\hat{q} = \frac{1}{\sqrt{2}}(\hat{a} + \hat{a}^{\dag})$ and $\hat{p} = \frac{i}{\sqrt{2}}(\hat{a}^{\dag} - \hat{a})$, it follows that $\hat{q}^{T} = \hat{q}$ and $\hat{p}^{T} = -\hat{p}$. If we take the partial transpose on the first mode, only $\sigma_{24} = \sigma_{42}$ related to measurements $\hat{p}_1$, $\hat{p}_2$ change the sign. Note that $\sigma_{23} = \sigma_{32}$ related to measurements $\hat{p}_1$, $\hat{q}_2$ remain 0. Then the temporal covariance matrix is equal to the spatial covariance matrix under the partial transpose and the high temperature approximation. This can be understood as a continuous-variable analogue on temporal and spatial correlations of bipartite pseudo-density matrices for the qubit case~\cite{zhao2018geometry}. Note that taking the partial trace of a two-qubit maximally entangled state $\frac{1}{2}\sum_{i, j = 0, 1}\ket{ii}\bra{jj}$ we get a one-qubit maximally mixed state $I$; the temporal analog of a two-qubit maximally entangled state $\frac{1}{2}\sum_{i, j = 0, 1}\ket{ii}\bra{jj}$ is the one-qubit maximally mixed state $I$ at two times under the identity evolution, that is represented by $\frac{1}{2}\sum_{i, j = 0, 1}\ket{ij}\bra{ji}$. They are invariant under the partial transpose as well. In the continuous variable context, the one-mode thermal state under the high temperature approximation is close to the maximally mixed state $I$. 
We will come back to this partial transpose again later via Choi-Jamiołkowski isomorphism. 

\section{Pseudo-density matrix formulation for general continuous variables}
Now we move on to define spacetime states for general continuous variables. 
We first define the spacetime Wigner function by generalising correlations to the spacetime domain, following the paradigm of pseudo-density matrices. Then demanding the one-to-one correspondence between a spacetime Wigner function and a spacetime density matrix, we gain the spacetime density matrix in continuous variables from the spacetime Wigner function. 
This spacetime density matrix in continuous variables can be regarded as the extension of the pseudo-density matrix to continuous variables. 
We further analyse the properties of this spacetime Wigner function based on the corresponding spacetime density matrix in continuous variables and rediscover the five properties of a uniquely-determined Wigner function. 

\subsection{Preliminaries}
The Wigner function is a convenient representation of non-relativistic quantum mechanics in continuous variables and is fully equivalent to the density matrix formalism. 
The one-to-one correspondence between the Wigner function and the density matrix~\cite{cahill1969ordered, cahill1969density} states that, 
\begin{align}
&\hat{\rho} = \int W(\alpha) T(\alpha) \pi^{-1} \textrm{d}^2\alpha, \label{wftodm} \\
&W(\alpha) = \Tr [\hat{\rho} T(\alpha)]. \label{dmtowf}
\end{align}
Here $T(\alpha)$ is defined as
\begin{equation}
T(\alpha) = \int D(\xi) \exp(\alpha\xi^* - \alpha^*\xi) \pi^{-1} \textrm{d}^2\xi,
\end{equation}
where $D(\xi)$ is the displacement operator defined as $D(\xi) = \exp(\xi \hat{a}^{\dag} - \xi^* \hat{a})$. It can be seen that $T(\alpha)$ is the complex Fourier transform of $D(\xi)$. 
Besides, $T(\alpha)$ can be reformulated as $T(\alpha) = 2U(\alpha)$ where $U(\alpha) = D(\alpha) (-1)^{\hat{a}^{\dag}\hat{a}}D^{\dag}(\alpha)$ is the displaced parity operator. 
$T(\alpha)$ is Hermitian, unitary, unit-trace, and an observable with eigenvalues $\pm 2$.

We can also see from Eqn.~(\ref{dmtowf}) that the Wigner function is the expectation value of $T(\alpha)$~\cite{royer1977wigner}. For an $n$-mode Wigner function, a straightforward generalisation is
\begin{equation}\label{nmwf}
W(\alpha_1, ..., \alpha_n) = \langle \bigotimes_{i=1}^n T(\alpha_i) \rangle,
\end{equation}
as Ref.~\cite{banaszek1998nonlocality} gives the two-mode version.

\subsection{Spacetime Wigner function}

Let us start to construct the Wigner function in spacetime. It seems a bit ambitious to merge position and momentum with time in a quasi-probability distribution at first sight, but we will see that it is possible to treat instances of time just as how we treat modes. 
Again we borrow the concept of events from the pseudo-density matrix in finite dimensions and consider $n$ events instead of $n$ modes. Notice that the only difference between a pseudo-density matrix and a standard density matrix in construction is the correlation measure. Here we change the correlation measures of an $n$-mode Wigner function given in Eqn.~(\ref{nmwf}) in a similar way.

\begin{definition}
Consider a set of events $\{E_1, E_2, ..., E_N\}$. At each event $E_i$, a measurement of $T(\alpha_i)$ operator on a single mode is made. 
Then for a particular choice of events with operators $\{ T(\alpha_i) \}_{i=1}^n$, the spacetime Wigner function is defined to be
\begin{equation}
\mathcal{W}(\alpha_1, ..., \alpha_n) = \langle \{ T(\alpha_i) \}_{i=1}^n \rangle,
\end{equation}
where $\langle \{ T(\alpha_i) \}_{i=1}^n \rangle$ is the expectation value of the product of the results of the measurements on these operators.
\end{definition}

For spatially separated events, the spacetime Wigner function reduces to the ordinary $n$-mode Wigner function, for the order of product and measurement does not matter and it remains the same after making a flip (remember that $n$-mode Wigner function is the expectation value of the measurement results of the tensor product of these operators). 
If the measurements are taken in time, then a temporal Wigner function is constructed under temporal correlations.  
Thus, it is a generalisation for the Wigner function to the spacetime domain.

It is easy to check that the spacetime Wigner function is real and normalised to 1. Since the measurement results of $T(\alpha_i) = 2U(\alpha_i)$ is $\pm 2$ (remember that $U(\alpha_i)$ is the displaced parity operator), the expectation value of the product of the measurement results is to make products of $\pm 2$ with certain probability distribution. Thus, $\mathcal{W}(\alpha_1, ..., \alpha_n)$ is real. 

For the normalisation, we give a proof for the bipartite case, i.e., 
\begin{equation}
\int W(\alpha, \beta) \pi^{-2}\textrm{d}^2\alpha \textrm{d}^2\beta = 1;
\end{equation} 
for $n$ events, it can be proven directly following the same logic.


As mentioned before, a bipartite spacetime Wigner function reduces to two-mode Wigner function for two spatially separated events. The normalisation obviously holds in this case.

For a spacetime Wigner function between two times $t_1$ and $t_2$, we assume the initial state $\hat{\rho}$ is arbitrary and the evolution between $t_1$ and $t_2$ is an arbitrary CPTP map from $\hat{\rho}$ to $\mathcal{E}(\hat{\rho})$. At the time $t_1$, we measure $T(\alpha)$. Note that $T(\alpha) = 2[ \Pi_2(\alpha) - \Pi_1(\alpha)]$ where $\Pi_2(\alpha) = \sum_{n=0}^{\infty} \ket{2n, \alpha}\bra{2n, \alpha}$ and $\Pi_1(\alpha) = \sum_{n=0}^{\infty} \ket{2n+1, \alpha}\bra{2n+1, \alpha}$. That is, we make projections $\Pi_1(\alpha)$ and $\Pi_2(\alpha)$ to the odd and even subspaces for the eigenvalues $-2$ and $+2$. According to the measurement postulation, we get the state $\hat{\rho}_1 = \Pi_i(\alpha) \hat{\rho} \Pi_i(\alpha) / \Tr[\Pi_i(\alpha) \hat{\rho} \Pi_i(\alpha)]$ with the probability $\Tr[\Pi_i(\alpha) \hat{\rho} \Pi_i(\alpha)]$ after making the measurement of $\Pi_i(\alpha)$ $(i = 1, 2)$. Note that projection operators $\Pi_i(\alpha) = \Pi_i^{\dag}(\alpha)$ and $\Pi_i^2(\alpha) = \Pi_i(\alpha)$.
Then from $t_1$ to $t_2$, $\hat{\rho}_1$ evolves to $\mathcal{E}(\hat{\rho}_1)$. At the time $t_2$, we measure $T(\beta)$. We make projections $\Pi_1(\beta)$ and $\Pi_2(\beta)$ for the eigenvalues $-2$ and $+2$ again. So the temporal Wigner function, or $\{T(\alpha), T(\beta)\}$ correlation, is given by
\begin{align}
& \mathcal{W}(\alpha, \beta) = \langle \{T(\alpha), T(\beta)\} \rangle \nonumber \\
= & 4 \sum_{i,j = 1,2} (-1)^{i+j} \Tr[ \Pi_i(\alpha) \hat{\rho}  \Pi_i(\alpha)] \Tr\left\{\Pi_j(\beta)\mathcal{E} \left[\frac{\Pi_i(\alpha) \hat{\rho} \Pi_i(\alpha)}{\Tr [\Pi_i(\alpha) \hat{\rho} \Pi_i(\alpha)] } \right] \Pi_j(\beta)\right\} \nonumber \\
= & 4 \sum_{i, j = 1,2} (-1)^{i + j} \Tr\{\Pi_j(\beta)\mathcal{E}[\Pi_i(\alpha) \hat{\rho} \Pi_i(\alpha)]\} \nonumber \\
= & 2 \sum_{i = 1,2} (-1)^{i} \Tr\{T(\beta)\mathcal{E}[\Pi_i(\alpha) \hat{\rho} \Pi_i(\alpha)]\}  
\end{align}


Now let us check the normalisation property. Note that $\int T(\beta) \pi^{-1}\textrm{d}^2\beta = \int T(\alpha) \pi^{-1}\textrm{d}^2\alpha= I$ and $\mathcal{E}$ is trace-preserving. Then we have
\begin{align}
& \iint \mathcal{W}(\alpha, \beta) \pi^{-2}\textrm{d}^2\alpha \textrm{d}^2\beta \nonumber \\
= & 2 \iint \sum_{i = 1,2} (-1)^{i} \Tr\{T(\beta)\mathcal{E}[\Pi_i(\alpha) \hat{\rho} \Pi_i(\alpha)]\}\pi^{-2}\textrm{d}^2\alpha \textrm{d}^2\beta  \nonumber \\
= & 2 \int \sum_{i = 1,2} (-1)^{i} \Tr\{\mathcal{E}[\Pi_i(\alpha) \hat{\rho} \Pi_i(\alpha)]\}\pi^{-1}\textrm{d}^2\alpha  \nonumber \\
= & 2 \int \sum_{i = 1,2} (-1)^{i} \Tr [\Pi_i(\alpha) \hat{\rho} \Pi_i(\alpha)] \pi^{-1}\textrm{d}^2\alpha  \nonumber \\
=  & \int \Tr [T(\alpha) \hat{\rho}]\pi^{-1}\textrm{d}^2\alpha  \nonumber \\
=  & 1.
\end{align}
Thus, the normalisation property holds.

\subsection{Spacetime density matrix in continuous variables}
Though it is not always convenient to use the density matrix formalism in continuous variables, we are still interested in the possible form of spacetime density matrices as it is the basic construction for states.
Remember that there is a one-to-one correspondence between the Wigner function and the density matrix.
Here we demand that a similar one-to-one correspondence holds for the spatio-temporal version. Then we can define a spacetime density matrix in continuous variables from the above spacetime Wigner function. 
\begin{definition}
A spacetime density matrix in continuous variables is defined as
\begin{equation}
\hat{R}  = \idotsint \mathcal{W}(\alpha_1, ..., \alpha_n) \bigotimes_{i=1}^n T(\alpha_i) \pi^{-n} \textrm{d}^2\alpha_1 \cdots \textrm{d}^2\alpha_n. \\
\end{equation}
\end{definition}

This follows the direction from a spacetime Wigner function to a spacetime density matrix in continuous variables just as Eqn.~(\ref{wftodm}). 
Analogous to Eqn.~(\ref{dmtowf}), the opposite direction from a spacetime density matrix in continuous variables to a spacetime Wigner function automatically holds: 
\begin{equation}\label{eqn: densitytowigner}
\mathcal{W}(\alpha_1, ..., \alpha_n) = \Tr \{[\bigotimes_{i=1}^n T(\alpha_i)] \hat{R}\} = \langle \{ T(\alpha_i) \}_{i=1}^n \rangle.
\end{equation}

Now we prove Eqn.~(\ref{eqn: densitytowigner}) as a transform from the spacetime density matrix in continuous variables to the spacetime Wigner function. 
Applying the definition of the spacetime density matrix in continuous variables to the middle hand side of Eqn.~(\ref{eqn: densitytowigner}), we get 
\begin{equation}
\Tr \left \{ \left[\bigotimes_{i=1}^n T(\alpha_i) \right] \hat{R} \right \} 
=  \Tr \bigg[ \idotsint \mathcal{W}(\beta_1, ..., \beta_n) \bigotimes_{i=1}^n T(\alpha_i)T(\beta_i) \pi^{-n} \textrm{d}^2\beta_1 \cdots \textrm{d}^2\beta_n\bigg].
\end{equation}

Note that 
\begin{equation}
T(\alpha)T(\beta) = 4 \exp[2(\alpha^*\beta - \alpha\beta^*)]D(2\alpha - 2\beta),
\end{equation}
\begin{equation}
\Tr D(\xi) = \pi \delta(\xi_I) \delta(\xi_R) = \pi \delta^{(2)}(\xi),
\end{equation}
and $\delta^{(2)}(2\xi) = \frac{1}{4}\delta^{(2)}(\xi)$.
\begin{align}
&\Tr \left\{ \left[\bigotimes_{i=1}^n T(\alpha_i)\right] \hat{R}\right\} \nonumber \\
= & \Tr \bigg \{  \idotsint \mathcal{W}(\beta_1, ..., \beta_n)  \bigotimes_{i=1}^n 4 \exp[2(\alpha_i^*\beta_i - \alpha_i\beta_i^*)] D(2\alpha_i - 2\beta_i) \pi^{-n} \textrm{d}^2\beta_1 \cdots \textrm{d}^2\beta_n \bigg\} \nonumber \\
= & \idotsint \mathcal{W}(\beta_1, ..., \beta_n)  \prod_{i=1}^n 4 \exp[2(\alpha_i^*\beta_i - \alpha_i\beta_i^*)] \delta^{(2)}(2\alpha_i - 2\beta_i) \textrm{d}^2\beta_1 \cdots \textrm{d}^2\beta_n \nonumber \\
= & \mathcal{W}(\alpha_1, ..., \alpha_n) \nonumber \\
= &  \langle \{ T(\alpha_i) \}_{i=1}^n \rangle.
\end{align}
Thus, Eqn.~(\ref{eqn: densitytowigner}) holds as 
$$\Tr \{ [\bigotimes_{i=1}^n T(\alpha_i) ] \hat{R}  \} =  \mathcal{W}(\alpha_1, ..., \alpha_n) =  \langle \{ T(\alpha_i) \}_{i=1}^n \rangle.$$

It is also convenient to define the spacetime density matrix in continuous variables directly from $T(\alpha)$ operators, without the introduction of a spacetime Wigner function. 
\begin{definition}
An equivalent definition of a spacetime density matrix in continuous variables is 
\begin{equation}
\hat{R} = \idotsint \langle \{ T(\alpha_i) \}_{i=1}^n \rangle \bigotimes_{i=1}^n T(\alpha_i) \pi^{-n} \textrm{d}^2\alpha_1 \cdots \textrm{d}^2\alpha_n.
\end{equation}
\end{definition}

If we compare this definition with the definition of the pseudo-density matrix in finite dimensions given as Eqn.~(\ref{pdm}) element by element, we will find a perfect analogue. This may suggest the possibility for a generalised continuous-variable version of pseudo-density matrices.


\subsection{Properties} 
Now we investigate the properties of the spacetime Wigner function and the spacetime density matrix for continuous variables.

It is easy to check the spacetime density matrix $\hat{R}$ is Hermitian and unit-trace. 
Since $T(\alpha_i)$ is Hermitian and $\mathcal{W}(\alpha_1, ..., \alpha_n)$ is real,  $\hat{R}$ is Hermitian. 
From the normalisation property of the spacetime Wigner function and the fact that $T(\alpha_i)$ has unit trace, we conclude that $\Tr \hat{R} = 1$. 

Analogous to the normal spatial Wigner function, we analyse the properties for the spacetime Wigner function. 
For example, the spacetime Wigner function can be used as a quasi-probability distribution in  calculating the expectation value of an operator from the spacetime density matrix. 
For an operator $\hat{A}$ in the Hilbert space $\mathcal{H}^{\otimes n}$,
\begin{equation}
\langle \hat{A} \rangle_R  = \Tr[\hat{R} \hat{A}]
= \iint \mathcal{W}(\alpha_1, ..., \alpha_n) A(\alpha_1, ..., \alpha_n) \pi^{-n} \textrm{d}^2\alpha_1 \cdots \textrm{d}^2\alpha_n,
\end{equation}
where
\begin{equation}
A(\alpha_1, ..., \alpha_n) = \Tr\{[\bigotimes_{i=1}^n T(\alpha_i)]\hat{A}\}.
\end{equation}

It is obvious that a spacetime Wigner function for a single event does not discriminate between space and time; that is, for a single event the spacetime Wigner function is the same as an ordinary one-mode Wigner function in space. From the following we consider a bipartite spacetime Wigner function and generalisation to arbitrary events is straightforward. 

The five properties to uniquely determine a two-mode Wigner function in Ref.~\cite{hillery1984distribution, o1981quantum} are:
(1) that it is given by a Hermitian form of the density matrix; 
(2) that the marginal distributions hold for $q$ and $p$ and it is normalised; 
(3) that it is Galilei covariant; 
(4) that it has corresponding transformations under space and time reflections; 
(5) that for two Wigner functions, their co-distribution is related to the corresponding density matrices. 
They all hold in a similar way for a bipartite spacetime Wigner function and the corresponding spacetime density matrix in continuous variables. 
For a bipartite spacetime Wigner function, the five properties are stated as follows:

\begin{property}
$\mathcal{W}(q_1, p_1, q_2, p_2)$ is given by a Hermitian form of the corresponding spacetime density matrix as 
\begin{equation}
\mathcal{W}(q_1, p_1, q_2, p_2) = \Tr[\hat{M}(q_1, p_1, q_2, p_2) \hat{R}]
\end{equation}
for 
\begin{equation}
\hat{M}(q_1, p_1, q_2, p_2) = \hat{M}^{\dag}(q_1, p_1, q_2, p_2).
\end{equation} Therefore, it is real.
\end{property}

\begin{property}
The marginal distributions of $q$ and $p$ as well as the normalisation property hold.
\begin{align}
\iint \textrm{d}p_1 \textrm{d}p_2  \mathcal{W}(q_1, p_1, q_2, p_2) = \bra{q_1, q_2}\hat{R}\ket{q_1, q_2}, \nonumber \\
\iint \textrm{d}q_1 \textrm{d}q_2  \mathcal{W}(q_1, p_1, q_2, p_2) = \bra{p_1, p_2}\hat{R}\ket{p_1, p_2}, \nonumber \\
\iiiint \textrm{d}q_1 \textrm{d}q_2 \textrm{d}p_1 \textrm{d}p_2 \mathcal{W}(q_1, p_1, q_2, p_2) = \Tr \hat{R} = 1.
\end{align}
\end{property}

\begin{property}
$\mathcal{W}(q_1, p_1, q_2, p_2)$ is Galilei covariant
\footnote{The original paper~\cite{hillery1984distribution} uses the word ``Galilei invariant''.}
, that is,
if $$\bra{q_1, q_2}\hat{R}\ket{q'_1, q'_2} \rightarrow \bra{q_1+a, q_2+b}\hat{R}\ket{q'_1+a, q'_2+b} $$ 
then 
$$\mathcal{W}(q_1, p_1, q_2, p_2) \rightarrow \mathcal{W}(q_1+a, p_1, q_2+b, p_2)$$ and if 
$$
\bra{q_1, q_2}\hat{R}\ket{q'_1, q'_2} \rightarrow \exp\{[ip'_1(-q_1+q'_1)+ip'_2(-q_2+q'_2)]/\hbar\}\bra{q_1, q_2}\hat{R}\ket{q'_1, q'_2},
$$
then $$\mathcal{W}(q_1, p_1, q_2, p_2) \rightarrow \mathcal{W}(q_1, p_1-p'_1, q_2, p_2-p'_2).$$
\end{property}

\begin{property}
$\mathcal{W}(q_1, p_1, q_2, p_2)$ has the following property under space and time reflections
\footnote{Again the original paper~\cite{hillery1984distribution} uses the word ``invariant under space and time reflections''.}
: if $$\bra{q_1, q_2}\hat{R}\ket{q'_1, q'_2} \rightarrow \bra{-q_1, -q_2}\hat{R}\ket{-q'_1, -q'_2}$$ then $$\mathcal{W}(q_1, p_1, q_2, p_2) \rightarrow \mathcal{W}(-q_1, -p_1, -q_2, -p_2)$$ and if $$\bra{q_1, q_2}\hat{R}\ket{q'_1, q'_2} \rightarrow \bra{q'_1, q'_2}\hat{R}\ket{q_1, q_2}$$ then $$\mathcal{W}(q_1, p_1, q_2, p_2) \rightarrow \mathcal{W}(q_1, -p_1, q_2, -p_2).$$
\end{property}

\begin{property}
Two spacetime Wigner functions are related to the two corresponding spacetime density matrices as 
\begin{equation}
\Tr(R_1R_2) = (2\pi\hbar) \iint \textrm{d}q \textrm{d}p \mathcal{W}_{R_1}(q, p) \mathcal{W}_{R_2}(q, p),
\end{equation}
for $\mathcal{W}_{R_1}(q, p)$ and $\mathcal{W}_{R_2}(q, p)$ are spacetime Wigner functions for spacetime density matrices in continuous variables $\hat{R}_1$ and $\hat{R}_2$ respectively.
\end{property}

All these six properties (five plus the previous one for the expectation value of an operator in this subsection) are proven in Appendix B.

\section{Generalised measurements for pseudo-density matrix}
Here we go beyond the pseudo-density matrix formulation, in the sense that we generalise spatial correlations to the spacetime domain. Nevertheless, we still follow the idea to build spacetime states upon measurements. We consider position measurements for a special diagonal case. To reduce the additional effects caused by measurement processes, we discuss weak measurements and construct spacetime states from them. Here the connection with path integral is more obvious.

\subsection{Position measurements}
Besides quadratures and $T(\alpha)$ operators, it is also possible to expand a continuous-variable density matrix in the position basis since it is an orthogonal and complete basis. Here we consider a special case which is the diagonal matrix for convenience. 

In principle, a density matrix in the continuous variables can be diagonalised in the position basis as 
\begin{equation}
\hat{\rho} = \int_{-\infty}^{\infty} \textrm{d}x \ p(x) \ket{x}\bra{x},
\end{equation}
where 
\begin{equation}
p(x) = \Tr[\ket{x}\bra{x} \hat{\rho}].
\end{equation}
In the standard theory of quantum mechanics, we assume that the measurement results are arbitrarily precise to get the probability density $p(x)$ with the state updated to $\ket{x}\bra{x}$ after the measurement of $\hat{x}$. It is hard to achieve in the actual setting and imprecise measurements will be employed in the following discussion.

Then we define the spacetime density matrix in exactly the same way with the probability density now in the spatio-temporal domain.
\begin{definition}
Consider a set of $N$ events labelled $\{E_1, \cdots, E_N\}$. At each event $E_i$, a measurement of the position operator $\hat{x}_i$ is made. For a particular choice of the event, for example, $\{E_i\}_{i=1}^n$, we can define the spacetime density matrix from the joint probability of all these measurements as 
\begin{equation}
\rho = \int_{-\infty}^{\infty}\cdots\int_{-\infty}^{\infty} \textrm{d}x_1\cdots\textrm{d}x_n p(x_1, \cdots, x_n) \ket{x_1}\bra{x_1} \otimes  \cdots \otimes \ket{x_n}\bra{x_n}.
\end{equation}
\end{definition}
The remaining problem is how to calculate the joint probability $p(x_1, \cdots, x_n)$. For spatially separated events, the problem reduces to results given by states in ordinary quantum mechanics. So we only need to consider how to formulate states in time. Successive position measurements have been discussed properly in the path integral formalism, effect and operation formalism and multi-time formalism~\cite{caves1986quantum1, caves1987quantum2}.

Based on the discussion in Ref.~\cite{caves1987quantum2}, 
we consider $n$ events of instantaneous measurements of $x(t)$ at times $t_1, \cdots, t_n$ ($t_1 < \cdots < t_n$). 
In reality, such a measurement cannot be arbitrarily precise; a conditional probability amplitude called resolution amplitude $\Upsilon(\bar{x} - x)$ is introduced for $\bar{x}$ as the measurement result with the initial position of the system at $x$. 
Denote the state of the system as $\ket{\psi(t)}$ with the wave function $\psi(x, t) = \bra{x} \psi(t)\rangle$.
For a meter prepared in the state $\ket{\Upsilon}$ with the wave function $\Upsilon(\bar{x}) = \bra{\bar{x}} \Upsilon \rangle$, the total system before the measurement will be $\ket{\Psi_i} = \ket{\Upsilon} \otimes \ket{\psi(t)}$ with the wave function $\bra{\bar{x}, x} \Psi_i \rangle = \Upsilon(\bar{x})\psi(x, t)$. 
Consider the interaction for the measurement process as $\hat{x}\hat{\bar{p}}$ at some particular time. 
The total system after the measurement will be 
$\ket{\Psi_f} = e^{-(i/\hbar)\hat{x}\hat{\bar{p}}} \ket{\Psi_i} = \int \textrm{d}x e^{-(i/\hbar)x\hat{\bar{p}}} \ket{\Upsilon} \otimes \ket{x} \psi(x, t)$, 
with the wave function 
$\bra{\bar{x}, x} \Psi_f\rangle = \Upsilon(\bar{x}-x)\psi(x, t) = \bra{x} \Upsilon(\bar{x} - \hat{x}) \ket{\psi(t)}$.
Following the calculation in Ref.~\cite{caves1987quantum2}, for the wave function of the system $\psi(x(t_1), t_1)$ at some initial time $t_1$, the joint probability for measurement results $(\bar{x}_1, \cdots, \bar{x}_n)$ is given by a path integral as 
\begin{equation}
p(\bar{x}_1, \cdots, \bar{x}_n) = \int_{t_1}^{t_n} \mathcal{D}x(t) \left[ \prod_{\nu=1}^n \Upsilon(\bar{x}_{\nu} - x(t_{\nu})) \right] e^{(i/\hbar) S[x(t)]} \psi(x(t_1), t_1),
\end{equation}
where 
\begin{equation}
\int_{t_1}^{t_n} \mathcal{D}x(t) = \lim_{N \rightarrow \infty} \left[ \prod_{k=1}^N \int_{-\infty}^{\infty} \textrm{d}x_k \right],
\end{equation}
with the insertion of $N-2$ times between the initial time $t_1$ and the final time $t_n = t_N$; and note that all the measurement times are included in the insertion. 
This integral sums over all path $x(t)$ from $x(t_1)$ to $x(t_n)$ with arbitrary initial values $x(t_1)$ and arbitrary final positions $x(t_n)$. Here
\begin{equation}
S[x(t)] = \int_{t_1}^{t_n} \textrm{d}t L(x, \dot{x}, t)
\end{equation}
is the action for the path $x(t)$ with the Lagrangian of the system as $L(x, \dot{x}, t)$.

Note that $p(\bar{x}_1, \cdots, \bar{x}_n)$ is normalised, i.e.,
\begin{equation}
\int_{-\infty}^{\infty}\cdots\int_{-\infty}^{\infty}  \textrm{d}\bar{x}_1 \cdots \textrm{d}\bar{x}_n p(\bar{x}_1, \cdots, \bar{x}_n) = 1;
\end{equation}
thus, the spacetime density matrix defined above has unit trace.

Here the diagonalised spacetime density matrix in the position basis is fully equivalent to the path integral formalism. Or we can take this definition as the transition from the path integral. 
Thus, this definition suggests a possible link between the pseudo-density matrix formulation and the path integral formalism.

\subsection{Weak measurements}

Weak measurements are the measurements that only slightly disturb the state, with POVM elements close to the identity. They are often continuous. It is particularly interesting here as weak measurements minimise the influence of measurements and maximally preserve the information of the original states. Via weak measurements, we do not need to worry about the change of marginal states at each time. 
There are several slightly different mathematical definitions for weak measurements. Here we follow the convention in the formulation of effects and operations~\cite{kraus1983states}. 

Recall that an effect $\hat{F}$ is defined as an operator which satisfies $\hat{F}^{\dag} = \hat{F}$ and $0 < \hat{F} < \mathbbm{1}$.
Similar to a projection, the probability of obtaining the result in the interval $I = (a, a+\Delta a)$ at time $t$ is writen as 
\begin{equation}
P(\rho | I, t) = \Tr\{ \hat{F}^{1/2}_H(I, t) \hat{\rho} \hat{F}^{1/2}_H(I, t)\},
\end{equation}
And the state  evolves to 
\begin{equation}
\rho' = \frac{\hat{F}^{1/2}_H(I, t) \hat{\rho} \hat{F}^{1/2}_H(I, t) }{\Tr\left\{ \hat{F}^{1/2}_H(I, t) \hat{\rho} \hat{F}^{1/2}_H(I, t) \right\}}
\end{equation}
Assume that the disturbance at time $t$ does not affect the discrimination for $I$ out of the whole range and the reduction postulate holds. At a later time $t'$, we have 
\begin{equation}
P(\rho' | I', t') = \Tr\left\{ \hat{F}^{1/2}_H(I', t')  \hat{\rho'} \hat{F}^{1/2}_H(I', t') \right\}.
\end{equation}
We have the joint probability as
\begin{equation}
P(\rho | I, t; I', t') = \Tr\left\{ \hat{F}^{1/2}_H(I', t') \hat{F}^{1/2}_H(I, t) \hat{\rho} \hat{F}^{1/2}_H(I, t) \hat{F}^{1/2}_H(I', t')\right\}.
\end{equation}
Consider the densities of effects
\begin{equation}
\textrm{d}\hat{F}(a) = \hat{f}(a) \textrm{d}\mu(a), \qquad \int_{-\infty}^{+\infty}  \textrm{d}\mu(a) \hat{f}(a) = \mathbbm{1},
\end{equation}
where $\textrm{d}\mu(a)$ is a measure for the function $\hat{f}(a)$. 
Then we have 
\begin{equation}
\hat{F}(\textrm{d}I, t; \textrm{d}I', t') = 	\hat{f}(a, t; a', t')\textrm{d}\mu(a)\textrm{d}\mu(a') = \hat{f}^{1/2}_H(a, t) \hat{f}_H(a', t') \hat{f}^{1/2}_H(a, t) \textrm{d}\mu(a)\textrm{d}\mu(a'),
\end{equation}
\begin{equation}
P(\rho | \textrm{d}I, t; \textrm{d}I', t') = p(\rho|a, t; a', t')\textrm{d}\mu(a)\textrm{d}\mu(a') = \Tr{\hat{f}(a, t; a', t')\hat{\rho}} \textrm{d}\mu(a)\textrm{d}\mu(a').
\end{equation}
In general, 
\begin{equation}
P(\rho | \textrm{d}I_1, t_1; \cdots; \textrm{d}I_n, t_n) = \Tr{\hat{f}(a_1, t_1; \cdots; a_n, t_n)\hat{\rho}} \textrm{d}\mu(a_1) \cdots \textrm{d}\mu(a_n),
\end{equation}
where
\begin{align}
\hat{f}(a_1, t_1; \cdots; a_{n-1}, t_{n-1}; a_n, t_n) =  \hat{f}^{1/2}_H&(a_1, t_1) \cdots \hat{f}^{1/2}_H(a_{n-1}, t_{n-1})\hat{f}_H(a_n, t_n) \nonumber \\
&\times \hat{f}^{1/2}_H(a_{n-1}, t_{n-1}) \cdots \hat{f}^{1/2}_H(a_1, t_1).
\end{align}

Now following the calculation in Ref.~\cite{barchielli1982model}, 
we can define a generalised observable corresponding to a simultaneous inaccurate measurement of position and momentum for a density matrix $\hat{\rho}$:
\begin{equation}
\hat{F}(T) = \int_T \frac{\textrm{d}x\textrm{d}p}{2\pi\hbar}	 \exp\left[\frac{i}{\hbar}(p\hat{q}-x\hat{p})\right] \hat{\rho} \exp\left[-\frac{i}{\hbar}(p\hat{q}-x\hat{p})\right].
\end{equation}
Take
\begin{equation}
\hat{\rho} = C \exp[-\alpha(\hat{q}^2 + \lambda \hat{p}^2)], \qquad \alpha, \lambda > 0,	
\end{equation}
where $C$ is some normalisation factor. 
We get the density of this generalised effect-valued measure as
\begin{equation}
\hat{f}(q, p) = C\exp\left[ -\alpha[(\hat{q}-q)^2 + \lambda (\hat{p}-p)^2] \right],
\end{equation}
where $\textrm{d}\hat{F}(q, p) = \hat{f}(q, p) \textrm{d}\mu(q, p)$.
We set
\begin{equation}
\alpha = \gamma \tau,
\end{equation}
where $\tau$ is the time interval between two subsequent measurements.
When $\alpha \rightarrow 0$, the measurement is continuous and we call it weak.
For an initial density matrix $\hat{\rho}$ at time $t = 0$, we make continuous measurements in time and find the probability density of obtaining measurement results $q, p$ at time $t = \tau$ is given by
\begin{equation}\label{stwffromweak}
p(q, p, \tau | \hat{\rho}) = \Tr \mathcal{F}(q ,p; \tau) \hat{\rho},
\end{equation}
where
\begin{align}\label{stweak}
\mathcal{F}(q ,p; \tau)  \hat{\rho} = & \int \textrm{d}\mu_G [q(t), p(t)] \delta \left( q - \frac{1}{\tau}\int_0^{\tau} \textrm{d}t q(t) \right) \delta \left( p - \frac{1}{\tau}\int_0^{\tau} \textrm{d}t p(t) \right) \exp[-\frac{i}{\hbar}\hat{H}\tau] \nonumber \\
& \mathcal{T} \exp \left[ -\frac{\gamma}{2} \int_{0}^{\tau} \textrm{d}t [(\hat{q}_H(t) - q(t))^2 + \lambda (\hat{p}_H(t) - p(t))^2 ] \right] \hat{\rho} \nonumber \\
& \mathcal{T}^* \exp \left[ -\frac{\gamma}{2} \int_{0}^{\tau} \textrm{d}t [(\hat{q}_H(t) - q(t))^2 + \lambda (\hat{p}_H(t) - p(t))^2 ] \right] \exp[\frac{i}{\hbar}\hat{H}\tau],
\end{align}
here
\begin{equation}
\textrm{d}\mu_G [q(t), p(t)] = \lim_{N \rightarrow \infty} \left( \frac{\gamma\tau\sqrt{\lambda}}{\pi N} \prod_{s=1}^{N} \textrm{d}q(t_s) \textrm{d}p(t_s) \right),
\end{equation}
and
\begin{align}
\hat{q}_H(t) = \exp\left[ \frac{i}{\hbar} \hat{H} t \right] \hat{q} \exp\left[ -\frac{i}{\hbar} \hat{H} t \right], \nonumber \\
\hat{p}_H(t) = \exp\left[ \frac{i}{\hbar} \hat{H} t \right] \hat{p} \exp\left[ -\frac{i}{\hbar} \hat{H} t \right].
\end{align}

\begin{definition}
A possible form for the temporal Wigner function $W (\bar{x}_1, \bar{p}_1, \bar{t}_1; \dots; \bar{x}_{\nu}, \bar{p}_{\nu}, \bar{t}_{\nu})$ is given by the probability density of simultaneous measurement results $\bar{x}_{i}, \bar{p}_{i}$ at the time $\bar{t}_{i}$ for $i = 1, \dots, \nu$ with $\hat{\rho}$ as the initial density matrix at the initial time $\bar{t}_1$ 
in Ref.~\cite{barchielli1982model}:
\begin{align}
&W (\bar{x}_1, \bar{p}_1, \bar{t}_1; \cdots; \bar{x}_{\nu}, \bar{p}_{\nu}, \bar{t}_{\nu}) \nonumber \\
= & \Tr \mathcal{F}(\bar{x}_{\nu}, \bar{p}_{\nu}; \bar{t}_{\nu}-\bar{t}_{\nu-1}) \mathcal{F}(\bar{x}_{\nu-1}, \bar{p}_{\nu-1}; \bar{t}_{\nu-1}-\bar{t}_{\nu-2}) \cdots  \mathcal{F}(\bar{x}_2, \bar{p}_2; \bar{t}_2 - \bar{t}_1) \mathcal{F}(\bar{x}_1, \bar{p}_1; 0)  \hat{\rho}.
\end{align}
\end{definition}
Here we employ the probability density in weak measurements to define a temporal Wigner function. 
This generalises the form of measurements to take. As shown in the next section, this temporal Wigner function turns out to be a desirable spacetime quantum state and expand the possibility for relating generalised measurement theory with spacetime. 
In general, a unified spacetime Wigner function defined from weak measurements is possible as well. For n-mode spatial Wigner function from weak measurements, it is defined as
\begin{equation}
W(q_1, p_1, \cdots, q_n, p_n) = \Tr \mathcal{F}(q_1, p_1; 0) \otimes \cdots \otimes \mathcal{F}(q_n, p_n ; 0) \hat{\rho}.
\end{equation}
Thus spacetime Wigner function is a mixture of product and tensor product of $\mathcal{F}$. We obtain the spacetime states from weak measurements. 
It follows the paradigm of pseudo-density matrix formalism that spacetime Wigner function is defined via measurement correlations. Specifically, we make simultaneous measurements of position and momentum; as a price to pay, we fixed the average positions and the average momentums for certain time periods. It is not the usual Wigner function but a generalised version in the average sense. 


\section{Experimental proposal for tomography}

Here we propose an experimental tomography for spacetime Gaussian states in quantum optics. Especially, we construct the temporal Gaussian states, in terms of measuring mean values and the temporal covariance matrix for two events in time. The covariance of quadratures are defined in terms of the correlation of quadratures and mean values. Thus, all we need to measure are mean values and correlations of quadratures.

With the balanced homodyne detection, we can measure the mean values of single quadratures $d_i = \langle x_i \rangle$, the correlation of the same quadrature $\langle x_i x_i \rangle$ (the diagonal terms of the covariance matrix), and the correlation of both position operators or both momentum operators at two times $\langle q_j q_k \rangle$ or $\langle p_j p_k \rangle$ ($j \neq k$ for this section). 
Mean values of single quadratures are measured by the balanced homodyne detection as usual.
For $\langle x_i x_i \rangle$, we can measure by almost the same method, only do an additional square for each measurement outcome of $\hat{x}_i$.
For $\langle q_j q_k \rangle$ or $\langle p_j p_k \rangle$, we record the homodyne results for a long time with small time steps and calculate the expectation values of the product the measurement results at two times to get the correlation.

It is a bit difficult to measure the correlation for a mixture of position and momentum operators. 
For such correlations at the same time $t_j$, the measurement of $q_j$ and $p_j$ cannot be precise due to the uncertainty principle.
An eight-port homodyne detector may be a suggestion; that is, we split the light into half and half by a 50/50 beam splitter, and measure each quadrature separately with a local oscillator which is split into two as well for homodyne detection. However, we cannot avoid the vacuum noise when we split the light and the local oscillator.
A better method for measuring $q_j$ and $p_j$ at time $t_j$ will be resort to quantum-dense metrology in Ref.~\cite{steinlechner2013quantum}.
For the correlation $\langle q_j p_k \rangle$, we use the same protocol as before. As the two-time correlation for the same quadrature, we record the homodyne results for a long time with small time steps and calculate the expectation values of the product of the measurement results at two times with a fixed time interval in between to get the correlation.


Then we gain all the correlations to construct the temporal covariance matrix. 
The corresponding temporal density matrix or temporal Wigner function is easily built with mean values and the temporal covariance matrix; thus, we achieve the experimental tomography.

\section{Comparison and comments}
The pseudo-density matrix for $n$ qubits is neatly defined and satisfies the properties listed in Ref.~\cite{horsman2017can}. These properties are: (1) that it is Hermitian; (2) that it represents probabilistic mixing; (3) that it has the right classical limit; (4) that it has the right single-time marginals; (5) for a single qubit evolving in time, composing different time steps is associative. 
For Gaussian spacetime states, the first four properties easily hold; for the fifth one, it remains true for the Gaussian evolution. 
For general continuous variables, except the one for single-time marginals, all the others hold. 
This property for single-time marginals is non-trivial. The correlation of a single Pauli operator for each single-time marginal is preserved after making the measurement of that Pauli operator. As each single-time marginal is just the spatial state at that time, the total correlation for all Pauli operators is independent of the measurement collapse. It is a perfect coincide.


The relation with the Choi-Jamiołkowski isomorphism is important in deriving the above properties. Consider a single qubit or mode evolving under a channel $\mathcal{E}_{B|A}$ from $t_A$ to $t_B$. Then define an operator $E_{B|A}$ as the Jamiołkowski isomorphism of $\mathcal{E}_{B|A}$:
\begin{equation}
E_{B|A} = (\mathcal{E}_{B|A} \otimes \mathcal{I}) (\ket{\Phi^+}\bra{\Phi^+}^{\Gamma})
\end{equation}
where $\ket{\Phi^+}$ is the unnormalised maximally entangled state on the double Hilbert space $\mathcal{H}_A \otimes \mathcal{H}_A$ at $t_A$ and $\Gamma$ denotes partial transpose. 
$\ket{\Phi^+} = \sum_{i = 0, 1} \ket{i} \otimes \ket{i}$ for the qubit case. 
$\ket{\Phi^+} = \sum_{n = 0}^{\infty} \ket{n, \alpha} \otimes \ket{n, \alpha}$ for continuous variables; in which $\ket{n, \alpha} = D(\alpha) \ket{n}$ with the displacement operator $D(\alpha)$ and the number eigenstates $\ket{n}$. 
Then the spacetime state in terms of pseudo-density matrix formulation is given as the Jordan product
\begin{equation}
R_{AB} = \frac{1}{2} \left[E_{B|A}( \rho_A \otimes I_B) + (\rho_A \otimes I_B) E_{B|A}\right].
\end{equation}
The qubit version is proved in Ref.~\cite{horsman2017can} and we can follow its argument for the continuous-variable version we defined above. 
It is particularly interesting when we consider temporal correlations for two times. The orders between $E_{B|A}$ and $\rho_A \otimes I_B$ automatically suggest a symmetrised order of operators in two-time correlations. 
For a special case that $\rho_A$ is maximally mixed as proportional to the identity $I$, $R_{AB} = E_{B|A}$. Consider the identity evolution $\mathcal{E}_{B|A}$ as $\mathcal{I}$, then $E_{B|A} =  \ket{\Phi^+}\bra{\Phi^+}^{\Gamma}$. The spatial and temporal analogue discussed in the Gaussian section is recovered by partial transpose again.

One thing of particular interest to look at in continuous variables is the relation between with the pseudo-density matrix formulation and the path integral formulation. In Ref.~\cite{zhang2019pseudo}, we establish the connection between pseudo-density matrix and decoherence functional in consistent histories. The only thing left unrelated in different spacetime approaches listed in the introduction is the path integral formulation. 
Here consider the propagator $\bra{y_2, t_2}\hat{U}\ket{y_1, t_1}$, or more specifically, the absolute square of this propagator as the probability for transforming $\ket{y_1}$ at $t_1$ to $\ket{y_2}$ at $t_2$. The initial state evolves under the unitary $\hat{U} = \exp(- \mathcal{T}\int_{t_1}^{t_2} i\hat{H}\textrm{d}t/\hbar)$. For the Gaussian case, $\ket{y_1}$ at the time $t_1$ and $\ket{y_2}$ at $t_2$ may be two eigenstates of $\hat{x}$ or $\hat{p}$ or a mixture of them over a period. For general continuous variables, they should be two eigenstates of $T(\alpha)$ and $T(\beta)$, that is, a mixture of $\ket{n, \alpha}$ and $\ket{m, \beta}$.Via this propagator, we can calculate the two-time correlation. It gives the same results as the pseudo-density matrix does, which suggests the two formulations may be equivalent.

Ref.~\cite{horsman2017can} suggests five criteria for a quantum state over time to satisfy as the analog of a quantum state over spatial separated systems. Here we also set up desirable properties of quantum states in the whole spacetime. The basic principle is that the statistics calculated using the spacetime state should be identical to those calculated using standard quantum theory. Note that Criterion 1, 2, 3, and 6 are adapted from Ref.~\cite{horsman2017can}. 

\begin{criterion}
A spacetime quantum state has a Hermitian form, that is, the spacetime density matrix is self-adjoint and the spacetime Wigner function is given by the expectation value of a Hermitian operator.
\end{criterion}
\begin{criterion}
The probability related to all the measurements at different spacetime events is normalised to one, that is, the spacetime density matrix is unit-trace and the spacetime Wigner function is normalised to one.
\end{criterion}
\begin{criterion}
A spacetime quantum state represents probabilistic mixing appropriately, that is, a spacetime state of different systems with a mixture of initial states is the corresponding mixture of spacetime states for each system, as well as the mixture of channel evolutions.
\end{criterion}
\begin{criterion}
A spacetime quantum state provides the right expectation values of operators. 
In particular, it gives the same expectation values of time-evolving operators as the Heisenberg picture does.
\end{criterion}
\begin{criterion}
A spacetime quantum state provides the right propagator/kernel which is the probability amplitude evolving from one time to another.
\end{criterion}
\begin{criterion}
A spacetime quantum state has the appropriate classical limit.
\end{criterion}

It is easy to check that the Gaussian characterisation satisfies Criterion 1, 2, 3, 5, 6 and the second half of Criterion 3; the first half of Criterion 3 does not hold since the mixture of Gaussian states is not necessarily Gaussian. 

For the Wigner function and corresponding density matrix representation, Criterion 1, 2, 3, 4, 6 hold. 
Criterion 5 remains to be further analysed. 

All of the Criteria 1-6 hold for position measurements and weak measurements, though the spacetime density matrix for position measurements assumes diagonalisation. It seems that the spacetime Wigner function from weak measurements is best-defined under these criteria. 

Note that we have considered whether the single time marginals of a spacetime quantum state reduce to the spatial state at that particular time. It unfortunately fails for Definition 2- 6 in general due to a property in the measurement theory which suggests the irreversibility of the time evolution in the repeated observations~\cite{barchielli1982model}; only the initial time marginal is reduced to the initial state. Thus, we prefer not to list it as one of the criteria.

\chapter{\label{ch:4-relation}Correlations from other spacetime formulations: relation and lesson}

\clearpage
\minitoc
\clearpage

\section{Introduction}


Now we have already generalised the pseudo-density matrix formalism to continuous variables and general measurement processes. There are several other approaches which also tends to treat space and time more equally but different from the pseudo-density matrix formalism. In this chapter, we identify the relationship among these spacetime approaches via quantum correlation in time~\cite{zhang2020quantum}. 

The problem of time~\cite{anderson2010problem} is especially notorious in quantum theory as time cannot be treated as an operator in contrast with space. 
Several attempts have been proposed to incorporate time into the quantum world in a more even-handed way to space, including: indefinite causal structures~\cite{chiribella2008quantum, chiribella2009theoretical, hardy2012operator, oreshkov2012quantum, pollock2018non, cotler2018superdensity}, consistent histories~\cite{griffiths1984consistent, griffiths2003consistent, gell2018quantum, gell1993classical, omnes1990hilbert}, generalised quantum games~\cite{buscemi2012all, rosset2018resource}, spatio-temporal correlation approches~\cite{maldacena2015bound, roberts2016chaos}, path integrals~\cite{feynman2010quantum, zinn2010path}, and pseudo-density matrices~\cite{fitzsimons2015quantum, zhao2018geometry, pisarczyk2019causal, zhang2020different}. 
Different approaches have their own advantages. Of particular interest here is the pseudo-density matrix approach for which one advantage is that quantum correlations in space and time are treated on an equal footing.
The present work is motivated by the need to understand how the different approaches connect via temporal correlations, so that ideas and results can be transferred more readily. 

We accordingly aim to identify mappings between these approaches and pseudo-density matrices. We ask what kind of relationship these space-time approaches hold in terms of temporal correlations. Are the allowed temporal correlations the same or different from each other? If the same, are they equal, or do they map with each other and what kind of mapping? If different, how different are they? 
More specifically, we take temporal correlations represented in different approaches and find that they are consistent with each other expect in the path integral formalism. 
Quantum correlations in time in these approaches are either exactly equal or operationally equivalent expect those used in the path integral formalism. 
By operational equivalence of two formalisms, we mean the correlations or the probabilities of possible measurement outcomes with given inputs in these two formalisms are equal. 
We find several mappings and relations between these approaches, including
(i) we map process matrices with indefinite causal order directly to pseudo-density matrices in three different ways; 
(ii) we show the diagonal terms of decoherence functionals in consistent histories are exactly the probabilities in temporal correlations of corresponding pseudo-density matrices; 
(iii) we show quantum-classical signalling games give the same probabilities as temporal correlations measured in pseudo-density matrices; 
(iv) the calculation of OTOCs reduces half numbers of steps by pseudo-density matrices; and 
(v) correlations in path integrals are defined as expectation values in terms of the amplitude measure rather than the probability measure as in pseudo-density matrices and are different from correlations in all the other approaches. 
A particular example via a tripartite pseudo-density matrix is presented to illustrate the unified picture of different approaches except path integrals. 
This applies to more complicated cases and provides a unified picture of these approaches. It also supports the further development of space-time formalisms in non-relativistic quantum theory. 
Difference in correlations between path integrals and other approaches also suggests the importance of measure choice in quantum theory. 

This chapter is based on Ref.~\cite{zhang2020quantum} and proceeds as follows. We introduce indefinite causal structures and compare the process matrix formalism with the pseudo-density matrix formalism in terms of correlation analysis, causality violation, and postselection in Section 4.2. In Section 4.3, we establish the relation between pseudo-density matrix and decoherence functional in consistent histories. We further explore generalised non-local games and build pseudo-density matrices from generalised signalling games in Section 4.4. In Section 4.5, we simplify the calculation of out-of-time-order correlations via pseudo-density matrices. We further argue that the path integral formalism defines correlations in a different way. Finally we provide a unified picture under a tripartite pseudo-density matrix except the path integral formalism and summarise our work and provide an outlook in Section 4.7.


\section{Indefinite causal structures}
The concept of indefinite causal structures was proposed as probabilistic theories with non-fixed causal structures as a possible approach to quantum gravity~\cite{hardy2007towards, hardy2009quantum}. There are different indefinite causal order approaches: quantum combs~\cite{chiribella2008quantum, chiribella2009theoretical}, operator tensors~\cite{hardy2012operator, hardy2018construction}, process matrices~\cite{oreshkov2012quantum, araujo2015witnessing}, process tensors~\cite{milz2017introduction, pollock2018non}, and super-density operators~\cite{cotler2018superdensity, cotler2019quantum}. 
Also, Several of the approaches are closely related~\cite{costa2018unifying}, for example, quantum channels with memories~\cite{kretschmann2005quantum}, general quantum strategies~\cite{gutoski2007toward}, multiple-time states~\cite{aharonov1964time, aharonov2009multiple, silva2017connecting}, general boundary formalism~\cite{oeckl2003general}, and quantum causal models~\cite{costa2016quantum, allen2017quantum}. General quantum strategies can be taken as a game theory representation; multiple-time states are a particular subclass of process matrices; quantum causal models just use the process matrix formalism.
Since there are clear maps among quantum combs, operator tensors, process tensors, and process matrices, we just take the process matrix formalism in order to learn from causality inequalities and postselection.
We will investigate its relation with the pseudo-density matrix and show what lessons we shall learn for pseudo-density matrices. 

\subsection{Preliminaries for process matrix formalism}

The process matrix formalism was originally proposed in Ref.~\cite{oreshkov2012quantum} as one of the indefinite causal structures assuming local quantum mechanics and well-defined probabilities. The process matrix was defined to take completely positive(CP) maps to linear probabilities. 
It is redefined in Ref.~\cite{araujo2017purification} in a more general way as high order transformations, where the definition is extended to take CP maps to other CP maps. 
Here we follow as Ref.~\cite{araujo2017purification}. We define bipartite processes first; the multipartite case is obtained directly or from Ref.~\cite{araujo2015witnessing}.

For the bipartite case, consider a global past $P$ and a global future $F$. Quantum states in the past are transformed to quantum states in the future through a causally indefinite structure. 
 A process is defined as a linear transformation take two CPTP maps $\mathcal{A}: A_I \otimes A_I' \rightarrow A_O \otimes A_O'$ and $\mathcal{B}: B_I \otimes B_I' \rightarrow B_O \otimes B_O'$ to a CPTP map $\mathcal{G}_{\mathcal{A},\mathcal{B}}: A_I' \otimes B_I' \otimes P \rightarrow A'_O \otimes B_O' \otimes F$ without acting on the systems $A_I'$, $A_O'$, $B_I'$, $B_O'$. Specifically, it is a transformation that act on $P \otimes A_I \otimes A_O \otimes B_I \otimes B_O \otimes F$.
 
 We introduce the Choi-Jamio\l{}kowski isomorphism~\cite{jamiolkowski1972linear, choi1975completely} to represent the process in the matrix formalism. 
 Recall that for a CP map $\mathcal{M}^A: A_I \rightarrow A_O$, its corresponding Choi-Jamio\l{}kowski matrix is given as $\mathfrak{C}(\mathcal{M}) \equiv [\mathcal{I} \otimes \mathcal{M}^A (|\mathbbm{1} \rrangle \llangle \mathbbm{1} |)]\in A_I \otimes A_O$ with $\mathcal{I}$ as the identity map and $|\mathbbm{1} \rrangle = |\mathbbm{1} \rrangle^{A_I A_I} \equiv \sum_j \ket{j}^{A_I} \otimes \ket{j}^{A_I} \in \mathcal{H}^{A_I} \otimes \mathcal{H}^{A_I}$ is the non-normalised maximally entangled state. The inverse is given as $\mathcal{M}(\rho^{A_I}) = \Tr[ (\rho^{A_I}\otimes \mathbbm{1}^{A_O}) M^{A_IA_O}]$ where $\mathbbm{1}^{A_O}$ is the identity matrix on $\mathcal{H}^{A_O}$.

Then $A = \mathfrak{C}(\mathcal{A})$, $B = \mathfrak{C}(\mathcal{B})$, and $G_{A, B} = \mathfrak{C}(\mathcal{G_{A,B}})$ are the corresponding CJ representations. We have
\begin{equation}
G_{A, B} = 	\Tr_{A_IA_OB_IB_O}[W^{T_{A_IA_OB_IB_O}}(A\otimes B)]
\end{equation}
where the process matrix is defined as $W \in P \otimes A_I \otimes A_O \otimes B_I \otimes B_O \otimes F$, $T_{A_IA_OB_IB_O}$ is the partial transposition on the subsystems $A_I$, $A_O$, $B_I$, $B_O$, and we leave identity matrices on the rest subsystems implicit. 
Note that we require that $G_{A, B}$ is a CPTP map for any CPTP maps $A$, $B$. This condition is equivalent to the followings:
\begin{align}
W \geq 0, \\
\Tr W = d_{A_O}d_{B_O}d_P, \\
W = L_V(W),
\end{align}
where $L_V$ is defined as a projector 
\begin{align}
L_V(W) = W - _FW +& _{A_OF}W	+ _{B_OF}W - _{A_OB_OF}W - _{A_IA_OF}W + _{A_IA_OB_OF}W \nonumber \\
& - _{B_IB_OF}W + _{A_IA_OB_OF}W - _{A_IA_OB_IB_OF}W + _{PA_IA_OB_IB_OF}W.
\end{align}

Terms that can exist in a process matrix include states, channels, channels with memory; nevertheless, local loops, channels with local loops and global loops are not allowed~\cite{oreshkov2012quantum}. 
A bipartite process matrix can be fully characterised in the Hilbert-Schmidt basis~\cite{oreshkov2012quantum}.
Define the signalling directions $\preceq$ and $\npreceq$ as follows: $A \preceq B$ means $A$ is in the causal past of $B$, $A \npreceq B$ means it is not; similarly for  $\succeq$ and $\nsucceq$. 
Any valid bipartite process matrix $W^{A_IA_OB_IB_O}$ can be given in the Hilbert-Schmidt basis as
\begin{equation}
W^{A_IA_OB_IB_O} = \frac{1}{d_{A_I}d_{B_I}} (\mathbbm{1} + \sigma_{A \preceq B} + \sigma_{A \succeq B} + \sigma_{A \npreceq \nsucceq B})
\end{equation}
where 
the matrices $\sigma_{A \preceq B}$, $\sigma_{A \succeq B}$, and $\sigma_{A \npreceq \nsucceq B}$ are defined by 
\begin{align}
\sigma_{A \preceq B} & \equiv \sum_{ij>0} c_{ij}\sigma_i^{A_O}\sigma_j^{B_I} + \sum_{ijk>0} d_{ijk}\sigma_i^{A_I}\sigma_j^{A_O}\sigma_k^{B_I}\\
\sigma_{A \succeq B} & \equiv \sum_{ij>0} e_{ij}\sigma_i^{A_I}\sigma_j^{B_O} + \sum_{ijk>0} f_{ijk}\sigma_i^{A_I}\sigma_j^{B_I}\sigma_k^{B_O}\\
\sigma_{A \npreceq \nsucceq B} & \equiv \sum_{i>0} g_i\sigma_i^{A_I} + \sum_{i>0} h_i\sigma_i^{B_I} + \sum_{ij>0} l_{ij}\sigma_i^{A_I}\sigma_j^{B_I} \\
\end{align}
Here $c_{ij}, d_{ijk}, e_{ij}, f_{ijk}, g_i, h_i,  l_{ij} \in \mathbb{R}$. 
That is, a bipartite process matrix of the system $AB$ is a combination of an identity matrix, the matrices where $A$ signals to $B$, where $B$ signals to $A$, and where $A$ and $B$ are causally separated. It is thus a linear combination of three possible causal structures. 

\subsection{Correlation analysis and causality inequalities}
In this subsection, we analyse correlations in both the process matrix formalism and the pseudo-density matrix formalism. We first take a special case with causal order and map correlations in two formalisms to each other. Then we consider the set of all possible causal correlations forms a causal polytope. The facets of the causal polytope are defined as causal inequalities and they are violated in the two formalisms with indefinite causal structures.

\subsubsection{Correlation analysis}

Now we analyse the relation between a process matrix and a pseudo-density matrix in the causal order. The basic elements in a process matrix are different laboratories, and the basic elements in a pseudo-density matrix are different events. We map a process matrix to a pseudo-density matrix in a way that each lab corresponds to each event. 

A process matrix with a single-qubit Pauli measurement taken at each laboratory is mapped to a finite-dimensional pseudo-density matrix.  
Compare them in the bipartite case as an illustration. 
In the simplest temporal case, a maximally mixed qubit evolves under the identity evolution between two times. 
The process matrix for this scenario is given as 
\begin{equation}
W = \frac{\mathbbm{1}^{A_I}}{2} \otimes [[\mathbbm{1}]]^{A_OB_I}, 
\end{equation}
where $[[\mathbbm{1}]]^{XY} = \sum_{ij}\ket{i}\bra{j}^X \otimes \ket{i}\bra{j}^Y = \frac{1}{2}(\mathbbm{1} \otimes \mathbbm{1} + X \otimes X - Y \otimes Y + Z\otimes Z)$. 
At the same time, the corresponding pseudo-density matrix is
\begin{equation}
R = \frac{1}{4} (I\otimes I + X \otimes X + Y \otimes Y + Z\otimes Z)= \frac{1}{2}[[\mathbbm{1}]]^{PT}= \frac{1}{2} S, 
\end{equation}
where the swap operator $S = \frac{1}{2}(\mathbbm{1}\otimes \mathbbm{1}  + X \otimes X + Y \otimes Y + Z\otimes Z) = [[\mathbbm{1}]]^{PT}$, here $PT$ is the partial transpose. 
For an arbitrary state $\rho$ evolving under the unitary evolution $U$, the process matrix is given as 
\begin{equation}
W = \rho^{A_I} \otimes [[U]]^{A_OB_I}, 
\end{equation}
where $[[U]] = (\mathbbm{1} \otimes U) [[\mathbbm{1}]] (\mathbbm{1} \otimes U^{\dag})$. The pseudo-density matrix is given from Ref.~\cite{zhao2018geometry} as
\begin{equation}
R = \frac{1}{2} (\mathbbm{1} \otimes U) (\rho^A \otimes \frac{\mathbbm{1}^B}{ 2} S + S\rho^A \otimes \frac{\mathbbm{1}^B}{2}) (\mathbbm{1} \otimes U^{\dag}) = \frac{1}{2}(\rho^A \otimes \frac{\mathbbm{1}^B}{2} [[U]]^{PT} + [[U]]^{PT} \rho^A \otimes \frac{\mathbbm{1}^B}{2}),
\end{equation}
where the partial transpose is taken on the subsystem $A$. 
Now we compare the correlations in the two formalisms and check whether they hold the same information.

The single-qubit Pauli measurement $\sigma_{i}$ for each event in the pseudo-density matrix has the Choi-Jamio\l{}kowski representation as 
\begin{equation}
\Sigma^{A_IA_O}_{i} = P^{+ A_I}_i \otimes P^{+ A_O}_i - P^{- A_I}_i\otimes P^{- A_O}_i  
\end{equation} 
where $P^{\pm}_i = \frac{1}{2}(\mathbbm{1} \pm \sigma_i)$; that is, to make a measurement $P^{\alpha}_i (\alpha = \pm 1)$ to the input state and project the corresponding eigenstate to the output system. 
It is equivalent to
\begin{equation}
\Sigma^{A_IA_O}_{i} = \frac{1}{2} (\mathbbm{1}^{A_I} \otimes \sigma_i^{A_O} + \sigma_i^{A_I} \otimes \mathbbm{1}^{A_O}).
\end{equation} 
In the example of a single qubit $\rho$ evolving under $U$, the correlations from the process matrix are given by
\begin{equation}
p( \Sigma^{A_IA_O}_{i}, \Sigma^{B_IB_O}_{j})  = \Tr [(\Sigma^{A_IA_O}_{i} \otimes \Sigma^{B_IB_O}_{j} )W] = \frac{1}{2} \Tr [\sigma_j U \sigma_i U^{\dag}];
\end{equation}
while the correlations from the pseudo-density matrix are given as 
\begin{equation}
\langle \{ \sigma_i, \sigma_j \} \rangle = \frac{1}{2}\left( \Tr[\sigma_j U \sigma_i \rho U^{\dag} ] + \Tr[\sigma_j U \rho \sigma_i U^{\dag} ] \right)= \frac{1}{2} \Tr [\sigma_j U \sigma_i U^{\dag}].
\end{equation}
The last equality holds as a single-qubit $\rho$ is decomposed into $\rho = \frac{\mathbbm{1}}{2} + \sum_{k=1, 2, 3} c_k \sigma_k$. 
The allowed spatio-temporal correlations given by the two formalisms are the same; thus, pseudo-density matrices and process matrices are equivalent in terms of encoded correlations. 
In a general case of bipartite systems on $AB$, this equivalence holds for each case with causal order as $A \preceq B$, $A \succeq B$, $A \npreceq \nsucceq B$. In principle, their superpositions for arbitrary process matrices will satisfy the correlation equivalence as well. 
The only condition here is that $A$ and $B$ make Pauli measurements in their local laboratories. 
Therefore, a process matrix where a single-qubit Pauli measurement is made at each laboratory corresponds to a finite-dimensional pseudo-density matrix since the correlations are equal. 

For generalised measurements, for example, arbitrary POVMs, a process matrix is fully mapped to the corresponding generalised pseudo-density matrix; thus, a process matrix can be always mapped to a generalised pseudo-density matrix in principle. 
 The process matrix and the corresponding generalised pseudo-density matrix just take the same measurement process in each laboratory or at each event. 
 The analysis for correlations is similar. 

For a given set of measurements, a process matrix where the measurement is made in each laboratory hold the same correlations as a generalised pseudo-density matrix with the measurement made at each event. 
Thus, a universal mapping from a process matrix to a pseudo-density matrix for general measurements is established. 
However, a pseudo-density matrix in finite dimensions is not necessarily mapped back to a valid process matrix. As mentioned before, a valid process matrix excludes the possibilities for post-selection, local loops, channels with local loops and global loops. Pseudo-density matrices are defined operationally in terms of measurement correlations and may allow these possibilities. 
We will come back to this point in the discussion for postselection and out-of-time-order correlation functions.

\subsubsection{Causal inequalities}
In the subsubsection, we introduce the causal polytope formed by the set of correlations with a definite causal order. Its facets are defined as causal inequalities~\cite{branciard2015simplest}. We show that the characterisation of bipartite correlations is consistent with the previous analysis in the pseudo-density matrix formalism. We show that causal inequalities can be violated in both of the process matrix formalism and the pseudo-density matrix formalism. 

We follow as Ref.~\cite{branciard2015simplest}. Recall that we denote Alice in the causal past of Bob as $A \preceq B$. Now for simplicity, we do not consider relativistic causality but normal Newton causality. We denote $A \prec B$ for events in Alice's system precedes those in Bob's system. Then Bob cannot signal to Alice, and the correlations satisfy that 
\begin{equation}\label{nosignalingtoAlice}
\forall x, y, y', a, \quad p^{A \prec B}(a|x,y) = p^{A \prec B}(a|x, y'),	
\end{equation}
where $p^{A \prec B}(a|x,y^{(')}) = \sum_b p^{A \prec B}(a,b|x,y^{(')})$.
Similarly, for $B \prec A$, Alice cannot signal to Bob that
\begin{equation}\label{nosignalingtoAlice}
\forall x, x', y, b, \quad p^{A \prec B}(b|x,y) = p^{A \prec B}(b|x', y),	
\end{equation}
where $p^{A \prec B}(b|x^{(')}, y) = \sum_a p^{A \prec B}(a,b|x^{(')}, y)$. 
 
Correlations of the order $A \prec B$ satisfy the properties of non-negativity and normalisation, and the no-signaling-to-Alice condition:
\begin{align}
p^{A \prec B}(a,b|x,y) \geq &0, \quad\forall x,y,a,b;\\
\sum_{a,b}p^{A \prec B}(a,b|x,y) = &1, \quad\quad\forall x,y;\\
p^{A \prec B}(a|x,y) = p^{A \prec B}(a|&x, y'), \quad\forall x,y,y',a.
\end{align}
Via these linear conditions, the set of correlations $p^{A \prec B}$ forms a convex polytope. 
Similarly for the set of correlations $p^{B \prec A}$.
The correlations are defined as causal if it is compatible with $A \prec B$ with probability $q$ and $B \prec A$ with probability $1-q$, that is, for $q \in [0,1]$, 
\begin{equation}
p(a,b|x,y) = q p^{A \prec B}(a,b|x,y) + (1-q) p^{B \prec A}(a,b|x,y),
\end{equation}
where $p^{A \prec B}$ and $p^{B \prec A}$ are non-negative and normalised to 1.
Then the set of causal correlations is the convex hull of the sets of correlations $p^{A \prec B}$ and $p^{B \prec A}$ and constitutes a causal polytope.

Suppose that Alice and Bob's inputs have $m_A$ and $m_B$ possible values, their outputs have $k_A$ and $k_B$ values respectively. 
The polytope of $p^{A \prec B}$ has $k_A^{m_A}k_B^{m_Am_B}$ vertices, of dimension $m_Am_B(k_Ak_B-1) - m_A(m_B-1)(k_A-1)$.
The polytope of $p^{B \prec A}$ has $k_A^{m_Am_B}k_B^{m_B}$ vertices, of dimension $m_Am_B(k_Ak_B-1) - (m_A-1)m_B(k_A-1)$.
The causal polytope has $k_A^{m_A}k_B^{m_Am_B} + k_A^{m_Am_B}k_B^{m_B} - k_A^{m_A}k_B^{m_B}$ vertices, of dimension $m_Am_B(k_Ak_B-1)$.
Consider the bipartite correlations where a qubit evolves between two times $t_A$ and $t_B$. 
We make a Pauli measurement at each time to record correlations. 
Given an initial state of the qubit, we have $m_A = m_B = 1$, $k_A = k_B = 2$. 
The polytope of $p^{A \prec B}$ has 4 vertices in 3 dimensions. 
The same as $p^{B \prec A}$ and the causal polytope. 
This result is consistent with the characterisation by the pseudo-density matrix formalism in Ref.~\cite{zhao2018geometry}. 

Now we characterise the causal polytope with $m_A = m_B = k_A = k_B = 2$. 
It has 112 vertices and 48 facets. 16 of the facets are trivial, which imply the non-negativity of the correlations $p(a,b|x,y)\geq 0$.
If we relabel the inputs and outputs of the systems, the rest of facets are divided into two groups, each with 16 facets:
\begin{equation}\label{gyni}
\frac{1}{4} \sum_{x,y,a,b} \delta_{a,y}\delta_{b,x}p(a,b|x,y) \leq \frac{1}{2},
\end{equation}
and 
\begin{equation}\label{lgyni}
\frac{1}{4} \sum_{x,y,a,b} \delta_{x(a\oplus y), 0}\delta_{y(b\oplus x), 0}p(a,b|x,y) \leq \frac{3}{4},
\end{equation}
where $\delta_{i,j}$ is the Kronecker delta function and $\oplus$ is the addition modulo 2.
They are interpreted into the bipartite "guess your neighbour's input" (GYNI) games and "lazy GYNI" (LGYNI) games~\cite{branciard2015simplest}.

Then we show the violation of causal inequalities via process matrix formalism and pseudo-density matrix formalism. 
In the process matrix formalism, we take the global past $P$, the global future $F$, Alice's ancilla systems $A_I'$, $A_O'$ and Bob's ancilla systems $B_I'$, $B_O'$ trivial. 
Then the process matrix correlations are given as 
\begin{equation}
p(a, b|x,y) = \Tr[ W^{T_{A_IA_OB_IB_O}} A_{a|x} \otimes B_{b|y}]	.
\end{equation}
Consider the process matrix 
\begin{equation}
W = \frac{1}{4}\left[ \mathbbm{1}^{\otimes 4} + \frac{Z^{A_I}Z^{A_O}Z^{B_I}\mathbbm{1}^{B_O} + Z^{A_I}\mathbbm{1}^{A_O}X^{B_I}X^{B_O}}{\sqrt{2}} \right].
\end{equation}
We choose the operations as (here slightly different from Ref.~\cite{branciard2015simplest}):
\begin{align}
A_{0|0} &= B_{0|0} = 0,\\
A_{1|0} &= B_{1|0} = (\ket{00}+\ket{11})(\bra{00}+\bra{11}), \\
A_{0|1} &= B_{0|1} = \frac{1}{2} \ket{0}\bra{0} \otimes \ket{0}\bra{0}+ \frac{1}{2}\ket{0}\bra{0} \otimes \ket{1}\bra{1}, \\
A_{1|1} &= B_{1|1} = \frac{1}{2}\ket{1}\bra{1} \otimes \ket{0}\bra{0}+ \frac{1}{2}\ket{1}\bra{1} \otimes \ket{1}\bra{1}.
\end{align}
Then 
\begin{align}
p_{GYNI} & = \frac{5}{16}(1+ \frac{1}{\sqrt{2}}) \approx 0.5335 > \frac{1}{2},\\
p_{LGYNI} & = \frac{5}{16}(1+ \frac{1}{\sqrt{2}}) + \frac{1}{4} \approx 0.7835 > \frac{3}{4}.
\end{align}

For a pseudo-density matrix, we consider a similar strategy.
Alice has two systems $X$ and $A$, where $X$ is the ancillary system prepare with $\ket{x}\bra{x}$. Bob has two systems $Y$ and $B$, where $Y$ is the ancillary system prepare with $\ket{y}\bra{y}$.
Given a pseudo-density matrix 
\begin{equation}
R = \frac{1}{4}\left[ \ket{x}\bra{x}^X \otimes \mathbbm{1}^A \otimes \ket{y}\bra{y}^Y \otimes \mathbbm{1}^B + \frac{Z^{X}Z^{A}Z^{Y}\mathbbm{1}^{B} + Z^{X}\mathbbm{1}^{A}X^{Y}X^{B}}{\sqrt{2}} \right],
\end{equation}
we choose the operations as before and gain the success probabilities as
\begin{align}
p_{GYNI} & = \frac{5}{16}(1+ \frac{1}{\sqrt{2}}) \approx 0.5335 > \frac{1}{2},\\
p_{LGYNI} & = \frac{5}{16}(1+ \frac{1}{\sqrt{2}}) + \frac{1}{4} \approx 0.7835 > \frac{3}{4}.
\end{align}

Again the causal inequalities are violated. 
This example also highlights another relationship for the mapping between a process matrix and a pseudo-density matrix. Instead of an input system and an output system in a process matrix, the corresponding pseudo-density matrix has an additional ancillary system for each event. 
Thus, a process matrix which makes a measurement and reprepares the state in one laboratory describes the same probabilities as a pseudo-density matrix with ancillary systems which makes a measurement and reprepares the state at each event. 
Another mapping from a process matrix to a pseudo-density matrix is established by introducing ancillary systems.

\subsection{Postselection and closed timelike curves}

Postselection is conditioning on the occurrence of certain event in probability theory, or conditioning upon certain measurement outcome in quantum mechanics. 
It allows a quantum computer to choose the outcomes of certain measurements and increases its computational power significantly.
In this subsection, we take the view from postselection and show that a particular subset of postselected two-time states correspond to process matrices in indefinite causal order. Postselected closed timelike curves are presented as a special case.

\subsubsection{Two-time quantum states}

In this subsubsection, we review the two-time quantum states approach~\cite{silva2017connecting} which fixes initial states and final states independent at two times. The two-time quantum state takes its operational meaning from postselection. Consider that Alice prepares a state $\ket{\psi}$ at the initial time $t_1$. Between the initial time $t_1$ and the final time $t_2$, she performs arbitrary operations in her lab. Then she measures an observable $O$ at the final time $t_2$.  The observable $O$ has a non-degenerate eigenstate $\ket{\phi}$. Taking $\ket{\phi}$ as the final state, Alice discards the experiment if the measurement of $O$ does not give the eigenvalue corresponding to the eigenstate $\ket{\phi}$. 

Consider that Alice makes a measurement by the set of Kraus operators $\{\hat{E}_a = \sum_{k,l} \beta_{a, kl}\ket{k}\bra{l}\}$ between $t_1$ and $t_2$. Note that $\{\hat{E}_a\}$ are normalised as $\sum_a \hat{E}_a^{\dag}\hat{E}_a = \mathbbm{1}$. The probability for Alice to gain the outcome $a$ under the pre- and post-selection is given as
\begin{equation}
p(a) = \frac{|\bra{\phi}\hat{E}_a\ket{\psi}|^2}{\sum_{a'} |\bra{\phi}\hat{E}_{a'}\ket{\psi}|^2}. 
\end{equation}
Now define the two-time state and the two-time version of Kraus operator as
\begin{align}
\Phi = _{\mathcal{A}_2}\bra{\phi} \otimes \ket{\psi}^{\mathcal{A}_1} & \in \mathcal{H}_{\mathcal{A}_2} \otimes \mathcal{H}^{\mathcal{A}_1}, \nonumber \\ 
E_a = \sum_{kl} \beta_{a, kl} \ket{k}^{\mathcal{A}_2} \otimes _{\mathcal{A}_1}\bra{l} & \in \mathcal{H}^{\mathcal{A}_2} \otimes \mathcal{H}_{\mathcal{A}_1},
\end{align}
where the two-time version of Kraus operator is denoted by $E_a$ without the hat. 
An arbitrary pure two-time state takes the form
\begin{equation}
\Phi = \sum \alpha_{ij} \ _{\mathcal{A}_2}\bra{i} \otimes \ket{j}^{\mathcal{A}_1} \in \mathcal{H}_{\mathcal{A}_2} \otimes \mathcal{H}^{\mathcal{A}_1}.
\end{equation}
Then the probability to obtain $a$ as the outcome is given as 
\begin{equation}
p(a) = \frac{|\Phi \cdot E_a |^2}{\sum_{a'} |\Phi \cdot E_{a'}|^2}. 
\end{equation}

A two-time density operator $\eta$ is given as 
\begin{equation}
\eta = \sum_r p_r \Phi_r \otimes \Phi_r^{\dag} \in \mathcal{H}_{\mathcal{A}_2} \otimes \mathcal{H}^{\mathcal{A}_1} \otimes \mathcal{H}_{\mathcal{A}_1^{\dag}} \otimes \mathcal{H}^{\mathcal{A}_2^{\dag}}.
\end{equation}
Consider a coarse-grained measurement 
\begin{equation}
J_a  = \sum_{\mu} E_a^{\mu} \otimes E_a^{\mu \dag} \in \mathcal{H}^{\mathcal{A}_2} \otimes \mathcal{H}_{\mathcal{A}_1}  \otimes \mathcal{H}^{\mathcal{A}_1^{\dag}} \otimes \mathcal{H}_{\mathcal{A}_2^{\dag}} 
\end{equation}
where the outcome $a$ corresponds to a set of Kraus operators $\{\hat{E}_a^{\mu}\}$. 
Then the probability to obtain $a$ as the outcome is given as 
\begin{equation}
p(a) = \frac{\eta \cdot J_a}{\sum_{a'} \eta \cdot J_{a'}}. 
\end{equation}

\subsubsection{Connection between process matrix and pseudo-density matrix under post-selection}


Now consider postselection applied to ordinary quantum theory. 
It is known that a particular subset of postselected two-time states in quantum mechanics give the form of process matrices within indefinite causal structures~\cite{silva2017connecting}. 
Here we first give a simple explanation for this fact and further analyse the relation between a process matrix and a pseudo-density matrix from the view of postselection. 

For an arbitrary bipartite process matrix $W \in \mathcal{H}^{A_I} \otimes \mathcal{H}^{A_O} \otimes \mathcal{H}^{B_I} \otimes \mathcal{H}^{B_O} $, we can expand it in some basis: 
\begin{equation}
W^{A_IA_OB_IB_O} = \sum_{ijkl, pqrs} w_{ijkl, pqrs} \ket{ijkl}\bra{pqrs}.
\end{equation}
For the elements in each Hilbert space, we map them to the corresponding parts in a bipartite two-time state. For example, we map the input Hilbert space of Alice to the bra and ket space of Alice at time $t_1$, and similarly for the output Hilbert space for $t_2$. That is, 
\begin{align}
\ket{i}\bra{p} & \in \mathcal{L}(\mathcal{H}^{A_I}) \rightarrow \bra{p} \otimes \ket{i} \in \mathcal{H}_{A_1^{\dag}} \otimes \mathcal{H}^{A_1}\\
\ket{j}\bra{q} & \in \mathcal{L}(\mathcal{H}^{A_O})\rightarrow \bra{q} \otimes \ket{j} \in \mathcal{H}_{A_2} \otimes \mathcal{H}^{A_2^{\dag}}
\end{align}
Thus, a two-time state $\eta_{W^{A_1A_2}} \in \mathcal{H}_{A_2} \otimes \mathcal{H}^{A_1} \otimes \mathcal{H}^{A_2^{\dag}} \otimes \mathcal{H}_{A_1^{\dag}}$ is equivalent to a process matrix for a single laboratory $W^{A_IA_O}$.

The connection with pre- and post-selection suggests one more interesting relationship between a process matrix and a pseudo-density matrix. 
For a process matrix, if we consider the input and output Hilbert spaces at two times, we can map it to a two-time state. That is, we connect a process matrix with single laboratory to a two-time state. 
A pseudo-density matrix needs two Hilbert spaces to represent two times. For a two-time state $\eta_{12}$, the corresponding pseudo-density matrix $R_{12}$ has the same marginal single-time states, i.e., $\Tr_1 \eta_{12} = \Tr_1 R_{12}$ and $\Tr_2 \eta_{12} = \Tr_2 R_{12}$.
Then we find a map between a process matrix for a single event and a pseudo-density matrix for two events. 
Note that in the previous subsections, we have mapped a process matrix for two events to a pseudo-density matrix with half Hilbert space for two events, and mapped a process matrix for two events to a pseudo-density matrix with two Hilbert spaces at each of two events. 
This suggests that the relationship between a process matrix and a pseudo-density matrix is non-trivial with a few possible mappings. 


One question arising naturally here concerns the pseudo-density matrices with postselection. 
The definitions for finite-dimensional and Gaussian pseudo-density matrices guarantee that under the partial trace, the marginal states at any single time will give the state at that time. In particular, tracing out all other times in a pseudo-density matrix, we get the final state at the final time. On the one hand, we may think that pseudo-density matrix formulation is kind of time-symmetric. On the other hand, the final state is fixed by evolution; that implies that we cannot assign an arbitrary final state, making it difficult for the pseudo-density matrix to be fully time-symmetric. For other generalisation of pseudo-density matrices like position measurements and weak measurements, the property for fixed final states does not hold. 
Nevertheless, we may define a new type of pseudo-density matrices with postselection. We assign the final measurement to be the projection to the final state and renormalise the probability. For example, a qubit in the initial state $\rho$ evolves under a CPTP map $\mathcal{E}: \rho \rightarrow \mathcal{E}(\rho)$ and then is projected on the state $\eta$. We may construct the correlations $\langle \{ \sigma_i , \sigma_j, \eta \} \rangle$ as 
\begin{equation}
\langle \{ \sigma_i , \sigma_j, \eta \} \rangle= \sum_{\alpha, \beta = \pm 1} \alpha \beta \Tr[\eta P^{\beta}_j \mathcal{E}(P^{\alpha}_i \rho P^{\alpha}_i )  P^{\beta}_j ] / p_{ij}(\eta),
\end{equation}
where $P^{\alpha}_i = \frac{1}{2}(\mathbbm{1} + \alpha \sigma_i)$ and $p_{ij}(\eta) = \sum_{\alpha, \beta = \pm 1} \Tr[\eta P^{\beta}_j \mathcal{E}(P^{\alpha}_i \rho P^{\alpha}_i )  P^{\beta}_j ]$. 
Then the pseudo-density matrix with postselection is given as 
\begin{equation}
R = \frac{1}{4} \sum_{i, j = 0}^{3} \langle \{ \sigma_i , \sigma_j, \eta \} \rangle \sigma_i \otimes \sigma_j \otimes \eta.
\end{equation}
We further conclude the relation between a process matrix and a pseudo-density matrices with postselection. 
A process matrix with postselection for a laboratory is operationally equivalent to a tripartite postselected pseudo-density matrix.

\subsubsection{Post-selected closed timelike curves}
We briefly discuss postselected closed timelike curves before we move on to a summary. 
Closed timelike curves (CTCs), after being pointed out by G\"{o}del to be allowed in general relativity~\cite{godel1949example}, have always been arising great interests. 
Deutsch~\cite{deutsch1991quantum} proposed a circuit method to study them and started an information theoretic point of view. 
Deustch's CTCs are shown to have many abnormal properties violated by ordinary quantum mechanics. For example, they are nonunitary, nonlinear, and allow quantum cloning~\cite{ahn2013quantum, brun2013quantum}. 
Several authors~\cite{bennett2005teleportation, svetlichny2011time, brun2012prefect, lloyd2011closed} later proposed a model for closed timelike curves based on postselected teleportation. 
It is studied that process matrices correspond to a particular linear version of postselected closed timelike curves~\cite{araujo2017quantum}. 
In pseudo-density matrices we can consider a system evolves in time and back; that is the case for calculating out-of-time-order correlation functions we will introduce later, and different from closed timelike curves as there is no loop. However, the black hole final state proposal in later section is very much related. Now we briefly introduce postselected closed timelike curves and its representation in pseudo-density matrices.

Postselected closed timelike curves can be seen as a ``chronology-respecting'' system $S$ and a CTC system $A$ evolving under a unitary $U_{SA}$. Consider the CTC system $A$ is part of the maximally entangled state $\ket{\Phi}_{AB} = \sum_{i=0}^{d-1} \frac{1}{\sqrt{d}} \ket{i}\ket{i}$. More specifically, the system $S$ and $A$ evolve under the unitary $U$ and then we project the two systems $AB$ onto the state $\ket{\Phi}$ and renormalise the probability. One assumes that this projection is certain with probability $1$. Then for the system $S$, $\rho_S$ goes in to the state $\frac{C \rho_S C^{\dag}}{\Tr[C \rho_S C^{\dag}]}$ where $C = \Tr_A U_{SA}$. In this way, we create a quantum channel from the future to the past and the CTC qubit goes back in time. 

Here we illustrate post-selected closed timelike curves by the pseudo-density matrices with postselection. It is a two-time process with a postselection. We assume that at the initial time systems $S$ and $AB$ are prepared. We make a measurement $P_i$. After the unitary evolution $U_{SA}$, we make another measurement $P_j$. Then we project the state to $\ket{\Phi}_{AB}$. The correlations are represented by 
\begin{equation}
\langle \{ P_i, P_j, \ket{\Phi}\bra{\Phi}_{AB} \} \rangle = \frac{ \sum_{\alpha, \beta} \alpha \beta p_{ij}^{\alpha \beta}}{\sum_{\alpha, \beta} p_{ij}^{\alpha \beta}},
\end{equation}
where 
\begin{equation}
	p_{ij}^{\alpha \beta} = \Tr[\mathbbm{1}_S \otimes \ket{\Phi}\bra{\Phi}_{AB} P_j^{\beta} (U_{SA}\otimes \mathbbm{1}_B) P_i^{\alpha} (\rho_S \otimes \ket{\Phi}\bra{\Phi}_{AB}) P_i^{\alpha \dag} (U^{\dag}_{SA}\otimes \mathbbm{1}_B) P_j^{\beta \dag} ].
\end{equation}
Here $P_i^{\alpha}$ is denoted for the measurement $P_i$ with the outcome $\alpha$ and $P_j^{\beta}$ for the measurement $P_j$ with the outcome $\beta$.
For simplicity, we consider $P_i = P_j = \mathbbm{1}$. Then 
\begin{align}
p & = \Tr[\mathbbm{1}_S \otimes \ket{\Phi}\bra{\Phi}_{AB} (U_{SA}\otimes \mathbbm{1}_B) (\rho_S \otimes \ket{\Phi}\bra{\Phi}_{AB}) (U^{\dag}_{SA}\otimes \mathbbm{1}_B) ]
& = \frac{1}{d^2} \Tr[C \rho_S C^{\dag}]
\end{align}
where $C = \Tr_A U_{SA}$. The result is consistent with Ref.~\cite{brun2013quantum}. However, the role of quantum correlations plays in the closed timelike curves is still an open problem.

\subsection{Summary of the relation between pseudo-density matrix and indefinite causal structures}
In this subsection, we have introduced the relation between pseudo-density matrices and indefinite causal structures. We argue that the pseudo-density matrix formalism belongs to indefinite causal structures. 
So far, all other indefinite causal structures to our knowledge use a tensor product of both input and output Hilbert spaces, while a pseudo-density matrix only assumes a single Hilbert space. 
For a simple example of a qudit at two times, the dimension used in other indefinite causal structures is $d^4$ but for pseudo-density matrix it is $2d^2$. 
Though other indefinite causal structures assume a much larger Hilbert space, pseudo-density matrix should not be taken as a subclass of any indefinite causal structures which already exist. There are certain non-trivial relation between pseudo-density matrices and other indefinite causal structures. 
As we can see from the previous subsections, it is possible to map a process matrix to a corresponding pseudo-density matrix in three different ways: one-lab to one-event direct map, one-lab to one-event with double Hilbert spaces map, and one-lab to two-event map. 
\begin{claim}
	A process matrix and the corresponding pseudo-density matrix allow the same correlations or probabilities in three different mappings.  
\end{claim}
One obvious difference between a process matrix and a pseudo-density matrix is that, for each laboratory, a process matrix measures and reprepares a state while a pseudo-density matrix usually only makes a measurement and the state evolves into its eigenstate for each eigenvalue with the corresponding probability. 
The correlations given by process matrices and pseudo-density matrices are also the same. 
Examples in postselection and closed time curves suggest further similarities. 
In general, we can understand that the pseudo-density matrix is defined in an operational way which does not specify the causal order, thus belongs to indefinite causal structures. 
We borrow the lessons from process matrices here to investigate pseudo-density matrices further. Maybe it will be interesting to derive a unified indefinite causal structure which takes the advantage of all existing ones. 

Nevertheless, the ultimate goal of indefinite causal order towards quantum gravity is still far reaching. So far, all indefinite causal structures are linear superpositions of causal structures; will that be enough for quantising gravity?
It is generally believed among indefinite causal structure community that what is lacking in quantum gravity is the quantum uncertainty for dynamical causal structures suggested by general relativity.  
The usual causal order may be changed under this quantum uncertainty and there is certain possibility for a superposition of causal orders and even beyond. 
Generalisation to relativistic quantum field theory and quantum gravity remains to be a very exciting open problem. 

\section{Consistent histories}

In this section we first review consistent histories and then explore the relation between pseudo-density matrices and consistent histories. 

\subsection{Preliminaries for consistent histories}
Consistent histories, or decoherent histories, is an interpretation for quantum theory, proposed by Griffiths~\cite{griffiths1984consistent, griffiths2003consistent}, Gell-Mann and Hartle~\cite{gell2018quantum, gell1993classical}, and Omnes~\cite{omnes1990hilbert}. 
The main idea is that a history, understood as a sequence of events at successive times, has a consistent probability with other histories in a closed system. The probabilities assigned to histories satisfy the consistency condition to avoid the interference between different histories and that set of histories are called consistent histories~\cite{dowker1992quantum, dowker1996consistent}. 


Consider a set of projection operators $\{P_{\alpha}\}$ which are exhaustive and mutually exclusive:
\begin{equation}
\sum_{\alpha} P_{\alpha} = \mathbbm{1}, \qquad P_{\alpha}P_{\beta} = \delta_{\alpha\beta}P_{\beta},
\end{equation}
where the range of $\alpha$ may be finite, infinite or even continuous. 
For each $P_{\alpha}$ and a system in the state $\rho$, the event $\alpha$ is said to occur if $P_{\alpha}\rho P_{\alpha} = \rho$ and not to occur if $P_{\alpha}\rho P_{\alpha}=0$. 
The probability of the occurrence of the event $\alpha$ is given by
\begin{equation}
p(\alpha) = \Tr[P_{\alpha}\rho P_{\alpha}].
\end{equation}
A projection of the form $P_{\alpha} = \ket{\alpha}\bra{\alpha}$ ($\{\ket{\alpha}\}$ is complete) is called completely fine-grained, which corresponds to the precise measurement of a complete set of commuting observables. Otherwise, for imprecise measurements or incomplete sets, the projection operator is called coarse-grained. Generally it takes the form $\bar{P}_{\bar{\alpha}} = \sum_{\alpha \in \bar{\alpha}} P_{\alpha}$.

In the Heisenberg picture, the operators for the same observables $P$ at different times are related by
\begin{equation}
P(t) = \exp(iHt/\hbar) P(0) \exp(-iHt/\hbar),
\end{equation}
with $H$ as the Hamiltonian of the system.
Then the probability of the occurrence of the event $\alpha$ at time $t$ is
\begin{equation}
p(\alpha) = \Tr[P_{\alpha}(t) \rho P_{\alpha}(t)].
\end{equation}

Now we consider how to assign probabilities to histories, that is, to a sequence of events at successive times. 
Suppose that the system is in the state $\rho$ at the initial time $t_0$. Consider a set of histories $[\alpha] = [\alpha_1,\alpha_2, \cdots, \alpha_n]$ consisting of $n$ projections $\{P^k_{\alpha_k}(t_k)\}_{k=1}^n$ at times $t_1 < t_2 < \cdots < t_n$. 
Here the subscript $\alpha_k$ allows for different types of projections, for example, a position projection at $t_1$ and a momentum projection at $t_2$.
Then the decoherence functional is defined as
\begin{equation}
D([\alpha], [\alpha']) = \Tr[P^n_{\alpha_n}(t_n)\cdots P^1_{\alpha_1}(t_1) \rho P^1_{\alpha'_1}(t_1) \cdots P^n_{\alpha'_n}(t_n)],
\end{equation}
where 
\begin{equation}
P^k_{\alpha_k}(t_k) = e^{i(t_k-t_0)H} P^k_{\alpha_k} e^{-i(t_k-t_0)H}.
\end{equation}
It is important in consistent histories because probabilities can be assigned to histories when the decoherence functional is diagonal. 
It is easy to check that
\begin{align}
& D([\alpha], [\alpha']) = D([\alpha'], [\alpha])^*,\\
& \sum_{[\alpha]} \sum_{[\alpha']} D([\alpha], [\alpha']) = \Tr \rho = 1.
\end{align}
The diagonal elements are the probabilities for the histories $(\rho, t_0) \rightarrow (\alpha_1, t_1) \rightarrow \cdots \rightarrow (\alpha_n, t_n)$:
\begin{equation}
p(\alpha_1, \alpha_2, \dots, \alpha_n) = D(\alpha_1,\alpha_2, \dots, \alpha_n | \alpha_1,\alpha_2, \dots, \alpha_n) = D([\alpha], [\alpha])
\end{equation}

Until now, we considered fine-grained projections $P^k_{\alpha_k}$ for fine-grained histories. The coarse-grained histories are characterised by the coarse-grained projections $\bar{P}^k_{\bar{\alpha}_k}$. To satisfy the probability sum rules, the probability for a coarse-grained history is the sum of the probabilities for its fine-grained histories. That is,
\begin{equation}
p(\bar{\alpha}_1, \bar{\alpha}_2, \dots, \bar{\alpha}_n) = \sum_{[\alpha]\in[\bar{\alpha}]} p(\alpha_1, \alpha_2, \dots, \alpha_n),
\end{equation}
where
\begin{equation}
\sum_{[\alpha]\in[\bar{\alpha}]} = \sum_{\alpha_1\in\bar{\alpha}_1}\sum_{\alpha_2\in\bar{\alpha}_2} \cdots \sum_{\alpha_n\in\bar{\alpha}_n}.
\end{equation}
On the other hand, we gain the decoherence functional for coarse-grained histories by directly summing over the fine-grained projections as
\begin{equation}
D([\bar{\alpha}], [\bar{\alpha}']) = \sum_{[\alpha]\in[\bar{\alpha}]} \sum_{[\alpha']\in[\bar{\alpha'}]} D([\alpha], [\alpha']).
\end{equation}
For the diagonal terms,
\begin{equation}
D([\bar{\alpha}], [\bar{\alpha}]) = \sum_{[\alpha]\in[\bar{\alpha}]} D([\alpha], [\alpha]) + \sum_{[\alpha]\neq [\alpha'], [\alpha]\in[\bar{\alpha}]} \sum_{[\alpha']\in[\bar{\alpha'}]} D([\alpha], [\alpha']),
\end{equation}
where $[\alpha] \neq [\alpha']$ means $\alpha_k \neq \alpha'_k$ for at least one $k$.

To obey the probability sum rules that all probabilities are non-negative and summed to $1$, the sufficient and necessary condition is 
\begin{equation}\label{consistencycond}
\Re[ D(\alpha_1,\alpha_2, \dots, \alpha_n | \alpha'_1,\alpha'_2, \dots, \alpha'_n) ] = p(\alpha_1, \alpha_2, \dots, \alpha_n) \delta_{\alpha_1\alpha'_1} \cdots \delta_{\alpha_n\alpha'_n}.
\end{equation}
Eqn.~\eqref{consistencycond} is called the consistency condition or decoherence condition. Sets of histories obeying the condition are referred to consistent histories or decoherent histories. A stronger version of consistency condition is
\begin{equation}\label{strongconsistencycond}
D(\alpha_1,\alpha_2, \dots, \alpha_n | \alpha'_1,\alpha'_2, \dots, \alpha'_n) = p(\alpha_1, \alpha_2, \dots, \alpha_n) \delta_{\alpha_1\alpha'_1} \cdots \delta_{\alpha_n\alpha'_n}.
\end{equation}

The decoherence functional has a path integral representation. With configuration space variables $q^i(t)$ and the action $S[q^i]$, 
\begin{equation}
D([\alpha], [\alpha']) = \int_{[\alpha]} \mathcal{D} q^i \int_{[\alpha']} \mathcal{D} q^{i'} \exp(iS[q^i] - iS[q^{i'}]) \delta(q_f^i - q_f^{i'}) \rho(q_0^i, q_0^{i'}),
\end{equation}
where the two paths $q^i(t)$, $q^{i'}(t)$ begin at $q^i_0$, $q^{i'}_0$ respectively at $t_0$ and end at $q^i_f = q^{i'}_f$ at $t_f$, and correspond to the projections $P^k_{\alpha_k}$, $P^k_{\alpha'_k}$ made at time $t_k$ ($k = 1, 2, \dots n$).

\subsection{Temporal correlations in terms of decoherence functional}

The relation with the $n$-qubit pseudo-density matrix is arguably obvious. For example, consider an $n$-qubit pseudo-density matrix as a single qubit evolving at $n$ times. For each event, we make a single-qubit Pauli measurement $\sigma_{i_k}$ at the time $t_k$. We can separate the measurement $\sigma_{i_k}$ into two projection operators $P_{i_k}^{+1} = \frac{1}{2}(I + \sigma_{i_k} )$ and $P_{i_k}^{-1} = \frac{1}{2}(I - \sigma_{i_k} )$ with its outcomes $\pm1$. 
Corresponding to the history picture, each pseudo-density event with the measurement $\sigma_{i_k}$ corresponds to two history events with projections $P_{i_k}^{\alpha_k} (\alpha_k = \pm 1)$.
A pseudo-density matrix is built upon measurement correlations $\langle \{\sigma_{i_k}\}_{k=1}^n \rangle$. Theses correlations can be given in terms of decoherence functionals as 
\begin{align}
\langle \{\sigma_{i_k}\}_{k=1}^n \rangle & = \sum_{\alpha_1, \dots, \alpha_n} \alpha_1\cdots \alpha_n \Tr[ P_{i_n}^{\alpha_n} U_{n-1} \cdots U_1 P_{i_1}^{\alpha_1} \rho P_{i_1}^{\alpha_1} U_1^{\dag} \cdots U_{n-1}^{\dag} P_{i_n}^{\alpha_n} ] \nonumber\\
& = \sum_{\alpha_1, \dots, \alpha_n} \alpha_1\cdots \alpha_n p(\alpha_1, \dots, \alpha_n) \nonumber\\
& =  \sum_{\alpha_1, \dots, \alpha_n} \alpha_1\cdots \alpha_n D([\alpha], [\alpha]),
\end{align}
where $D([\alpha], [\alpha])$ is the diagonal terms of decoherence functional with $[\alpha] = [\alpha_1, \dots, \alpha_n]$. Note that here only diagonal decoherence functionals are taken into account, which coincides with the consistency condition.

Similar relations hold for the Gaussian spacetime states. For each event, we make a single-mode quadrature measurement $\hat{q}_k$ or $\hat{p}_k$ at time $t_k$. We can separate the measurement $\hat{x}_k = \int x_k \ket{x_k}\bra{x_k} \textrm{d}x_k$ into projection operators $\ket{x_k}\bra{x_k}$ with outcomes $x_k$. Then each Gaussian event with the measurement $\hat{x}_k$ corresponds to infinite and continuous history events with projections $\ket{x_k}\bra{x_k}$.
\begin{align}
\langle \{x_k\}_{k=1}^n \rangle & = \int_{-\infty}^{\infty}\cdots\int_{-\infty}^{\infty} \textrm{d}x_1 \cdots \textrm{d}x_n x_1\cdots x_n \nonumber \\
& \qquad \qquad \Tr[ \ket{x_n}\bra{x_n}U_{n-1} \cdots U_1 \ket{x_1}\bra{x_1} \rho \ket{x_1}\bra{x_1} U_1^{\dag} \cdots U_{n-1}^{\dag} \ket{x_n}\bra{x_n} ]\nonumber \\
& = \int_{-\infty}^{\infty}\cdots\int_{-\infty}^{\infty} \textrm{d}x_1 \cdots \textrm{d}x_n x_1\cdots x_n p(x_1, \dots, x_n) \nonumber\\
& =  \int_{-\infty}^{\infty}\cdots\int_{-\infty}^{\infty} \textrm{d}x_1 \cdots \textrm{d}x_n x_1\cdots x_n D([x], [x]),
\end{align}
where $D([x], [x])$ is the diagonal terms of decoherence functional with $[x] = [x_1, \dots, x_n]$. 

For general spacetime states for continuous variables, we make a single-mode measurement $T(\alpha_k)$ at time $t_k$ for each event. It separates into two projection operators $P^{+1}(\alpha_k)$ and $P^{-1}(\alpha_k)$, then it follows as the $n$-qubit case.

The interesting part is to apply the lessons from consistent histories to the generalised pseudo-density matrix formulation with general measurements. 
We have argued that the spacetime density matrix can be expanded diagonally in terms of position measurements as 
\begin{equation}\label{stdmfromposition}
\rho = \int_{-\infty}^{\infty}\cdots\int_{-\infty}^{\infty} \textrm{d}x_1\cdots\textrm{d}x_n p(x_1, \cdots, x_n) \ket{x_1}\bra{x_1} \otimes  \cdots \otimes \ket{x_n}\bra{x_n}.
\end{equation}
It reminds us of the diagonal terms of the decoherence functional. It is possible to build a spacetime density matrix from all possible decoherence functionals as 
\begin{equation}
\rho = \int_{-\infty}^{\infty}\cdots\int_{-\infty}^{\infty} \textrm{d}x_1\textrm{d}x'_1\cdots\textrm{d}x_n\textrm{d}x'_n D(x_1, \dots, x_n | x'_1, \dots x'_n) \ket{x_1}\bra{x'_1} \otimes  \cdots \otimes \ket{x_n}\bra{x'_n}.
\end{equation}
Applying the strong consistency condition to the above equation, we gain Eqn.~\eqref{stdmfromposition} again. 
This argues why it is effective to only consider diagonal terms in position measurements. which is originally taken for convenience. 

Similarly, the spacetime Wigner function from weak measurements is easily taken as a generalisation for the diagonal terms of the decoherence functional allowing for general measurements. 
Recall that a generalised effect-valued measure is represented by 
\begin{equation}
\hat{f}(q, p) = C\exp\left[ -\alpha[(\hat{q}-q)^2 + \lambda (\hat{p}-p)^2] \right].
\end{equation}
The generalised decoherence functional for weak measurements is then given by
\begin{equation}
D(q, p, q', p', \tau | \hat{\rho}) = \Tr \left[ \mathcal{F}(q ,p, q', p'; \tau)  \hat{\rho} \right],
\end{equation}
where
\begin{align}
\mathcal{F}(q ,p, q',p'; \tau)  \hat{\rho} = & \int \textrm{d}\mu_G [q(t), p(t)] \int \textrm{d}\mu_G [q'(t), p'(t)] \delta \left( q - \frac{1}{\tau}\int_0^{\tau} \textrm{d}t q(t) \right) \nonumber \\
&  \delta \left( p - \frac{1}{\tau}\int_0^{\tau} \textrm{d}t p(t) \right) \delta \left( q' - \frac{1}{\tau}\int_0^{\tau} \textrm{d}t q'(t) \right) \delta \left( p' - \frac{1}{\tau}\int_0^{\tau} \textrm{d}t p'(t) \right) \nonumber \\
& \exp[-\frac{i}{\hbar}\hat{H}\tau] \mathcal{T} \exp \left[ -\frac{\gamma}{2} \int_{0}^{\tau} \textrm{d}t [(\hat{q}_H(t) - q(t))^2 + \lambda (\hat{p}_H(t) - p(t))^2 ] \right] \hat{\rho} \nonumber \\
& \mathcal{T}^* \exp \left[ -\frac{\gamma}{2} \int_{0}^{\tau} \textrm{d}t [(\hat{q'}_H(t) - q'(t))^2 + \lambda (\hat{p'}_H(t) - p'(t))^2 ] \right] \exp[\frac{i}{\hbar}\hat{H}\tau],
\end{align}
here
\begin{align}
&\textrm{d}\mu_G [q(t), p(t)] = \lim_{N \rightarrow \infty} \left( \frac{\gamma\tau\sqrt{\lambda}}{\pi N} \prod_{s=1}^{N} \textrm{d}q(t_s) \textrm{d}p(t_s) \right),\\
&\textrm{d}\mu_G [q'(t), p'(t)] = \lim_{N \rightarrow \infty} \left( \frac{\gamma\tau\sqrt{\lambda}}{\pi N} \prod_{s=1}^{N} \textrm{d}q'(t_s) \textrm{d}p'(t_s) \right),
\end{align}
and
\begin{align}
&\hat{q}_H(t) = \exp\left[ \frac{i}{\hbar} \hat{H} t \right] \hat{q} \exp\left[ -\frac{i}{\hbar} \hat{H} t \right], 
\qquad
\hat{q'}_H(t) = \exp\left[ \frac{i}{\hbar} \hat{H} t \right] \hat{q'} \exp\left[ -\frac{i}{\hbar} \hat{H} t \right], 
\nonumber \\
&\hat{p}_H(t) = \exp\left[ \frac{i}{\hbar} \hat{H} t \right] \hat{p} \exp\left[ -\frac{i}{\hbar} \hat{H} t \right], \qquad
\hat{p'}_H(t) = \exp\left[ \frac{i}{\hbar} \hat{H} t \right] \hat{p'} \exp\left[ -\frac{i}{\hbar} \hat{H} t \right].
\end{align}
The diagonal terms under the strong consistency condition reduce to the form in the previous chapter: 
\begin{equation}\label{stwffromweak}
p(q, p, \tau | \hat{\rho}) = \Tr \mathcal{F}(q ,p; \tau)  \hat{\rho},
\end{equation}
where
\begin{align}\label{stweak}
\mathcal{F}(q ,p; \tau)  \hat{\rho} = & \int \textrm{d}\mu_G [q(t), p(t)] \delta \left( q - \frac{1}{\tau}\int_0^{\tau} \textrm{d}t q(t) \right) \delta \left( p - \frac{1}{\tau}\int_0^{\tau} \textrm{d}t p(t) \right) \exp[-\frac{i}{\hbar}\hat{H}\tau] \nonumber \\
& \mathcal{T} \exp \left[ -\frac{\gamma}{2} \int_{0}^{\tau} \textrm{d}t [(\hat{q}_H(t) - q(t))^2 + \lambda (\hat{p}_H(t) - p(t))^2 ] \right] \hat{\rho} \nonumber \\
& \mathcal{T}^* \exp \left[ -\frac{\gamma}{2} \int_{0}^{\tau} \textrm{d}t [(\hat{q}_H(t) - q(t))^2 + \lambda (\hat{p}_H(t) - p(t))^2 ] \right] \exp[\frac{i}{\hbar}\hat{H}\tau].
\end{align}

Now we conclude the relation between decoherence functionals in consistent histories and temporal correlations in pseudo-density matrices. 	
\begin{claim}
The decoherence functional in consistent histories is the probabilities in temporal correlations of pseudo-density matrices. 	
\end{claim}
Thus, we establish the relationship between consistent histories and all possible forms of pseudo-density matrix. 
From the consistency condition, we also have a better argument for why spacetime states for general measurements are defined in the diagonal form. It is not a coincide.

\section{Generalised non-local games}
Game theory studies mathematical models of competition and cooperation under strategies among rational decision-makers~\cite{myerson2013game}. Here we give an introduction to nonlocal games, quantum-classical nonlocal games, and quantum-classical signalling games. Then we show the relation between quantum-classical signalling games and pseudo-density matrices, and comment on the relation between general quantum games and indefinite causal order. 

\subsection{Introduction to non-local games}
The interests for investigating non-local games start from interactive proof systems with two parties, the provers and the verifiers. They exchange information to verify a mathematical statement. A nonlocal game is a special kind of interactive proof system with only one round and at least two provers who play in cooperation against the verifier. In nonlocal games, we refer to the provers as Alice, Bob, $\dots$, and the verifier as the referee. In Ref.~\cite{cleve2004consequences}, nonlocal games were formally introduced with shared entanglement and used to formulate the CHSH inequality~\cite{clauser1969proposed}. Here we introduce the CHSH game as an example and then give the general form of a non-local game. 

The CHSH game has two cooperating players, Alice and Bob, and a referee who asks questions and collects answers from the players. The basic rules of the CHSH game are as the following: 

1) There are two possible questions $x \in \{0, 1\}$ for Alice and two possible questions $y \in \{0, 1\}$ for Bob. Each question has an equal probability as $p(x, y) = \frac{1}{4}, \forall x, \forall y$. 

2) Alice answers $a \in \{0, 1\}$ and Bob $b \in \{0, 1\}$. 

3) Alice and Bob cannot communicate with each other after the game begins. 

4) If $a \oplus b = x \cdot y$, then they win the game, otherwise they lose. 

For a classical strategy, that is, Alice and Bob use classical resources, they win with the probability at most $\frac{3}{4}$. 
Alice and Bob can also adopt a quantum strategy. If they prepare and share a joint quantum state $\ket{\Phi^+} = \frac{1}{\sqrt{2}}(\ket{00}+\ket{11})$ and make local measurements based on the questions they receive separately, then they can achieve a higher winning probability $\cos^2(\pi/8) \approx 0.854$. 

In general, a non-local game $G$ is formulated by $(\pi, l)$ on 
\begin{equation}
\overrightarrow{nl} = \langle \mathcal{X}, \mathcal{Y}; \mathcal{A}, \mathcal{B}; l \rangle,
\end{equation}
where $\mathcal{X}$, $\mathcal{Y}$ are question spaces of Alice and Bob and $\mathcal{A}$, $\mathcal{B}$ are answer spaces of Alice and Bob.
Here $\pi(x, y)$ is a probability distribution of the question spaces for Alice and Bob in the form
$\pi: \mathcal{X} \times \mathcal{Y} \rightarrow [0, 1]$. 
$l(a, b | x, y)$ is a function of question and answer spaces for Alice and Bob to decide whether they win or lose in the form
$l:  \mathcal{X} \times \mathcal{Y} \times \mathcal{A} \times \mathcal{B} \rightarrow [0, 1];$
for example, if they win, $l = 1$; otherwise lose with $l = 0$. 
For any strategy, the probability distribution for answers $a, b$ of Alice and Bob given questions $x, y$, respectively, is referred to as the correlation function $p(a, b | x, y)$ of the form 
\begin{equation}
p:  \mathcal{X} \times \mathcal{Y} \times \mathcal{A} \times \mathcal{B} \rightarrow [0, 1].
\end{equation}
with the condition
$\sum_{a,b} p(a, b | x, y) = 1.$
With a classical source,
\begin{equation}
p_c(a, b | x, y) = \sum_{\lambda}  \pi(\lambda) d_A(a | x, \lambda) d_B(b | y, \lambda),
\end{equation}
where $d_A(a | x, \lambda)$ is the probability of answering $a$ given the parameter $\lambda$ and the question $b$ and similar for $d_B(b | y, \lambda)$;
with a quantum source,
\begin{equation}
p_q(a, b | x, y) = \Tr [\rho_{AB} (P_A^{a|x} \otimes Q_B^{b|y})],
\end{equation}
where $\rho_{AB}$ is the quantum state shared by Alice and Bob, $P_A^{a|x}$ is the measurement made by Alice with the outcome $a$ given $x$, $Q_B^{b|y}$ is the measurement made by Bob with the outcome $b$ given $y$. 
Then the optimal winning probability is given by 
\begin{equation}
\mathbb{E}_{\overrightarrow{nl}} [*] \equiv \max \sum_{x,y} \pi(x, y) \sum_{a,b} l(a, b | x, y) p_{c/q}(a, b | x, y).
\end{equation}



\subsection{Quantum-classical non-local \& signalling games}

First we introduce a generalised version of non-local games where the referee asks quantum questions instead of classical questions (therefore this type of non-local games are refereed to quantum-classical)~\cite{buscemi2012all}. Then we give the temporal version of these quantum-classical non-local games as quantum-classical signalling games~\cite{rosset2018resource}. 

\subsubsection{Quantum-classical non-local games}
We now recap the model of quantum-classical non-local games~\cite{buscemi2012all}, in which the questions are quantum rather than classical. More specifically, the referee sends quantum registers to Alice and Bob instead of classical information.

For a non-local game, with the question spaces $\mathcal{X} =  \{ x \} $ and $\mathcal{Y} =  \{ y \}$, the referee associates two quantum ancillary systems $X$ and $Y$ such that $\dim \mathcal{H}_{X} \geq |\mathcal{X}|$, $\dim \mathcal{H}_{Y} \geq |\mathcal{Y}|$, the systems are in the states $\tau^{x}_{X} = \ket{x}\bra{x}$ and $\tau^{y}_{Y} = \ket{y}\bra{y}$ with the questions $x \in \mathcal{X}$ and $y \in \mathcal{Y}$. 
Assume that Alice and Bob share a quantum state $\rho_{AB}$. 
Given the answer sets $\mathcal{A} = \{ a \}$ and $\mathcal{B} = \{ b \}$ and quantum systems $XA$ and $YB$, Alice and Bob can make the corresponding POVMs $P^a_{XA}$ and $Q^b_{YB}$ in the linear operators on the Hilbert space $\mathcal{H}_{XA}$ and $\mathcal{H}_{YB}$, such that $\sum_a P^a_{XA} = \mathbbm{1}_{XA}$ and $\sum_b Q^b_{YB} = \mathbbm{1}_{YB}$. 
Then the probability distribution for the questions and answers of Alice and Bob, that is, the correlation function $P(a, b | x, y)$, is given by
\begin{equation}
P(a, b | x, y) = \Tr [(P^a_{XA} \otimes Q^b_{YB}) (\tau^{x}_{X} \otimes \rho_{AB} \otimes \tau^{y}_{Y}) ].
\end{equation}

Quantum-classical non-local games replace classical inputs with quantum ones, formulated by $(\pi(x, y), l(a, b | x, y))$ on 
\begin{equation}
\overrightarrow{qcnl} = \langle \{\tau^x\}, \{\omega^y\}; \mathcal{A}, \mathcal{B}; l \rangle.
\end{equation}
The referee picks $x \in \mathcal{X}$ and $y \in \mathcal{Y}$ with the probability distribution $\pi(x, y)$ as the classical-classical non-local game. 
With a classical source,
\begin{equation}
p_c(a, b | x, y) = \sum_{\lambda} \pi(\lambda) \Tr [ (\tau^x_X \otimes \omega^y_Y) (P_X^{a|\lambda} \otimes Q_Y^{b|\lambda})];
\end{equation}
with a quantum source,
\begin{equation}
p_q(a, b | x, y) = \Tr [ (\tau^x_X \otimes \rho_{AB} \otimes \omega^y_Y)  (P_{XA}^{a} \otimes Q_{BY}^{b})].
\end{equation}
The optimal winning probability is, again, given by
\begin{equation}
\mathbb{E}_{\overrightarrow{qcnl}} [*] \equiv \max \sum_{x,y} \pi(x, y) \sum_{a,b} l(a, b | x, y) p_{c/q}(a, b | x, y).
\end{equation}

\subsubsection{Quantum-classical signalling games}

In quantum-classical signalling games~\cite{rosset2018resource}, instead of two players Alice and Bob, we consider only one player Abby at two successive instants in time.
Then quantum-classical signalling games change the Alice-Bob duo to a timelike structures of single player Abby with
\begin{equation}
\overrightarrow{qcsg} = \langle \{\tau^x\}, \{\omega^y\}; \mathcal{A}, \mathcal{B}; l \rangle.
\end{equation}
With unlimited classical memory,
\begin{equation}
p_c(a, b | x, y) = \sum_{\lambda} \pi(\lambda) \Tr [ \tau^x_X P_X^{a|\lambda} ] \Tr [ \omega^y_Y Q_Y^{b|a, \lambda}].
\end{equation}
For admissible quantum strategies, suppose Abby at $t_1$ receives $\tau^x_X$ and makes a measurement of instruments $\{\Phi_{X\rightarrow A}^{a|\lambda}\}$, and gains the outcome $a$. Then the quantum output goes through the quantum memory $\mathcal{N}: A \rightarrow B$. The output of the memory and $\omega^y_Y$ received by Abby at $t_2$ are fed into a measurement $\{\Psi_{BY}^{b|a, \lambda}\}$, with outcome b. 
Then
\begin{equation}
p_q(a, b | x, y) =  \sum_{\lambda} \pi(\lambda) \Tr [ (\{ (\mathcal{N}_{A \rightarrow B} \circ \Phi_{X \rightarrow A}^{a | \lambda})(\tau^x_X) \} \otimes \omega^y_Y)\Psi_{BY}^{b | a, \lambda}].
\end{equation}
The optimal payoff function is, again, given by
\begin{equation}
\mathbb{E}_{\overrightarrow{qcsg}} [*] \equiv  \max \sum_{x,y} \pi(x, y) \sum_{a,b} l(a, b | x, y) p_{c/q}(a, b | x, y).
\end{equation}

\subsection{Temporal correlations from signalling games}

To compare quantum-classical signalling games with pseudo-density matrices, first we generalise the finite-dimensional pseudo-density matrices from Pauli measurements to general positive-operator valued measures(POVMs). Recall that a POVM is a set of Hermitian positive semi-definite operator $\{E_i\}$ on a Hilbert space $\mathcal{H}$ which sum up to the identity $\sum_i E_i = \mathbbm{1}_{\mathcal{H}}$. Instead of making a single-qubit Pauli measurement at each event, we make a measurement $E_i = M_i^{a \dag}M_i^a$ with the outcome $a$. For each event, there is a measurement $\mathcal{M}_i: \mathcal{L}(\mathcal{H}^X) \rightarrow \mathcal{L}(\mathcal{H}^A), \tau^x_X \mapsto \sum_i M_i^a \tau^x_X M_i^{a \dag}$ with $\sum M_i^{a \dag} M_i^a = \mathbbm{1}_{\mathcal{H}^X}$.

Now we map the generalised pseudo-density matrices to quantum-classical signalling games. Assume $\omega_Y^y$ to be trivial. For Abby at the initial time and the later time, we consider $\Phi_{X \rightarrow A}^{a}: \tau^x_X \rightarrow \sum_i M_i^a \tau^x_X M_i^{a \dag}$, where $\sum M_i^{a \dag} M_i^a = \mathbbm{1}_{\mathcal{H}^A}$. 
Between two times, the transformation from $A$ to $B$ is given by $\mathcal{N}: \rho_A \rightarrow \sum_j N_j \rho_A N_j^{\dag}$ with $\sum_j N_j^{\dag}N_j = \mathbbm{1}_{\mathcal{H}^A}$. Then 
\begin{align}
p_q(a, b | x, y) & =  \Tr [ \{ (\mathcal{N}_{A \rightarrow B} \circ \Phi_{X \rightarrow A}^{a})(\tau^x_X) \} \Psi_{B}^{b | a}] \nonumber \\
& =  \sum_{ik} \Tr [ \mathcal{N} \{ M_i^{a} \tau^x_X M_i^{a \dag} \} \Psi_{B}^{b | a}] \nonumber \\
& = \sum_{ijk} \Tr [ N_j M_i^{a} \tau^x_X M_i^{a \dag} N_j^{\dag} \Psi_{B}^{b | a}] \\
 \langle \{ \Phi, \Psi \} \rangle & = \sum_{a, b}  ab p_q(a, b | x, y)
\end{align}
It is the temporal correlation given by pseudo-density matrices. 
That is, a quantum-classical signalling game with a trivial input at later time corresponds to a pseudo-density matrix with quantum channels replacing measurements for events. 
\begin{claim}
The probability in a quantum-classical signalling game with a trivial input at later time corresponds to the probability in a pseudo-density matrix where the state goes through quantum channels instead of measurements. 
\end{claim}

It is also convenient to establish the relation between generalised games in time and indefinite causal structures with double Hilbert spaces for each event. For completeness, we also mention that Gutoski and Watrous~\cite{gutoski2007toward} proposed a general theory of quantum games in terms of the Choi-Jamio\l{}kowski representation, which is an equivalent formulation of indefinite causal order.

\section{Out-of-time-order correlations (OTOCs)}
In this section we introduce out-of-time-order correlation functions, find a simple method to calculation these temporal correlations via the pseudo-density matrix formalism, and apply the out-of-time-order correlation functions into the black hole final state projection proposal as one of the proposals for black hole information paradox. 

\subsection{Brief introduction to OTOCs}

Consider local operators $W$ and $V$. With a Hamiltonian $H$ of the system, the Heisenberg representation of the operator $W$ is given as $W(t) = e^{iHt}We^{-iHt}$. 
Out-of-time-order correlation functions (OTOCs)~\cite{maldacena2015bound, roberts2016chaos} are usually defined as 
\begin{equation}
\langle VW(t)V^{\dag}W^{\dag}(t) \rangle = \langle VU(t)^{\dag}WU(t)V^{\dag}U^{\dag}(t)W^{\dag}U(t) \rangle, 
\end{equation}
where $U(t) = e^{-iHt}$ is the unitary evolution operator and the correlation is evaluated on the thermal state $\langle \cdot \rangle = \Tr[e^{-\beta H} \cdot]/ \Tr[e^{-\beta H}]$. 
Note that OTOC is usually defined for the maximally mixed state $\rho = \frac{\mathbbm{1}}{d}$. 
Consider a correlated qubit chain. Measure $V$ at the first qubit and $W$ at the last qubit. Since the chain is correlated in the beginning, we have OTOC as $1$ at the early time. As time evolves and the operator growth happens, OTOC will approximate to $0$ at the later time.


\subsection{Calculating OTOCs via pseudo-density matrices}
In this subsection we make a connection between OTOCs and the pseudo-density matrix formalism. Consider a qubit evolving in time and backward, we can get a tripartite pseudo-density matrix. 
In particular, we consider measuring $A$ at $t_1$, $B$ at $t_2$ and $A$ again at $t_3$ and assume the evolution forwards is described by $U$ and backwards $U^{\dag}$. 
Then the probability is given by 
\begin{equation}\label{otocrel}
\Tr[A U^{\dag} B U A \rho A^{\dag} U^{\dag} B^{\dag} U A^{\dag}] = \Tr[A B(t) A \rho A^{\dag} B^{\dag}(t) A^{\dag}] 
\end{equation}
If we assume that $AA^{\dag} = A$, $\rho = \frac{\mathbbm{1}}{d}$, then Eqn.~\eqref{otocrel} will reduce to the OTOC $\langle AB(t)AB(t) \rangle$. 
\begin{claim}
OTOCs can be represented as temporal correlations in pseudo-density matrices with half numbers of steps for calculation; for example, a four-point OTOC, usually calculated by evolving forwards and backwards twice, is represented by a tripartite pseudo-density matrix with only once evolving forwards and backwards. 
\end{claim}

\subsection{Black hole final state proposal}
In this subsection, we briefly review black hole information paradox and final state projection proposal, and use the relation between OTOCs and pseudo-density matrices to analyse OTOC in the final state proposal. 

\subsubsection{Review of black hole information paradox}
Hawking showed that black holes emit exactly thermal radiations~\cite{hawking1975particle}. Consider that a black hole initially in a pure state evolves unitarily. The fact that the radiation emitted outside the black hole is in a mixed state is not surprising when we take the black hole interior and the outside radiation as the whole system. However, the problem appears when the black hole fully evaporates and only thermal radiation is left. The final state is a mixed state. We find that a pure state evolves into a mixed state in the black hole evaporation; that is, in a closed system the unitarity is violated. This is the black hole information paradox~\cite{hawking1976breakdown, preskill1992black}. 

A few possible solutions have been proposed for the information paradox. For example, some people believe that there is fundamental non-unitarity in the universe and the information is just lost. Another possibility might be that information is stored in a Planck-sized remnant~\cite{banks1992horned} and we need to apply an unknown quantum gravity theory to solve it. Also, information might be stored in a baby universe~\cite{hawking1988wormholes, hawking1990baby} which carries away the collapsing matter as well as the information. Or, information is encoded in the correlations between the early and late radiation~\cite{page1993average, page1993information}.

\subsubsection{Final state projection proposal}
To solve the black hole information paradox, one possible proposal is the black hole final state projection proposal~\cite{horowitz2003black, gottesman2003comment, lloyd2013unitarity, bousso2014measurements}. 
The matter that forms the black hole lives in the Hilbert space $\mathcal{H}_M$ with the dimension $N = e^{S_{BH}}$ where $S_{BH}$ is the black hole entropy. The evaporation of the black hole, usually formulated by a semiclassical approximation to field fluctuations, divides the fluctuation fields into $\mathcal{H}_{in}$ and $\mathcal{H}_{out}$, inside and outside the event horizon respectively. Each of them has dimension $N = e^{S_{BH}}$ as well. The (Unruh) state $\ket{\Phi}_{in \otimes out}$ on $\mathcal{H}_{in} \otimes \mathcal{H}_{out}$ is the maximally entangled state
\begin{equation}
\ket{\Phi}_{in \otimes out} = \frac{1}{\sqrt{N}}\sum_{i=1}^N \ket{i}_{in}\ket{i}_{out}
\end{equation}
where $\ket{i}_{in}$, $\ket{i}_{out}$ are orthonormal bases in $\mathcal{H}_{in}$ and $\mathcal{H}_{out}$. 
In the final state projection proposal, Horowitz and Maldacena attempt to construct the unitary evaporation 
\begin{equation}
\ket{m}_M \rightarrow S_{jm} \ket{j}_{out}, 
\end{equation}
to solve the problem of information paradox. In particular, they impose a final state boundary condition at the singularity and project the state in $\mathcal{H}_M \otimes \mathcal{H}_{in}$ to a super-normalised maximally entangled state 
\begin{equation}
\bra{BH} = N^{1/2} \sum_{m,i} S_{im}\bra{m}_M\ket{i}_{in} = N \bra{\Phi}_{M \otimes in} (S \otimes \mathbbm{1}).
\end{equation}
The whole process is formulated as 
\begin{align}
\ket{m} &\rightarrow \ket{m}_M \ket{\Phi}_{in \otimes out} \nonumber\\
& \rightarrow \bra{BH}_{M, in} \left( \ket{m}_M \ket{\Phi}_{in \otimes out} \right) \nonumber\\
& = S_{jm}\ket{j}.
\end{align}
Thus, a unitary process for evaporation is achieved with this final state projection. 

\subsubsection{OTOC analysis for final state proposal}

Now we apply the above OTOC analysis to the final state proposal. 
First, we write the initial state as 
\begin{equation}
\rho = \ket{\psi}\bra{\psi}_M \otimes \ket{\Phi}\bra{\Phi}_{in \otimes out}, 
\end{equation}
and the final state as 
\begin{equation}
\sigma = \ket{\Phi}\bra{\Phi}_{M \otimes in} \otimes S\ket{\psi}\bra{\psi}S^{\dag}_{out}. 
\end{equation}
From the initial state to the final state, there is an evolution described by $U = S_{M} \otimes \mathbbm{1}_{in} \otimes \mathbbm{1}_{out}$, and a projection $P = \ket{BH}\bra{BH}_{M,in} \otimes \mathbbm{1}_{out}$.
Suppose $S$ is a Haar random unitary, we have 
\begin{equation}
\Tr[P U \rho U^{\dag} P^{\dag}] = 1.
\end{equation}
Here we assume $S_M = S^{\dag}$. 
Now we consider the OTOC between the initial time and the final time. It can be computed in the pseudo-density matrix formulation by assuming the evolution forwards and backwards. Thus we measure $A$ at $t_1$, let the state evolve under $U$, after that we make the final state projection $P$ at $t_2$, then the state evolves under $U^{\dag}$, and we measure $A$ at $t_3$. 
That is, 
\begin{equation}
OTOC = \langle A P(t) A^{\dag} P^{\dag}(t) \rangle = \Tr[ A U^{\dag} P U A \rho A^{\dag} U^{\dag} P^{\dag} U].
\end{equation}
For simplicity, we take $A$ as the identity operator. Then again we have
\begin{equation}
OTOC = \Tr[P U \rho U^{\dag} P^{\dag}] = 1.
\end{equation}
Consider the measurement $A$ is acted on the outside radiation part as $A = \mathbbm{1}_{M \otimes in} \otimes \ket{\psi}\bra{\psi}_{out}$, then
\begin{equation}
OTOC = \Tr[A P U A \rho A^{\dag} U^{\dag} P^{\dag}] = 1.
\end{equation}
Note that $[A, P] = 0$. $P$ is acted on the matter and inside radiation while $A$ is acted on the outside radiation. The out-of-time-order correlation remains unchanged. This suggests there is no operator growth from the interior of the black hole to outside. This is consistent with the preservation of the information and unitarity. However, we notice that the projection is onto a supernormalised state; it remains doubt whether this is physical enough to be achieved.  


\section{Path integrals}

The path integral approach~\cite{feynman2010quantum} is a representation of quantum theory, not only useful in quantum mechanics but also quantum statistical mechanics and quantum field theory. It generalises the action principle of classical mechanics and one computes a quantum amplitude by replacing a single classical trajectory with a functional integral of infinite numbers of possible quantum trajectories. 
Here we argue that the path integral approach of quantum mechanics use amplitude as the measure in correlation functions rather than probability measure in the above formalisms.

\subsection{Introduction to path integrals}
Now we briefly introduce path integrals and correlation functions in this formalism~\cite{zinn2010path}.
Consider a bound operator in a Hilbert space $U(t_2, t_1) (t_2 \geq t_1)$ as the evolution from time $t_1$ to $t_2$, which satisfies the Markov property in time as
\begin{equation}
U(t_3, t_2)U(t_2, t_1) = U(t_3, t_1), \forall\   t_3 \geq t_2 \geq t_1 \qquad U(t, t) = \mathbbm{1}.
\end{equation}
We further assume that $U(t, t')$ is differentiable and the derivative is continuous:
\begin{equation}
\left.\frac{\partial U(t, t')}{\partial t}\right	|_{t=t'} = -H(t)/\hbar
\end{equation}
where $\hbar$ is a real parameter, and later identified with Planck's constant; $H = i \tilde{H}$ where $\tilde{H}$ is the quantum Hamiltonian. 
Then
\begin{equation}
U(t'', t') = \prod_{m=1}^n U[t'+m\epsilon, t'+(m-1)\epsilon], \qquad n\epsilon = t"-t'.
\end{equation}
The position basis for $\hat{q}\ket{q} = q\ket{q}$ is orthogonal $\langle q' \ket{q} = \delta(q-q')$, and complete $\int \textrm{d}q \ket{q}\bra{q} = \mathbbm{1}$. 
We have 
\begin{equation}
\bra{q''}U(t'', t')\ket{q'} = \int \prod_{k=1}^{n-1} \textrm{d}q_k \prod_{k=1}^n \bra{q_k} U(t_k, t_{k-1})\ket{q_{k-1}}
\end{equation}
with $t_k = t' + k\epsilon, q_0 = q', q_n = q''$. 
Suppose that the operator $H$ is identified with a quantum Hamiltonian of the form
\begin{equation}
H = \hat{\bm{p}}^2/2m + V(\hat{\bm{q}}, t)	
\end{equation}
where $\bm{p}, \bm{q} \in \mathbb{R}^d$. We have 
\begin{equation}
\bra{\bm{q}}U(t, t')\ket{\bm{q'}} = \left( \frac{m}{2\pi\hbar(t-t')} \right)^{d/2} \exp[-\mathcal{S}(\bm{q})/\hbar]
\end{equation}
where 
\begin{equation}
\mathcal{S}(\bm{q}) = \int_{t'}^t \textrm{d}\tau [\frac{1}{2}m\dot{\bm{q}}^2(\tau) + V(\bm{q}(\tau), \tau)] + O((t-t')^2)	,
\end{equation}
and 
\begin{equation}
\bm{q}(\tau) = \bm{q}'+\frac{\tau - t'}{t- t'}(\bm{q}-\bm{q}').
\end{equation}
We consider short time slices, then
\begin{equation}
\bra{\bm{q''}}U(t'', t')\ket{\bm{q'}} = \lim_{n\rightarrow\infty} \left( \frac{m}{2\pi\hbar\epsilon} \right)^{dn/2} \int \prod_{k=1}^{n-1} \textrm{d}^d q_k \exp[-\mathcal{S}(\bm{q}, \epsilon)/\hbar],
\end{equation}
with 
\begin{equation}\label{sss}
\mathcal{S}(\bm{q}, \epsilon) =\sum_{k=0}^{n-1} \int_{t_k}^{t_{k+1}} \textrm{d}t [\frac{1}{2}m\dot{\bm{q}}^2(t) + V(\bm{q}(t), t)] + O(\epsilon^2).
\end{equation}
Introducing a linear and continuous trajectory 
\begin{equation}
\bm{q}(t) = \bm{q}_k + \frac{t- t_k}{t_{k+1} - t_k}(\bm{q}_{k+1} - \bm{q}_k) \ \ \text{for} \ \ t_k\leq t \leq t_{k+1},
\end{equation}
we can rewrite Eqn.~\eqref{sss} as
\begin{equation}
\mathcal{S}(\bm{q}, \epsilon) = \int_{t'}^{t''} \textrm{d}t [\frac{1}{2}m\dot{\bm{q}}^2(t) + V(\bm{q}(t), t)] + O(n\epsilon^2).
\end{equation}
Taking $n \rightarrow \infty$ and $\epsilon \rightarrow 0$ with $n\epsilon = t''-t'$ fixed, we have
\begin{equation}
\mathcal{S}(\bm{q}) = \int_{t'}^{t''} \textrm{d}t [\frac{1}{2}m\dot{\bm{q}}^2(t) + V(\bm{q}(t), t)]
\end{equation}
as the Euclidean action. 
The path integral is thus defined as 
\begin{equation}
\bra{\bm{q}''}U(t'',t')\ket{\bm{q}'} = \int_{\bm{q}(t') = \bm{q}'}^{\bm{q}(t'') = \bm{q}''}[\textrm{d}\bm{q}(t)] \exp(-\mathcal{S}(\bm{q})/ \hbar),
\end{equation}
where a normalisation of $\mathcal{N} = ( \frac{m}{2\pi\hbar\epsilon} )^{dn/2}$ is hidden in $[\textrm{d}\bm{q}(t)]$. 

The quantum partition function $\mathcal{Z}(\beta) = \Tr e^{-\beta H}$ ($\beta$ is the inverse temperature) can be written in terms of path integrals as 
\begin{align}
\mathcal{Z}(\beta) & = \Tr e^{-\beta H} = \Tr U(\hbar\beta, 0) = \int \textrm{d}q''\textrm{d}q'\delta(\bm{q}''-\bm{q}')\bra{\bm{q}''}U(\hbar\beta, 0)\ket{\bm{q}'} \nonumber\\
& = \int_{\bm{q}(0) = \bm{q}(\hbar\beta)} [\textrm{d}q(t)]\exp[-\mathcal{S}(\bm{q})/\hbar].
\end{align}
The integrand $e^{-\mathcal{S}(\bm{q})/\hbar}$ is a positive measure and defines the corresponding expectation value as 
\begin{equation}
\langle \mathcal{F}(q) \rangle = \mathcal{N}\int [\textrm{d}q(t)] \mathcal{F}(q) \exp[-\mathcal{S}(\bm{q})/\hbar],
\end{equation}
where $\mathcal{N}$ is chosen for $\langle 1 \rangle = 1$.
Moments of the measure in the form as 
\begin{equation}
\langle q(t_1)q(t_2) \cdots q(t_n) \rangle = \mathcal{N}\int [\textrm{d}q(t)]q(t_1)q(t_2) \cdots q(t_n)\exp[-\mathcal{S}(\bm{q})/\hbar]
\end{equation}
are the $n$-point correlation function. 
Suppose for the finite time interval $\beta$ periodic boundary conditions hold as $q(\beta/2) = q(-\beta/2)$. The normalisation is given as $\mathcal{N} = \mathcal{Z}^{-1}(\beta)$. 
Then we define 
\begin{equation}
Z^{(n)}(t_1, \cdots, t_n) = 	\langle q(t_1) \cdots q(t_n) \rangle.
\end{equation}
The generating functional of correlation functions is 
\begin{align}
\mathcal{Z}(f) & = \sum_{n=0} \frac{1}{n!}\int \textrm{d}t_1 \cdots \textrm{d}t_n 	Z^{(n)}(t_1, \cdots, t_n) f(t_1) \cdots f(t_n) \nonumber \\
& = \sum_{n=0} \frac{1}{n!}\int \textrm{d}t_1 \cdots \textrm{d}t_n \langle q(t_1) \cdots q(t_n) \rangle f(t_1) \cdots f(t_n) \nonumber \\
& = \left \langle \exp\left[ \int\textrm{d}t q(t)f(t)\right] \right \rangle
\end{align}
Note that the $n$-point quantum correlation functions in time also appear as continuum limits of the correlation functions of $1D$ lattice in classical statistical models. 
The path integral formalism represents a mathematical relation between classical statistical physics on a line and quantum statistical physics of a point-like particle at thermal equilibrium. 
This is the first example of the quantum-classical correspondence which maps between quantum statistical physics in $D$ dimensions and classical statistical physics in $D + 1$ dimensions~\cite{zinn2010path}.

\subsection{Temporal correlations in path integrals are different}
Here we take two-point correlations functions:
\begin{equation}\label{tpcf}
\langle q(t_1)q(t_2) \rangle  = \frac{\int [\textrm{d}q(t)]q(t_1)q(t_2)\exp[-\mathcal{S}(\bm{q})/\hbar]}{\int [\textrm{d}q(t)]\exp[-\mathcal{S}(\bm{q})/\hbar]}
\end{equation}
In the Gaussian representation of pseudo-density matrices, temporal correlation for $q_1$ at $t_1$ and $q_2$ at $t_2$ with the evolution $U$ and the initial state $\ket{q_1}$ is given as
\begin{align}
\langle \{q_1, q_2\} \rangle & = \int\textrm{d}q_1 \textrm{d}q_2 q_1 q_2 |\bra{q_2}U\ket{q_1}|^2 \nonumber\\ 
&=  \frac{\int\textrm{d}q_1 \textrm{d}q_2 q_1 q_2 \left| \int_{q(t_1) = q_1}^{q(t_2) = q_2} [\textrm{d}q(t)] \exp[-\mathcal{S}(\bm{q})/\hbar]\right|^2}{\left|\int [\textrm{d}q(t)]\exp[-\mathcal{S}(\bm{q})/\hbar]\right|^2}
\end{align}
Correlations are defined as the expectation values of measurement outcomes. However, path integrals and pseudo-density matrices use different positive measure to calculate the expectation values. 
The correlations in path integrals use the amplitude as the measure, while in pseudo-density matrices the measure is the absolute square of the integrated amplitudes, or we say the probability. 

To see the difference, we consider a quantum harmonic oscillator. The Hamiltonian is given as $H = \hat{p}^2/2m + m\omega^2\hat{q}^2/2$. 
Note that the quantum amplitude of a quantum harmonic oscillator is given as
\begin{equation}
\bra{q_2}U(t_2, t_1)\ket{q_1} = \left(\frac{m\omega}{2\pi\hbar\sinh \omega \tau}	\right)^{1/2}\exp\left\{ -\frac{m\omega}{2\hbar\sinh \omega\tau} [(q_1^2+q_2^2)\cosh \omega \tau - 2q_1q_2] \right\},
\end{equation}
where $\tau = t_2 - t_1$.
In the Gaussian representation of pseudo-density matrices, temporal correlations are represented as
\begin{equation}
\langle \{ q_1, q_2 \} \rangle = \int\textrm{d}q_1 \textrm{d}q_2 q_1 q_2 |\bra{q_2}U\ket{q_1}|^2 = \frac{\hbar}{8m\omega\sinh^2\omega\tau}.
\end{equation}
However, in the path integral formalism, we consider 
\begin{equation}
\Tr U_G(\tau/2, -\tau/2; b) = \int [\textrm{d}q(t)] \exp[-\mathcal{S}_G(q, b)/\hbar]	
\end{equation}
with  
\begin{equation}
\mathcal{S}_G(q, b) = \int_{-\tau/2}^{\tau/2} \textrm{d}t [\frac{1}{2}m\dot{q}^2(t) + \frac{1}{2}m\omega^2q^2(t) - b(t)q(t)]
\end{equation}
and periodic boundary conditions $q(\tau/2) = q(-\tau/2)$. 
We have 
\begin{equation}
\mathcal{Z}_G(\beta, b) = \Tr U_G(\hbar\beta/2, -\hbar\beta/2; b) = \mathcal{Z}_0(\beta) \left \langle \exp\left[ \frac{1}{\hbar}\int_{-\hbar\beta/2}^{\hbar\beta/2}\textrm{d}t b(t)q(t)\right] \right \rangle_0
\end{equation}
where $\langle \cdot \rangle_0$ denotes the Gaussian expectation value in terms of the distribution $e^{-\mathcal{S_0}/\hbar}/\mathcal{Z}_0(\beta)$ and periodic boundary conditions. Here $\mathcal{Z}_0(\beta)$ is the partition function of the harmonic oscillator as 
\begin{equation}
\mathcal{Z}_0(\beta) = \frac{1}{2\sinh(\beta\omega/2)} = \frac{e^{-\beta\hbar\omega/2}}{1-e^{-\beta\hbar\omega}}.
\end{equation}
Then two-point correlations functions are given as
\begin{equation}
\langle q(t_1)q(t_2) \rangle 
= \mathcal{Z}_0^{-1}(\beta)\hbar^2 \left.\frac{\delta^2}{\delta b(t) \delta b(u)}\mathcal{Z}_G(\beta, b)\right|_{b=0} 
= \frac{\hbar}{2\omega \tanh(\omega\tau/2)}.
\end{equation}
It is no surprise that the temporal correlations are different in path integrals and in pseudo-density matrices. 
\begin{claim}
In general, temporal correlations in path integrals do not have the same operational meaning as those in pseudo-density matrices since they use different measures, with exception of path-integral representation for spacetime states and decoherence functionals.
\end{claim}
This indicates a fundamental difference of temporal correlations in path integrals and other spacetime approaches, and raises again the question whether the probability or the amplitude serves as the measure in quantum theory. 
It is natural that amplitudes interferes with each other in field theory and expectation values of operators are defined with the amplitude interference. Thus temporal correlations in path integrals cannot be operationally represented as pseudo-density matrices. 
However, spacetime states defined via position measurements and weak measurements in pseudo-density matrix formulation~\cite{zhang2020different} are motivated by the path integral formalism and have a path-integral representation naturally. In addition, consistent histories also have a path-integral representation of decoherence functionals as we mentioned earlier.  

\section{Conclusion and discussion}

In this section, we unify these spacetime approaches in non-relativistic quantum mechanics and summarise all the claims.

We consider a unified picture in which temporal correlations serve as a connection for indefinite causal structures, consistent histories, generalised quantum games and OTOCs. Given a tripartite pseudo-density matrix, a qubit in the state $\rho$ evolves in time under the unitary evolution $U$ and then back in time under $U^{\dag}$. The correlations in the pseudo-density matrix are given as 
\begin{equation}\label{tripdm}
\langle \sigma_i, \sigma_j, \sigma_k \rangle = \sum_{\alpha, \beta, \gamma = \pm 1} \alpha \beta \gamma \Tr[P^{\gamma}_k U^{\dag} P^{\beta}_j U P^{\alpha}_i \rho P^{\alpha}_i U^{\dag} P^{\beta}_j U]	
\end{equation}
where $P^{\alpha}_i = \frac{1}{2}(\mathbbm{1}+\alpha\sigma_i)$, $P^{\beta}_j = \frac{1}{2}(\mathbbm{1}+\beta\sigma_j)$ and $P^{\gamma}_k = \frac{1}{2}(\mathbbm{1}+\gamma\sigma_k)$. 
As the pseudo-density matrix belongs to indefinite causal structures, we won't discuss the transformation for other formalisms of indefinite causal structures. 

For consistent histories, we assume the state in $\rho$ at the initial time and construct a set of histories $[\chi] = [\alpha \rightarrow \beta \rightarrow \gamma]$ with projections $\{P^{\alpha}_i, P^{\beta}_j, P^{\gamma}_k\}$. Then the decoherence functional is given as 
\begin{equation}
D([\xi], [\xi']) =\Tr[P^{\gamma}_k U^{\dag} P^{\beta}_j U P^{\alpha}_i \rho P^{\alpha'}_i U^{\dag} P^{\beta'}_j U P^{\gamma'}_k ] 	
\end{equation}
When we apply the consistency conditions, it is part of Eqn.~\eqref{tripdm} as 
\begin{equation}
D([\xi], [\xi]) =\Tr[P^{\gamma}_k U^{\dag} P^{\beta}_j U P^{\alpha}_i \rho P^{\alpha}_i U^{\dag} P^{\beta}_j U P^{\gamma}_k ],
\end{equation}
\begin{equation}
\langle \sigma_i, \sigma_j, \sigma_k \rangle = \sum_{\alpha, \beta, \gamma = \pm 1} \alpha \beta \gamma D([\xi], [\xi]). 
\end{equation}

A quantum-classical signalling game is described in terms of one player Abby at two times in a loop, or one player Abby at three times with evolution $U$ and $U^{\dag}$. 
The quantum-classical signalling game is formulated by $(\pi(x,y), l(a, b|x, y))$ on 
\begin{equation}
\overrightarrow{qcsg} = \langle \{\tau^x\}, \{\omega^y\}, \{\eta^z\}; \mathcal{A}, \mathcal{B}, \mathcal{C}; l \rangle.
\end{equation}
The referee associates three quantum systems in the states $\tau^x$, $\omega^y$ and $\eta^z$ with the questions chosen from the question spaces $x \in \mathcal{X}$, $y \in \mathcal{Y}$, and $z \in \mathcal{Z}$. Suppose Abby at $t_1$ receives $\tau^x_X$ and makes a measurement of instruments $\{M_i^a\}_i$ with the outcome $a$. From $t_1$ to $t_2$, the quantum output evolves under the unitary quantum memory $U: A \rightarrow B$. After that, Abby receives the output of the channel and $\omega^y$, and makes a measurement of instruments $\{N_j^b\}_j$ with the outcome $b$. Then, we can consider that either the quantum memory goes backwards to $t_1$ or evolves under $U^{\dag}: B \rightarrow C$ to $t_3$. Abby receives the output of the channel again and $\eta^z$, and makes a measurement of instruments $\{O_k^c\}_k$ with the outcome $c$.
Then we have 
\begin{equation}
p_q(a, b, c | x, y, z) = \sum_{\lambda, i, j, k} \pi(\lambda) \Tr[O_k^c U^{\dag} N_j^b U M_i^a \rho M_i^a U^{\dag} N_j^b U O_k^c].
\end{equation}
If we properly choose the measurements, we will have the decoherence functionals and the probabilities in the correlations of pseudo-density matrix.

What is more, the tripartite pseudo-density matrix we describe is just the one we used to construct OTOC. 
Thus, through this tripartite pseudo-density matrix, we gain a unified picture for indefinite causal order, consistent histories, generalised quantum games and OTOCs in which temporal correlations are the same or operationally equivalent. Thus all these approaches are mapping into each other directly in this particular case via temporal correlations. Generalisation to more complicated scenarios are straightforward.

Now we conclude that there is not much difference in different spacetime approaches for non-relativistic quantum mechanics under this comparison of temporal correlations except path integrals. They are closely related compared with pseudo-density matrices and formulate temporal correlations in the same way or operationally equivalent. However, the path integral approach of quantum mechanics give temporal correlation in a different way. 
Via the pseudo-density matrix formalism, we establish the relations among different spacetime formulations like indefinite causal structures, consistent histories, generalised nonlocal games, out-of-time-order correlation functions, and path integrals. 
As we can see, all these relations are rather simple. The big surprise we learn from these relations is that almost everything we know about space-time in non-relativistic quantum mechanics so far is connected with each other but path integrals are not.
Thus, it shows the possibility of a unified picture of non-relativistic quantum mechanics in spacetime and a gap to relativistic quantum field theory. 
We claim: \\ 
(1) A process matrix and the corresponding pseudo-density matrix allow the same correlations or probabilities in three different mappings. \\
(2) The decoherence functional in consistent histories is the probabilities in temporal correlations of pseudo-density matrices. \\ 
(3) The probability in a quantum-classical signalling game with a trivial input at later time corresponds to the probability in a pseudo-density matrix with quantum channels as measurements. \\
(4) OTOCs can be represented as temporal correlations in pseudo-density matrices with half numbers of steps for calculation; for example, a four-point OTOC, usually calculated by evolving forwards and backwards twice, is represented by a tripartite pseudo-density matrix with only once evolving forwards and backwards. \\
(5) In general, temporal correlations in path integrals do not have the operational meaning as those in pseudo-density matrices since they use different measures, with exception of path-integral representation for spacetime states and decoherence functionals. \\
A unified theory for non-relativistic quantum mechanics is suggested; nevertheless, how to move on to relativistic quantum information, or further to quantum gravity, is still a big gap worth exploring.

\chapter{\label{ch:5-tc}Time crystals as long-range order in time}
\clearpage
\minitoc
\clearpage
\section{Literature review for time crystals}

In this section we review spontaneous symmetry breaking, time translation symmetry breaking, and a few mathematical definitions for time crystals. 

\subsection{Spontaneous symmetry breaking}

Spontaneous symmetry breaking~\cite{strocchi2005symmetry} occurs when the ground state does not hold the symmetry which the equation of motion or the Lagrangian holds. 
Phases of matter are described by spontaneous symmetry breaking. 
For example, the spontaneous breaking of continuous space translation symmetry gives a normal spatial crystal with periodic structures; spin rotational symmetry is spontaneously broken with a net magnetisation along certain direction in ferromagnets, in contrast that spins are uncorrelated in a paramagnetic phase without a net magnetisation. 

There are two diagnostics for spontaneous symmetry breaking~\cite{khemani2019brief}. Note that in equilibrium the expectation values of the order parameter are zero and cannot serve as a measure for spontaneous symmetry breaking. However, we can use two-point correlation functions, when taken the long distance range, to be long-range order. 
\begin{equation}
\lim_{|\bm{r}-\bm{r'}|\rightarrow \infty} \lim_{V \rightarrow \infty} \langle C(\bm{r}, \bm{r'}) \rangle = \lim_{|\bm{r}-\bm{r'}|\rightarrow \infty} \lim_{V \rightarrow \infty} \langle O(\bm{r})O(\bm{r'})\rangle - \langle O(\bm{r})\rangle \langle O(\bm{r'})\rangle \neq 0,
\end{equation}
where $O(\bm{r})$ is a local order parameter and $\langle \cdot \rangle $ is the expectation value in the equilibrium Gibbs states (or eigenstates). This is the standard diagnostic for spontaneous symmetry breaking. Another diagnostic is to add a small symmetry breaking field with strength $h$ and compute the expectation value of the global order parameter $\langle O \rangle_h$ which turns into non-zero.
\begin{equation}
	\lim_{h \rightarrow 0} \lim_{V \rightarrow \infty} \frac{1}{V} \langle O \rangle_h \neq 0. 
\end{equation}
One variant is to apply a small symmetry breaking field at the boundaries and evaluate how the expectation value of order parameter have influence on the bulk.


The Goldstone theorem~\cite{nambu1960quasi, goldstone1961field, goldstone1962broken} states that at least one massless bosonic state exists in the spectrum when the theory allows a universal symmetry to be spontaneously broken. 
The Mermin-Wagner theorem~\cite{mermin1966absence, hohenberg1967existence, gelfert2001absence} concludes that, in one or two dimensions, continuous symmetries cannot be spontaneously broken at finite temperature in systems with sufficiently short-range interaction. Hohenberg~\cite{hohenberg1967existence} shows no phase transition at finite temperature for one- and two-dimensional superfluid systems; Mermin and Wagner~\cite{mermin1966absence} further exclude the possibility for spontaneous magnetisation in the Heisenberg model. 

\subsection{Time translation symmetry breaking}

Time translation symmetry breaking is associated with the emergence of time crystals, as an analogue to ordinary spatial crystals. 
In the following context, we only focus on quantum time crystals. 
We know that time-independent systems preserve the continuous time translation symmetry, and when the continuous time translation symmetry is broken, the system displays certain time-dependent properties. For an operator $O$ without the intrinsic time dependence, we have 
\begin{equation}
\bra{\Psi} \dot{O} \ket{\Psi} = i \bra{\Psi} [H, O] \ket{\Psi} = 0  \qquad \text{for} \ \Psi = \Psi_E
\end{equation}
where $\Psi_E$ is the eigenstate of the Hamiltonian $H$ of the system. 
It seems impossible for the breaking of even an infinitesimal time translation symmetry. 
However, in the spatial analogue, one-point expectation values do not serve as a proper diagnostic for spontaneous symmetry breaking as well. 
Wilczek initially proposes a model with periodic motion in the ground state~\cite{wilczek2012quantum}. Later it is pointed out that periodic motion are exhibited in some excites state instead and the actual ground state does not show any time crystallinity~\cite{bruno2013comment}. Further the possibilities of any spontaneous rotating time crystals are excluded~\cite{bruno2013impossibility}. 

In general, continuous time crystals are proved to be impossible in the ground state and in the equilibrium~\cite{watanabe2015absence}. 
More specifically from Ref.~\cite{watanabe2015absence}, two-point temporal correlation functions do not have a period to break the continuous time translation symmetry but tend to be time independent under the large volume limit for the system.  Note that Ref.~\cite{khemani2019brief} pointed out some errors in the original proofs. 

Instead of continuous time translation symmetry, we may also consider whether discrete time translation symmetry can be broken down in periodically driven systems. These are referred to Floquet time crystals~\cite{else2016floquet}, or discrete time crystals~\cite{yao2017discrete}. 
In many-body localised Floquet systems with a period-$T$ driving, the temporal correlations exhibit a period of $nT ( n > 1)$ or a Fourier peak at $k/n$-frequency $(k = 1, 2, \dots, n)$ and show robustness of the perturbation. 
Experimental verification for discrete time crystals has been conducted in trapped ions~\cite{zhang2017observation}, nitrogen-vacancy centres in diamond~\cite{choi2017observation}, and NMR~\cite{rovny2018observation, rovny2018p}.

There are also other variants of time crystals, like prethermal continuous time crystals~\cite{else2017prethermal}, boundary time crystals~\cite{iemini2018boundary}, cosmological time crystals~\cite{das2018cosmological, feng2018cosmological}, time quasi-crystals~\cite{autti2018observation}, and so on. 

\subsection{Mathematical definitions of time crystals}

There are a few mathematical definitions for time crystals. They are consistent with each other but exhibit in different forms. As the temporal analogue of crystals, time crystals are expected to break time translation symmetry and exhibit long-range correlations in time. 

These definition use two-point correlations functions in space and time, take the large volume limit and show symmetry-breaking properties in time. We introduce Watanabe and Oshikawa's definitions first via local and integrated order parameters respectively for continuous time translation symmetry. Then we offer the corresponding definitions for discrete time translation symmetry. We also give a practical definition for experimental use. We further illustrate the definitions in the representation theory.

The mathematical definition of time crystals is firstly given by Watanabe and Oshikawa~\cite{watanabe2015absence} via time-dependent long-range order. In this way, they argue that time crystals cannot exist in the ground state or in the equilibrium. Long-range order exists if the spatial correlation of a local order parameter $\hat{\phi}(\vec{x}, t)$ has a non-zero limit 
\begin{equation}
\lim_{V \rightarrow \infty} \langle \hat{\phi}(\vec{x}, 0) \hat{\phi}(\vec{x}', 0) \rangle \rightarrow c \neq 0,
\end{equation}
for $|\vec{x} - \vec{x}'|$ very large compared to microscopic scales we are considering. 
In the representation of integrated order parameter $\hat{\Phi} \equiv \int_V \mathrm{d}^dx \hat{\phi}(\vec{x}, 0)$, the long-range order is defined to exist when $\lim_{V \rightarrow \infty} \langle \hat{\Phi}^2 \rangle / V^2 = c \neq 0$. 
As crystals are characterised by long-range order, time crystals are defined analogously in terms of the temporal version of long-range correlations; that is, 
\begin{equation}
\lim_{V \rightarrow \infty} \langle \hat{\phi}(\vec{x}, t) \hat{\phi}(0, 0) \rangle \rightarrow f(t),
\end{equation}
$f(t)$ is a non-vanishing periodic function for large $|\vec{x}|$, or, 
\begin{equation}
\lim_{V \rightarrow \infty} \langle e^{i\hat{H}t}\hat{\Phi}e^{-i\hat{H}t}\hat{\Phi} \rangle / V^2 \rightarrow f(t).
\end{equation}

In the above we consider continuous time translation symmetry. For discrete time translation breaking~\cite{khemani2017defining}, we denote the local operator $O_i$ with the subscript $i$ for the position. Then we have the long-range order in time as 
\begin{equation}
\lim_{|i-j| \rightarrow \infty} \lim_{L \rightarrow \infty} \langle O_i(t) O_j \rangle = f(t),
\end{equation}
where $L$ is the system size. 
Or consider the superposition of local operators $O = \frac{1}{L} \sum_i c_i O_i$, then long-range order in time can be written as 
\begin{equation}
\lim_{L \rightarrow \infty} \langle O^{\dag}(t) O \rangle = f(t).
\end{equation}
With a Floquet unitary $U(T)$, the system exhibits the temporal correlations in the limit of large system size when $f(t)$ has a period $t = n T$, $n \in \mathbb{Z}$; this is a special case for the so-called discrete time crystals. 
In particular, time translation symmetry breaking is defined in Ref.~\cite{else2016floquet} when the expectation values of a local operator are different in a period of the Floquet system for every state with short-range correlations. The short-range correlations exist in a state $\ket{\psi}$ when $\bra{\psi} \phi(x) \phi(x') \ket{\psi} - \bra{\psi} \phi(x) \ket{\psi} \bra{\psi} \phi(x') \ket{\psi} \rightarrow 0$ for any local operator $\phi(x)$.

However, in practice, it is hard to measure 
this long-range order in time for experiments. Thus, a adapted definition frequently used in experiments is given by 
\begin{equation}
\lim_{t \rightarrow \infty} \lim_{V \rightarrow \infty} \bra{\psi_0} O_i(t) \ket{\psi_0} = f(t)
\end{equation}
with $\ket{\psi_0}$ as a generic short-range correlated initial state. 

Another definition uses the representation theory~\cite{khemani2017defining}. 
Suppose that a family of local order parameters $\Phi_{i, \alpha}$, labeled by the position $i$ and the irreducible representation $\alpha$ of the time translation symmetry (either continuous $\mathbb{R}$ or discrete $\mathbb{Z}$), transform under nontrivial irreducible representations as $U^{\dag}(t)\Phi_{i, \alpha} U(t) = e^{i \Delta_{\alpha}t} \Phi_{i, \alpha}$. By $\alpha$ nontrivial, we mean that $\bra{n} \Phi_{i, \alpha} \ket{n} = 0$ for all eigenstates $\ket{n}$ of either the Hamiltonian $H$ or the Floquet unitary $U(T)$. Then continuous or discrete symmetry $\mathbb{R}$ or $\mathbb{Z}$ is spontaneously broken into a discrete subgroup $H$, if (1) 
\begin{equation}
\lim_{|i-j| \rightarrow \infty} \lim_{L \rightarrow \infty} |\bra{n} \Phi_{i, \alpha} \Phi_{j, \bar{\alpha}} \ket{n} - \bra{n} \Phi_{i, \alpha} \ket{n} \bra{n}  \Phi_{j, \bar{\alpha}} \ket{n} | = c_0 \neq 0
\end{equation}
for $\Phi_{i, \alpha}$ transforming trivially under $H$ but nontrivially under $\mathbb{R}$ or $\mathbb{Z}$; and (2) 
\begin{equation}
\lim_{|i-j| \rightarrow \infty} \lim_{L \rightarrow \infty} |\bra{n} \Phi_{i, \alpha} \Phi_{j, \bar{\alpha}} \ket{n} - \bra{n} \Phi_{i, \alpha} \ket{n} \bra{n}  \Phi_{j, \bar{\alpha}} \ket{n} | = 0
\end{equation}
for $\Phi_{i, \alpha}$ transforming nontrivially under $H$.

%
%
%
%
%
%
%
%
%
%
%

\section{Definition: time crystals as long-range order in time}
In this section, we propose a definition for time crystals in the pseudo-density matrix formalism. It is consistent with all other definitions proposed so far. Before that, we add a bit more discussion for long-range order. 

%
%

\subsection{Long-range order}
A crystal or crystalline solid is defined as a solid material 
whose constituents are arranged in a periodic array on the microscopic level. Specifically, in a unit cell, the arrangement of atoms or other constituents is repeated again and again under the translation invariance. This lattice periodicity as the defining property of a crystal, implies long-range order: the orderliness over long distances can be predicted with the knowledge of one cell and the translation symmetry. In Ref.~\cite{yang1962concept}, it is explained that the solid phase is characterised by the existence of a long-range correlation. In other words, it is known that a solid is crystalline if it has long-range order. Thus a crystal is characterised by the existence of long-range order. A crystal has long-range order in space; a time crystal, however it is defined, should have long-range order in time. 

To define a time crystal, the off-diagonal long-range order might be interesting as well. 
The long-range order in the solid is exhibited in the quantum mechanics in the diagonal element of the reduced density matrix $\rho_2$ in the coordinate space~\cite{yang1962concept}. 
For a density matrix $\rho$ with $\Tr \rho = 1$, reduced density matrices $\rho_1$, $\rho_2$, $\cdots$ are defined as $\bra{j} \rho_1 \ket{i} = \Tr (a_j\rho a_i^{\dag})$, $\bra{kl} \rho_2 \ket{ij} = \Tr (a_k a_l \rho a_j^{\dag} a_i^{\dag})$, etc., where $a_i$, $a_j$ represent annihilation operators for the one-particle states $\ket{i}$, $\ket{j}$. 
Then the off-diagonal long-range order exists if $\bra{\mathbf{x'}}\rho_1\ket{\mathbf{x}}$ does not vanish as $|\mathbf{x} - \mathbf{x'}| \rightarrow \infty$. 
Yang~\cite{yang1962concept} also defines the off-diagonal long-range order in a many-body system of bosons or fermions with annihilation operators on different particles; the order characterises the existence of a Bose-Einstein condensation in the phases He II and superconductors.
Thus, we expect that a time crystal is characterised by the existence of long-range order in time and take it as the definition of a time crystal.

\subsection{Time crystals in terms of temporal correlations}
In the pseudo-density matrix formulation, the measure of long-range order in time~\cite{zhang2019long} is expressed as the two-point temporal correlation at times $t_1$ and $t_N$:
\begin{equation}
\langle \sigma^{(1)}, \sigma^{(N)} \rangle =  \Tr\{(\sigma^{(1)}\otimes \sigma^{(N)})R_{1,N}[\rho, \Phi]\},
\end{equation}
where $R_{1,N}$ is the pseudo-density matrix between $t_1$ and $t_N$, $\rho$ is the initial state and $\Phi$ is the channel evolution between different times.
$\sigma^{(1)}$ and $\sigma^{(N)}$, for example, can be chosen as Pauli operators measured at $t_1$ and $t_N$ in the spin chains. 
For $\rho$ with multiple qubits, $\sigma^{(1)}$ and $\sigma^{(N)}$ are usually acted on different qubits with a large separation in space. 

As a simple example, we consider the long-range order in time for a single spin under the unitary evolution. The temporal correlation is always preserved and no symmetry is broken here.
Consider a qubit evolving unitarily in time. 
Suppose the qubit is in an arbitrary state $\rho = \frac{1}{2}\mathbbm{1} + \sum_{i=1, 2, 3} c_i \sigma_i$.
From the time $t_k$ to $t_{k+1}$, $k= 1, 2, \dots$, the qubit evolves under the same unitary evolution $U$. 
Consider the temporal correlation from $t_1$ to $t_n$,
\begin{align}\label{unitaryevo}
\langle \sigma_i, \sigma_j \rangle & = \Tr [\sigma_i \otimes \sigma_j R_{1n} ] \nonumber\\
& = \sum_{\alpha, \beta = \pm 1} \Tr[P^{\beta}_j U^{n-1} P^{\alpha}_i \rho P^{\alpha}_i (U^{\dag})^{n-1}] \nonumber\\
& = \frac{1}{2} (\Tr[\sigma_j U^{n-1} \rho \sigma_i (U^{\dag})^{n-1}] + \Tr[\sigma_j U^{n-1}\sigma_i \rho (U^{\dag})^{n-1}]) \nonumber\\
& = \frac{1}{2} \Tr[\sigma_j U^{n-1} \sigma_i (U^{\dag})^{n-1}].
\end{align}
Here $P^{\alpha}_i = \frac{1}{2}(\mathbbm{1} + \alpha \sigma_i)$. For the last equality, we use $ \rho \sigma_i +\sigma_i \rho = 2c_i I+ \sigma_i$ with $\rho = \frac{1}{2}I + \sum_{i=1, 2, 3} c_i \sigma_i$.
Take $i=j$, Eqn.~\eqref{unitaryevo} is equivalent to 
\begin{equation}
\langle \sigma_i, (U^{\dag})^{n-1} \sigma_i U^{n-1} \rangle = 1.
\end{equation}
We conclude that long-range order in time is preserved for the unitary evolution. 

\section{Continuous time translation symmetry}
In this section, we discuss continuous time translation symmetry in terms of general decoherent processes and the Mermin-Wagner theorem. 


%
%

\subsection{General decoherent process}
We have considered a single spin under the unitary evolution in the previous section. 
In practice, interaction with the environment is unavoidable and noise is always present.
For a qubit evolving under a generic decohering channel evolution, $\Phi$, we prove that there is no long-range order in time for whatever strength of the decoherence. Specifically, there exists an effective rate $\gamma <1$ from one time to the next 
\footnote{Note that in the case where only ($Z$) dephasing noise is present in the system, the (unrealistic) exact initial state preparation of $\rho =(I + rZ)/2$, could lead to long-range order in time.\label{ftn1}}, for which the long-range order in time is bounded by 
\begin{equation}
\Tr\{(X^{(1)}\otimes X^{(N)})R_{1,N}[\rho, \Phi]\} \leq \gamma^{N-1},
\end{equation}
which tends to $0$ exponentially as $N\rightarrow \infty$.

For the depolarising noise~\cite{nielsen2002quantum}, suppose the evolution between two times $t_{k}$ and $t_{k+1}$ (k = 1, 2, ... , N-1) is
\begin{equation}\label{depolarisingChannel}
\Phi: \rho \rightarrow (1-p)\rho + p \frac{I}{2}.
\end{equation}
For an arbitrary initial state,
the two-time correlation function $\langle XX \rangle$ between $t_1$ and $t_N$ is
\begin{equation}\label{depoloarisingnoise}
\langle X^{(1)}, X^{(N)} \rangle = \Tr[(X\otimes X) R_{1N}] = (1-p)^{N-1}.
\end{equation}
It goes down exponentially with N, which suggests that the temporal long-range order vanishes and no possible existence for time crystals. 

Dephasing corresponds to the Bloch vector transformation
\begin{equation}
\Phi: \vec{r} = (r_x, r_y, r_z) \rightarrow (r_x\sqrt{1-\lambda}, r_y\sqrt{1-\lambda}, r_z),
\end{equation}
where $\vec{r}$ is a three component real vector and the state of a single qubit is written in the Bloch representation $\rho = \frac{I + \vec{r} \cdot \vec{\sigma}}{2}$, $\vec{\sigma} = (X, Y, Z)$; $e^{-t/2T_2} = \sqrt{1-\lambda}$ with the dephasing as a `$T_2$' (or `spin-spin') relaxation process~\cite{nielsen2002quantum}.
For an arbitrary initial state,
suppose the evolution between two times $t_{k}$ and $t_{k+1}$ (k = 1, 2, ... , N-1) is $\Phi$, 
the two-time correlation function $\langle XX \rangle$ between $t_1$ and $t_N$ is
\begin{equation}\label{dephasingnoise}
\langle X^{(1)}, X^{(N)} \rangle = \Tr[(X\otimes X) R_{1N}] = (\sqrt{1-\lambda})^{N-1}.
\end{equation}
It also goes down exponentially with N, so that the temporal long-range order vanishes and no time-crystalline phase can exist.

Another example may be a spin-echo unitary in an open system. 
Suppose the Hamiltonian is given by $H = \frac{1}{2}\omega\sigma_z$. Putting it into the Lindblad master equation, we have
\begin{equation}
\frac{\partial \rho}{\partial t} = \frac{i\omega}{2} [\sigma_z, \rho] + \frac{\gamma}{2} (\sigma_z \rho \sigma_z - \rho)
\end{equation}
The solution $\Phi: \rho(0) \rightarrow \rho(t)$ is given in terms of the matrix elements
\begin{equation}
\begin{split}
\rho_{00}(t) & = \rho_{00}(0) \\
\rho_{01}(t) & = \rho_{01}(0)e^{-i\omega t - \gamma t} \\
\rho_{10}(t) & = \rho_{10}(0)e^{i\omega t - \gamma t} \\
\rho_{11}(t) & = \rho_{11}(0)
\end{split}
\end{equation}
For $\rho = \frac{1}{2}(I+X)$,
So
\begin{equation}
\langle \sigma_i, \sigma_j  \rangle = \sum_{\alpha, \beta = \pm 1} \alpha\beta\Tr (\frac{1+\beta \sigma_j}{2} \Phi(\frac{1+\alpha \sigma_i}{2})  ) \Tr(\frac{1+\alpha \sigma_i}{2} \rho) = \cos(\omega t)e^{-\gamma t}.
\end{equation}
Similarly, long-range order in time vanishes under the dephasing noise.

Now we consider a general decohering evolution. Instead of a particular kind of noise, we assume the evolution under a completely positive trace-preserving map $\mathcal{E}: \rho \rightarrow \sum_k E_k \rho E_k^{\dag}$ with $\sum_k E_kE_k^{\dag} = I$. For every $E_k$, $E_k < I$ in the decohering case; then there exists $\gamma < 1$ such that $E_k \leq \gamma I$ for all $k$. Then, for one round of evolution, 
\begin{equation}
Tr [(\sigma_i \otimes \sigma_j )R] = \sum_k \Tr[\sigma_j E_k \sigma_i E_k^{\dag}]
\end{equation}
For $n$ rounds of evolution,
\begin{align}
Tr [(\sigma_i^{(1)} \otimes \sigma_i^{(n+1)})R] & = \sum_{k_1, \dots, k_n} \Tr[\sigma_i E_{k_n} \cdots E_{k_1} \sigma_i E_{k_1}^{\dag} \cdots E_{k_n}^{\dag}] \nonumber\\
& =  \sum_{k_1, \dots, k_n} \Tr[E_{k_n} \cdots E_{k_1} \sigma_i (E_{k_n} \cdots E_{k_1} \sigma_i)^{\dag}] \nonumber\\
& \leq \gamma^{2n}.
\end{align}
The long-range temporal correlations decay exponentially in time, suggesting that the order vanishes. Thus under arbitrary decoherent evolutions in terms of CPTP maps, a single spin has no long-range order in time. This result and the discussion on unitary evolutions, exclude the possibility of time crystals in 0+1 dimension unless we take the definition too trivial.

\subsection{Generalised Mermin-Wagner theorem}

We mentioned the Goldstone theorem and the Mermin-Wagner theorem in the literature review part. Here we discuss how to apply them to continuous time translation symmetry breaking and time crystals.

In one of the early papers on the Goldstone theorem, it states that if the Lagrangian of the system is invariant under the continuous symmetry transformation, then either spinless particles with zero mass exist, or the vacuum state is invariant~\cite{goldstone1962broken}. That is, for a local scalar field $\phi$ and a local conserved vector current $j_{\nu}$ with $\partial^{\nu} j_{\nu} = 0$, either Goldstone bosons exist, or the expectation value of $\delta \phi$ vanishes in the vacuum state, where the scalar field $\delta \phi$ is defined by
$
\delta \phi(y) = i \int \textrm{d}^3 \boldsymbol{x} [j_0(x_0, \boldsymbol{x}), \phi(y)].
$
In Ref.~\cite{coleman1973there}, it is argued that the vacuum expectation value of $\delta \phi$ always vanishes for two-dimensional spacetime:
\begin{equation}
\bra{0} \delta \phi (0) \ket{0} = i \bra{0} \int \textrm{d}x_1 [j_0(x_0, x_1), \phi(0)] \ket{0} = 0.
\end{equation}
This suggests continuous symmetry cannot be broken in 1+1 dimensions for the ground state. 
The proof is straightforward for continuous time translation symmetry as the Hamiltonian $H$ is conserved. In 1+1 dimensions, $H \delta t = \int \textrm{d}x j_0(t, x)$; thus, the expectation value of $\delta \phi$ vanishes in the vacuum state. For a general continuous symmetry, consider the integrals $F_{\nu}(k_0, k_1) = \iint \textrm{d}x_0\textrm{d}x_1 e^{i(k_0x_0+k_1x_1)}\bra{0}j_{\nu}(x_0, x_1)\phi(0)\ket{0}$ in the momentum space. After solving the integrals from conservation conditions, we find the only contribution to $\bra{0} \delta \phi (0) \ket{0}$ vanishes to avoid a singularity. The proof is given similarly as in Ref.~\cite{coleman1973there}. 



Further, we apply the Mermin-Wagner theorem~\cite{mermin1966absence, hohenberg1967existence} to 1+1 dimensional spacetime and argue that no continuous time translation symmetry breaking occurs for finite temperature due to lack of long-range temporal order. 
More specifically, consider a system in the equilibrium, that is, in the thermal state $\rho = e^{-\beta H} / \Tr(e^{-\beta H})$. The expectation value of an operator $A$ is given by $\langle A \rangle = \lim_{V \rightarrow \infty} \Tr(e^{-\beta H} A) / \Tr(e^{-\beta H})$.
Under the continuous time translation symmetry, the Hamiltonian $H$ serves as the generator and is invariant. From statistical mechanics, we learn that even for an operator $B$ that does not commute with the generator $H$ ($[B, H] = C \neq 0$), the expectation value of $[B, H] = C$ is still 0. 
However, in the spontaneous symmetry breaking, when we add a small perturbation to the Hamiltonian, the expectation value of $[B, H] = C$ does not vanish anymore. Here we use the long-range temporal correlations as the indicator and argue that they vanish as the perturbation parameter goes smaller to exclude the possibility of spontaneous continuous time translation symmetry breaking. 
We take the experimental definition of time crystals and investigate the temporal correlations in the Heisenberg model. We prove that under the finite temperature, when we add a small perturbation to the Hamiltonian, the long-range temporal correlation vanishes for the perturbation goes smaller and smaller. Thus, we conclude that 
there can be no spontaneous breaking for continuous time translation symmetry in 1+1 dimensions for finite temperature. 
The proof is given via the Bogoliubov inequality in Appendix C.

In this subsection, we apply the Goldstone theorem and the Mermin-Wagner theorem to 1+1 dimensional spacetime, and argue that there is no continuous time translation symmetry breaking for the ground states and the equilibrium. This result is consistent with general absence of continuous time crystals and provides a different understanding which might be useful to discuss space-time crystals in relativistic field theory. We leave it for the future work. 

So far, we investigate the possibilities of continuous time translation symmetry breaking in 0+1 and 1+1 dimensions. As a result of the lack of long-range order in time, continuous time translation symmetry cannot be spontaneously broken in these cases. 

\section{Discrete time translation symmetry}
In this section, we investigate discrete time translation symmetry. One possible suggestion from quantum information is to apply periodic stabilisation of quantum computation and quantum error correction to counteract the decoherence to preserve long-range order in time. In this case, discrete time translation symmetry is preserved for the single-qubit case. We further turn on to one-dimensional spin chains under many-body localisation and Floquet driving. That is the usual model considered for discrete time crystals. We apply the pseudo-density matrix formulation to simplify the calculation for temporal correlations and use group theory to gain a better understanding of how the subharmonic periodicity emerges. 

\subsection{Stabilisation of quantum computation}

In this subsection we discuss temporal correlations under the stabilisation of quantum computation. For simplicity, only single-qubit error is considered here. 

\begin{figure}[h]
	\centering
	\includegraphics[scale=0.38]{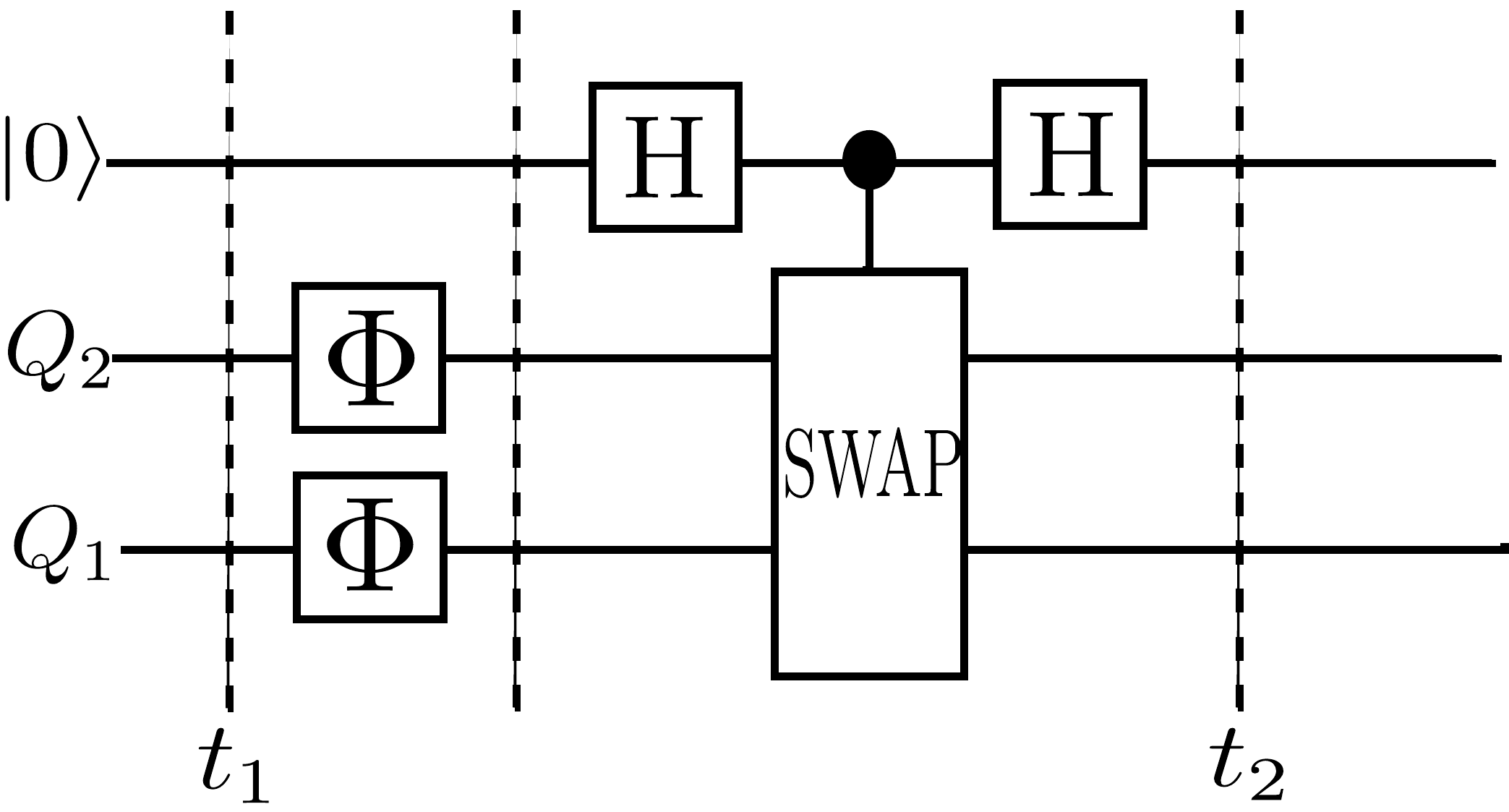}
	\caption{Quantum circuit for symmetrisation error correction as a time crystal. If the auxiliary qubit is found in state $\ket{0}$, the symmetrisation has been successful.}
	\label{fig: symmetrisation}
\end{figure}

Let us recall the principle of stabilisation of quantum computation via the projection onto the symmetric subspace~\cite{barenco1997stabilization}. 
The key idea is that a pure state $\ket{\phi}$ can be protected against decoherence by encoding it redundantly in $N$ qubits and projecting their overall state onto the symmetric subspace (i.e., the minimal subspace containing all the states $\ket{\phi}^{\otimes N}$). For the sake of simplicity, let us use two qubits (see Fig.~\ref{fig: symmetrisation}). This can be generalised to $N$ qubits easily. Suppose the two qubits, initialised in an arbitrary pure state $\ket{\psi} \otimes \ket{\psi}$
, undergo the noisy channel evolution $\Phi\otimes\Phi$, where $\Phi$ is the depolarising noise~\cite{nielsen2002quantum}. 
Suppose the evolution between two times $t_{k}$ and $t_{k+1}$ (k = 1, 2, ... , N-1) is
\begin{equation}\label{depolarisingChannel}
\Phi: \rho \rightarrow \rho_p (1-p)\rho + p \frac{I}{2},
\end{equation}
so that the two qubits evolve into some mixed state $\rho_p$. 
Now, project each of them onto the symmetric subspace 
by measuring the auxiliary qubit and discarding outcomes of 1. This is represented by the operator:
\begin{equation}
\Sigma_{12} = \frac{1}{2}(I_{12} + S_{12}),
\end{equation}
where $S_{12}$ is the SWAP operator acting on the two qubits.
It completes the effective evolution caused by the error correction protocol (which is a probabilistic procedure). 
The action of the projection on a single qubit starting in the state $\rho_p$ is
\begin{equation}
\Sigma:  \rho_p \rightarrow \Tr_2\left[\frac{\Sigma_{12}(\rho_p\otimes\rho_p)\Sigma_{12}^{\dag}}{\Tr(\Sigma_{12}(\rho_p\otimes\rho_p)\Sigma_{12}^{\dag})}\right]  = \frac{\rho_p + \rho_p^2}{\Tr(\rho_p + \rho_p^2)} \equiv \rho',
\end{equation}
where
\begin{equation}
\Tr(\rho'^2) > \Tr(\rho_p^2).
\end{equation}
Thus, error correction by symmetrisation makes the state purer. The convergence to a pure state is improved by acting on a larger number of qubits.

\begin{figure}[h]
	\centering
	\includegraphics[scale=.8]{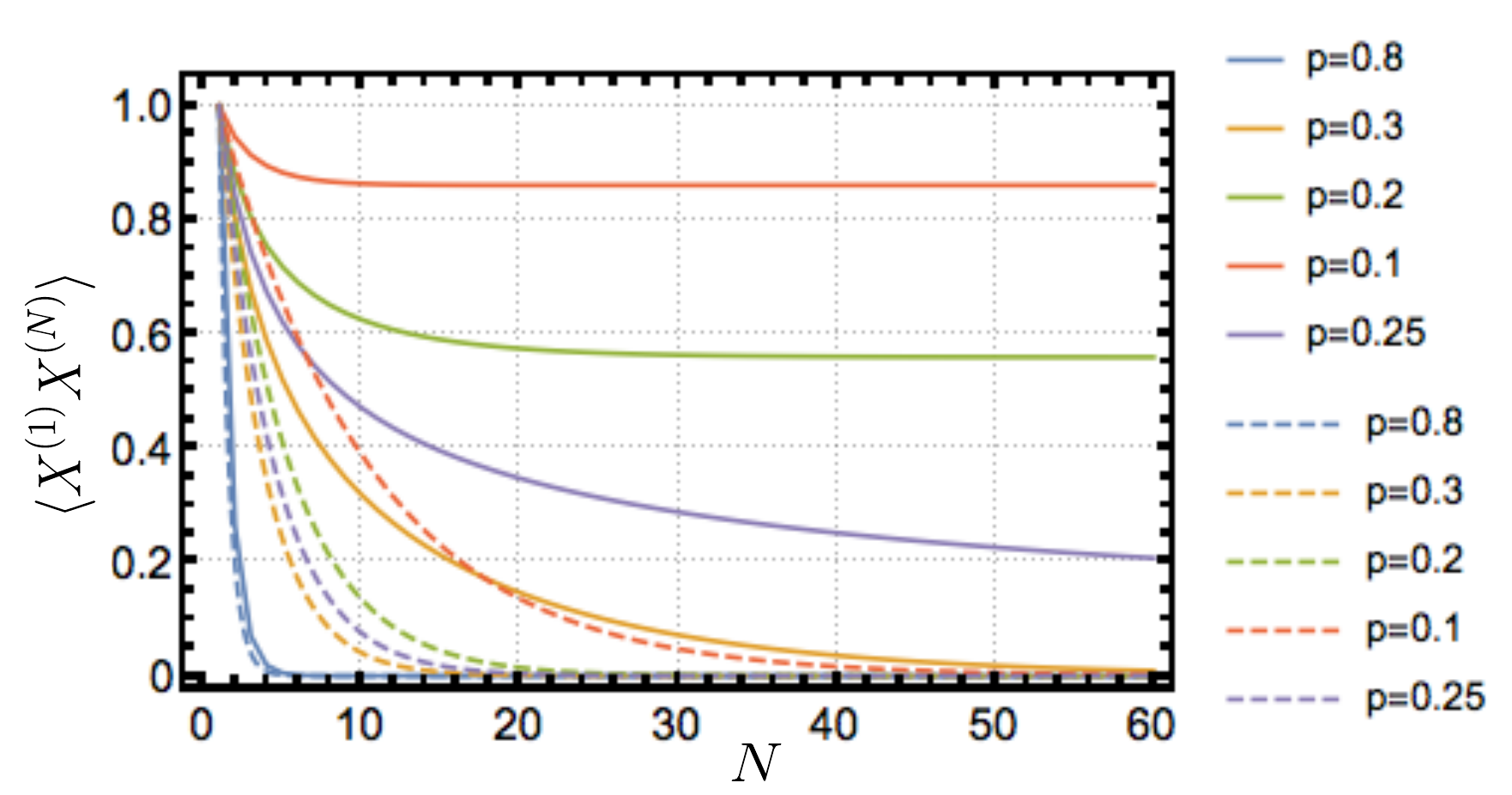}
	\caption{$\langle X^{(1)}X^{(N)} \rangle$ vs.~N for depolarising noise with and without error correction (solid and dashed lines, resp.).}
	\label{fig: correct}
\end{figure}

We can apply the pseudo-density matrix description to the evolution outlined above. 
For an arbitrary initial state, only with depolarising noise $\Phi$ and without symmetrisation error correction $\Sigma$, recall that 
the two-time correlation function $\langle XX \rangle$ between $t_1$ and $t_N$ is
\begin{equation}\label{depoloarisingnoise}
\langle X^{(1)}, X^{(N)} \rangle = \Tr[(X\otimes X) R_{1N}] = (1-p)^{N-1}.
\end{equation}
With both of depolarising noise $\Phi$ and symmetrisation error correction $\Sigma$,
\begin{align}
\langle X^{(1)}, X^{(N)} \rangle = \Tr[(X\otimes X) R_{1N}] = a_N,\nonumber\\
\text{where}\ a_{n+1} = \frac{4a_n(1-p)}{3+a_n^2(1-p)^2}, \ a_1 = 1.
\end{align}
For $p \leq 1/4$, $\langle X^{(1)}X^{(N)} \rangle$ converges to a constant $\frac{\sqrt{1-4p}}{1-p}$ as $N$ becomes large. For $1/4 < p <1$, it decays to 0 with a smaller rate than in the case of uncorrected noise
(cf. Fig.~\ref{fig: correct}).

The analysis is similar for dephasing noise mentioned before. 
For an arbitrary initial state,
suppose the evolution between two times $t_{k}$ and $t_{k+1}$ (k = 1, 2, ... , N-1) is $\Phi$, 
the two-time correlation function $\langle X, X \rangle$ between $t_1$ and $t_N$ is
\begin{equation}\label{dephasingnoise}
\langle X^{(1)}, X^{(N)} \rangle = \Tr[(X\otimes X) R_{1N}] = (\sqrt{1-\lambda})^{N-1}.
\end{equation}
With the full protocol applied,
the two-time correlation function for the first qubit reads
\footnote{As mentioned in footnote~\ref{ftn1}, initial states of form $\rho = (I + rZ)/2$  already exhibit long-range order in time and do not require the protocol for dephasing noise only.}
\begin{align}
\langle X^{(1)}, X^{(N)} \rangle = \Tr[(X\otimes X) R_{1N}] = b_N,\nonumber\\
\text{where}\ b_{n+1} = \frac{4b_n\sqrt{1-\lambda}}{3+b_n^2(1-\lambda)}, \ b_1 = 1.
\end{align}

Comparing the two results with and without error correction, 
we find that the two-time correlation function with error correction converges to a finite value for $p < 1/4$ or $\lambda < 7/16$ and decays at a slower rate otherwise; this implies that (finite) long-range order can in principle be restored. 
Furthermore, if we keep applying the error correction scheme on a larger number $N$ of qubits, the long-range order in time will be fully restored.


\subsection{Quantum error correction of phase flip codes}

\begin{figure}[h]
\centering
\includegraphics[scale=.65]{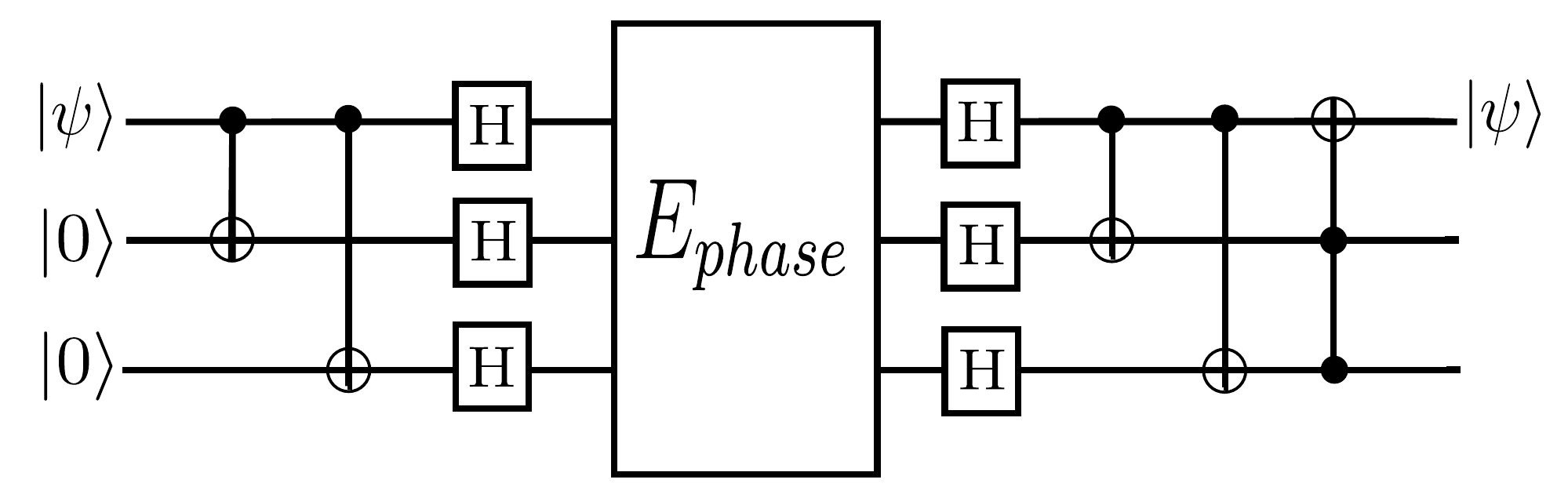}
\caption{Quantum circuit of the phase flip code. The noise $E_{\text{phase}}$ flips $\ket{+}$ to $\ket{-}$ for one qubit and vice versa. In the error model, a qubit is left alone with probability $1 − p$, and with probability $p$ the relative phase of the $\ket{0}$ and $\ket{1}$ states is flipped. That is, the initial state $\alpha \ket{0} + \beta \ket{1}$ goes to the state $\alpha \ket{0} - \beta \ket{1}$ after the phase flip $Z$.} 
\label{fig: phaseflip}
\end{figure}

Now we consider the quantum error correction of phase flip codes. Let $\ket{\psi} = \alpha \ket{0} + \beta \ket{1}$ be an arbitrary qubit. Suppose the only noise is one single phase flip $Z$ on one of the three qubits in Figure~\ref{fig: phaseflip}. 
This occurs with probability $(1-p)^3 + 3p(1-p)^2 = 1 -3p^2 + 2p^3$. 

For single flip and no flip, after the error correction, the state remain unchanged as $\ket{\psi} = \alpha \ket{0} + \beta \ket{1}$ with probability $1 -3p^2 + 2p^3 = 1-q$.
For two or three flips, after the error correction, the state becomes $\ket{\psi'} = \alpha \ket{1} + \beta \ket{0}$ with probability $3p^2 - 2p^3 = q$.
Now apply the protocol for $N$ times. Consider the two-time correlation functions for the first qubit at the initial time $t_1$ and the final time $t_N$\footnote{Consider small $p$. For $N$ odd, the probability $p_c$ to change the state is $C_N^0 q^N + C_N^2 q^{N-2}(1-p)^2 + \cdots + C_N^{N-1}q(1-q)^{N-1} = Nq + O(q^2)$;
for N even, $p_c = C_N^1 q^{N-1}(1-q) + C_N^3 q^{N-3}(1-q)^3 + \cdots + C_N^{N-1}q(1-q)^{N-1} = Nq + O(q^2)$.
\mbox{$\langle Z^{(1)}Z^{(N)} \rangle = 1 - 2p_c$}.}:
\begin{align}
\langle X^{(1)}, X^{(N)} \rangle &= 1,\nonumber\\
\langle Z^{(1)}, Z^{(N)} \rangle &= 1- 2N(3p^2-2p^3)+O(p^4).
\end{align}
In this case, the $\langle XX \rangle$ correlation is always 1 and long-range correlation in time along the $X$ direction is preserved. For small $p$ and finite $N$, long-range correlation in time along the $X$ direction remains close to 1 and is almost preserved. 
We will not discuss a full quantum error correction scheme here as it is very similar with the $X$ direction here, or the $Z$ direction with $p=0$. 
In principle, long-range order in time can be fully restored under a full error correction scheme. 


\subsection{Floquet many-body localisation}
In this subsection, we consider many-body localised systems with Floquet driving. Discrete time translation symmetry is broken and thus the model constitutes a discrete time crystal. We formulate these discrete time crystals in the language of pseudo-density matrices and group theory.

\subsubsection{Temporal correlations in pseudo-density matrix formulation}
Here we calculate temporal correlations from the pseudo-density matrix formulation. In such a particular Floquet many-body localised system, discrete time translation symmetry of a period $T$ is broken to discrete time translation symmetry of a period $nT (n \in \mathbb{Z}, n>1)$. 
In particular, we consider a one-dimensional spin-$\frac{1}{2}$ chain under the binary stroboscopic Floquet Hamiltonian for a period $T = T_1 + T_2$. 
\begin{equation}
H_f(t) = \left\{
\begin{array}{lcl}
H_1 = (g - \epsilon) \sum_i \sigma_i^x & & {0 < t < T_1}\\
H_2 = \sum_i J_i\sigma_i^z \sigma_{i+1}^z + h_i^z\sigma_i^z + h_i^x\sigma_i^x & & {T_1 < t < T}
\end{array}
\right.
\end{equation}
Without the loss of generality, we assume that $T_1 = T_2 = 1$. Then the Floquet unitary is given by 
\begin{equation}
U_f = U_2 U_1 = e^{-i H_2} e^{-i H_1}
\end{equation}
Take $g = \pi/2$. 
For small perturbations with $\epsilon > 0$, the periodicity does not hold. 

The simplest case takes $\epsilon = J_i = h_i^x = 0$. We take an arbitrary state in $z$-basis $\ket{\psi_0} = \ket{\{ s_i \}}$ with $s_i = \pm 1$ and $\sigma_k^z \ket{\{ s_i \}} = s_k \ket{\{ s_i \}}$. After the spin-echo unitary $U_1 = e^{i \pi/2 \sum_i \sigma_i^x} = \prod_i i\sigma_i^x$, the state evolves to $\ket{\psi_1} = \ket{\{ -s_i \}}$. Then $U_2 = \sum_i h_i^z\sigma_i^z$ gives a global phase that $\ket{\psi_2} = e^{i\phi} \ket{\{ -s_i \}}$. 
In the pseudo-density matrix formulation, the temporal correlation for odd periods is $-e^{i\phi}$. For even periods, temporal correlation remains equal to 1. 
However, decoupled spins under spin echos cannot be taken as a discrete time crystal. The reason is that for small $\epsilon > 0$, the $\omega/2$ Fourier peak is split and $2T$-periodicity is broken down. 

Now we turn on the interaction $J_i > 0$. 
Take $h_i^x = 0$. Eigenstates of $H_2$ are eigenstates of $\sigma_i^z$ in the form of $\ket{\{ s_i \}}$ as before:
\begin{equation}
H \ket{\{ s_i \}} = [E^+(\{s_i\}) + E^-(\{s_i\})] \ket{\{ s_i \}}
\end{equation}
with $E^+(\{s_i\}) = \sum_i J_is_is_{i+1}$ and $E^-(\{s_i\}) = \sum_i h_i^zs_i$. 
Consider $\epsilon = 0$ first. Again $U_1 = e^{i \pi/2 \sum_i \sigma_i^x} = \prod_i i\sigma_i^x$. Then the Floquet eigenstates of the Floquet unitary $U_f$ are $e^{iE^-(\{s_i\})/2}\ket{\{s_i\}} \pm e^{-iE^-(\{s_i\})/2}\ket{\{-s_i\}}$. The Floquet eigenvalues are $\pm \exp[iE^+(\{s_i\})]$. 
In the pseudo-density matrix formulation, for the arbitrary initial state $\ket{\{ s_i \}}$, the temporal correlation on $\sigma_z^k$ of a particular spin $s_k$ is $-e^{iE^+(\{s_i\})-iE^-(\{s_i\})}$ in one period. For double periods, it will be $e^{2iE^+(\{s_i\})}$ with the absolute value 1. For all even periods, the absolute value of the temporal correlation remains equal to 1. 
Note here we consider for temporal correlations for a single spin instead of a superposition of all spins. An arbitrary superposition will give no correlations; for certain particular superpositions, the temporal correlations are the same as single-spin temporal correlations. 
When $\epsilon > 0$, $U_1 = e^{i (\pi/2-\epsilon) \sum_i \sigma_i^x} = \prod_i I \sin\epsilon + i\sigma_i^x \cos\epsilon$. In the pseudo-density matrix formulation, the temporal correlation is a bit complicated but the absolute value still converges to 1 for even periods without the half-frequency peak splitting. 
The robustness guarantees the model to be taken as a discrete time crystal.

Consider $h_i^x \neq 0$. For simplicity, assume that $h_i^z = 0$. In this particular case, the model exhibits a hidden emergent Ising symmetry $\tilde{S} = U^{\dag}_{FD} \prod_i \sigma_i^x U_{FD} = \prod_i \sigma_i^x$. Here a finite depth unitary transformation $U_{FD}$ satisfies
\begin{equation}
U_{FD} U(T) U_{FD}^{\dag} = e^{-i \tilde{H} T} \prod_i \sigma_i^x,
\end{equation}
with $\tilde{H} = \sum_i \tilde{J}_i^z \sigma_i^z \sigma_{i+1}^z + \tilde{h}_i^x \sigma_i^x$. 
Then we have 
\begin{equation}
U(2T) = U_{FD}^{\dag} e^{-2i\tilde{H}T} U_{FD} = e^{-2i U_{FD}^{\dag} \tilde{H} U_{FD} T} = e^{-2iH_{eff}T}.
\end{equation}
Referring back to the results in unitary evolution part, this suggests the $2T$-periodicity of temporal correlations in the model. Noisy perturbations won't split the half-frequency peak. Thus, the model constitutes a discrete time crystal. With $h_i^z \neq 0$, the results are similar but with a different hidden Ising symmetry $\tilde{S}$.

\subsubsection{Group representation}
Here we consider discrete time translation symmetry breaking in terms of group representation. It is more clear how multiple periods come into existence in the Floquet many-body localisation. 

Consider the Hamiltonians $H(t)$ have an onsite symmetry group $G$ and a discrete time translation symmetry $\mathbb{Z}$ that $H(t+T) = H(t)$. 
Based on the discussion in Ref.~\cite{von2016phase}, the Floquet phases are characterised by a central element of the group, that is, Floquet unitary takes the form 
\begin{equation}
U_f = u_{\{B\}}(z_0) V(z_0),
\end{equation}
where $z_0$ is an element of the centre of the group denoted $Z(G)$.
The onsite symmetry group $G$ has, for example, an irreducible representation $\chi$ with operators $\mathfrak{g}_{ij}^{\xi}$.
An initial state in a singlet evolves under the global symmetry in this irreducible representation $\chi$. 
Remember that in any irreducible representation of a finite group G, all the elements of $Z(G)$ are represented by $\lambda I$ where $\lambda$ is a constant and $I$ is unit matrix~\cite{robinson2012course}.
Then we have
\begin{equation}
U_f \mathfrak{g}_{ij}^{\chi} U_f^{\dag} = \frac{\chi(z)}{\chi(1)} \mathfrak{g}_{ij}^{\chi} 
\end{equation}
where $\chi(z)$ is the shifted constant at $z \in Z(G)$ under the irreducible representation $\chi$. 
Apply $U_f$ for $n$ times. For $z \neq 1$,
\begin{equation}
\mathfrak{g}_{ij}^{\chi}(nT) = \left[ \frac{\chi(z)}{\chi(1)} \right]^n \mathfrak{g}_{ij}^{\chi}(0).
\end{equation}
For a one-dimensional spin chain under Floquet many-body localisation as in the previous subsection, $Z(nT) = (-1)^n Z(0)$ shows a period of $2T$ for the order parameter. Thus, it constitutes a discrete time crystal.


\subsection{Possible sufficient conditions for general open systems}
A straightforward generalisation to time crystals in open systems is to formulate the Hamiltonian in terms of annihilation and creation operators and solve the Lindblad equation. The difficulty lies in the exact solution of Lindblad equations. 
Here, we attempt to reformulate the evolution into Kraus operators. 
In general, it is hard to see what kind of Kraus operators will work for arbitrary initial states and arbitrary periods, as it is unknown what kind of physical evolution a general Kraus operator is represented for and its physical meaning. 
We only give a simple illustration on the mathematical conditions for the initial state $\ket{\{s_i\}}$ and $2T$-periodicity in the pseudo-density matrix formulation. The general cases work in the similar way. 

For an initial state $\ket{\{s_j\}}$, we measure $\sigma_i^z$ to gain the eigenvalue $s_i$ with probability $1$ and leave the state unchanged. Assume the evolution is given by a set of Kraus operators $\{ E_k \}$, then the state evolves to $\sum_k E_k \ket{\{s_j\}}\bra{\{s_j\}} E_k^{\dag}$. We  measure for $\sigma_i^z$ again. The temporal correlation given by the pseudo-density matrix formulation is 
\begin{equation}
\langle \sigma_i^z, \sigma_i^z \rangle =  \Tr[\sigma_i^z \otimes \sigma_i^z R]  = s_i \Tr[\sigma_i^z \sum_k E_k \ket{\{s_j\}}\bra{\{s_j\}} E_k^{\dag}]
\end{equation}
For $\langle \sigma_i^z, \sigma_i^z \rangle$ to have $2T$-periodicity, a sufficient condition might be
\begin{equation}
\sum_k E_k \ket{\{s_j\}}\bra{\{s_j\}} E_k^{\dag} = - \ket{\{-s_j\}}\bra{\{-s_j\}}.
\end{equation}
For $k=1$, it reduces to the temporal correlation which is the same as in one-dimensional spin chain under Floquet many-body localisation. 


\section{An algebraic point of view}
In this section, we apply the algebraic tools to analyse spontaneous time translation symmetry breaking. The algebraic approach of symmetry breaking offer a representation with clear mathematics. Note that spontaneous time translation symmetry breaking can only be exhibited in the thermodynamic limit where the number of particles $N \rightarrow \infty$, the volume of the system $V \rightarrow \infty$ and the ratio $n = N/V$ fixed. For infinite degrees of freedom, the algebraic approach does not distinguish relativistic quantum field theory and quantum mechanics for continuous variables and will naturally offer a relativistic treatment. Here we attempt to treat space and time more equally in the pseudo-density matrix formalism, and it is interesting to investigate the algebraic approach of symmetry breaking for further generalisation to the relativistic context. 

In the following context, we review the algebraic criterion on spontaneous symmetry breaking and later apply them to explore time crystals. We further discuss the possibility to classify and understand temporal correlations from operator algebra.

\subsection{Preliminaries}
In this subsection we review the preliminaries for the algebraic approach. Specifically, we introduce the Weyl algebra, the concept of states, the Gelfand-Naimark-Segal(GNS) construction, and the algebraic symmetry of an algebra. This part is based on Ref.~\cite{strocchi2005symmetry}.

Instead of the canonical variables $q$, $p$ and the Heisenberg algebra $\mathcal{A}_H$, we construct the Weyl operators
$U(\alpha) \equiv e^{i \alpha q}$, $V(\beta) \equiv e^{i \beta p}$, 
where $\alpha q \equiv \sum_i \alpha_iq_i$, $\beta p \equiv \sum_i \beta_ip_i$, $\alpha_i, \beta_i \in \mathbb{R}$, and the corresponding Weyl algebra $\mathcal{A}_W$. The Heisenberg commutation relations ($\hbar = 1$) given as 
$[q_i, p_j] = i \delta_{ij}$, $[q_i. q_j] = 0 = [p_i, p_j]$, $i,j =1, 2, \cdots, N,$
turns into
\begin{align}
U(\alpha)U(\alpha') = U(\alpha + \alpha'), & \qquad V(\beta)V(\beta') = V(\beta+\beta') \nonumber\\
U(\alpha)V(\beta) = &e^{-i \alpha \beta} V(\beta)U(\alpha).
\end{align}
The conditions of $q = q^{\dag}$ and $p = p^{\dag}$ give
$U(\alpha)^* = U(-\alpha)$, $V(\beta)^* = V(-\beta)$.
We introduce a norm $\norm{\cdot}$ for elements in $\mathcal{A}_W$ that 
$\norm{A^*A} = \norm{A}^2$, $\forall A \in \mathcal{A}_W	,$
then the Weyl algebra $\mathcal{A}_W$ becomes a $C^*$-algebra. 


A state $\Omega$ of the system is characterised by the set of expectation values $\{\Omega(A), A \in \mathcal{A}\}$ where $\Omega(A) \equiv \langle A \rangle_{\Omega}$. That is, $\Omega$ is a functional $\Omega: \mathcal{A} \rightarrow \mathbb{C}$ satisfying the linearity $\Omega(\alpha A + \beta B) = \alpha \Omega(A) + \beta \Omega(B)$, the positivity $\Omega(A^* A) \geq 0, \forall A \in \mathcal{A}$, and the normalisation $\Omega(1) = 1$. 
In $C^*$-algebra, any state which cannot be decomposed into any other two states as $\Omega = \lambda \Omega_1 + (1-\lambda) \Omega_2, 0<\lambda<1$ is a pure state; otherwise, it is mixed. 
In a Hilbert space $\mathcal{H}$, a representation $\pi$ of a $C^*$-algebra is a $^*$-homomorphism $\pi$ of $\mathcal{A}$ preserving all the algebraic operations, into the $C^*$-algebra of bounded linear operators in $\mathcal{H}$ .
The Gelfand-Naimark-Segal(GNS) construction uses a representation $\pi_{\Omega}$ of $A (A \in \mathcal{A})$ which is uniquely determined by the state $\Omega$ in terms of its expectations on $\mathcal{A}$ up to isometries:
\begin{equation}
(\Psi_{\Omega}, \pi_{\Omega}(A)\Psi_{\Omega}) = \Omega(A), \forall A \in \mathcal{A},
\end{equation}
where $\Psi_{\Omega}$ is a reference vector in the Hilbert space $\mathcal{H}_{\Omega}$. 

The algebraic symmetry of an algebra $\mathcal{A}$ is then defined by an invertible mapping $\beta$ of the algebra into itself, preserving all the algebraic relations including the $*$-automorphism of $\mathcal{A}$. For a state $\omega$ on $\mathcal{A}$,
\begin{equation}
(\beta^*\omega)(A) \equiv \omega(\beta(A))
\end{equation}
is a state on $\mathcal{A}$ as well. 
The algebraic symmetry $\beta$, under a representation $\pi_{\omega}$ of $\mathcal{A}$, has a Wigner symmetry in $\mathcal{H}_{\omega}$ under a unitary operator $U_{\beta}$ such that
\begin{equation}
U_{\beta}\pi_{\omega}(A)U_{\beta}^{\dag} = \pi_{\omega}(\beta(A)) = \pi_{\beta^*\omega}(A).
\end{equation}
This is, $\pi_{\beta^*\omega}$ is unitarily equivalent to $\pi_{\omega}$. We will say $\{ \pi_{\omega}, \mathcal{H}_{\omega} \}$ is $\beta$-symmetric. However, when $\pi_{\beta^*\omega}$ is not unitarily equivalent to $\pi_{\omega}$, the symmetry $\beta$ is spontaneously broken. 

\subsection{Existence of time crystals}
In this subsection we review the criteria on spontaneous symmetry breaking and apply them to time crystals. 

\subsubsection{Criteria on spontaneous symmetry breaking}
Here we review the criteria on spontaneous symmetry breaking for the ground state~\cite{strocchi2005symmetry} and the equilibrium~\cite{heissenberg2019generalized}. 

Given the following conditions for a representation $\pi$ of the algebra $\mathcal{A}$:

(1) The space and time translations, represented by different continuous groups for unitary operators, guarantee the existence of energy and momentum in the representation space $\mathcal{H}_{\pi}$; 

(2) The energy remains nonnegative; that is, all the possible values for the Hamiltonian are bounded;

(3) The ground state $\omega$ is uniquely invariant under translations in $\mathcal{H}_{\pi}$, and it is represented by a cyclic vector locally. 

For an algebraic symmetry $\beta$ which commutes with space translations and time translations, $\beta$ is unbroken in $\pi$ if and only if correlation functions for all the ground states are invariant under $\beta$:
\begin{equation}
\omega(\beta(A)) \equiv \langle \beta(A) \rangle_0 = \langle A \rangle_0 = \omega(A), \forall A \in \mathcal{A},
\end{equation}
where $\omega$ is the ground state. 
Conditions (1)-(3) imply the cluster property: the correlations of two operators factorise when one of them goes to the spacial infinity
\begin{equation}
\lim_{|\mathbf{x} \rightarrow \infty|} [\langle AB_{\mathbf{x}} \rangle_0 - \langle A \rangle_0 \langle B \rangle_0] = 0.
\end{equation}

Similar for the equilibrium. 
If we change the condition (3) into

(3') The state $\omega$ is invariant under a subgroup $\mathcal{T}$ of spatial translations in $\mathcal{H}_{\pi}$, and it satisfies $\mathcal{T}$-asymptotic abelianess:
\begin{equation}
\lim_{n \rightarrow \infty} [T^n(A) , B] = 0, \forall A, B \in \mathcal{A}, T\in\mathcal{T},
\end{equation}

and calculate correlation functions for the thermal states, 
then we have the criteria for the equilibrium as Ref.~\cite{heissenberg2019generalized}. These criteria for spontaneous symmetry breaking are equivalent to the existence of long-range order in infinitely extended systems with local structures and asymptotic abelianess; that is,
\begin{equation}
\lim_{n \rightarrow \infty} \omega(T^n(\Delta A)B) = \omega(\Delta A)\omega(B) \neq 0
\end{equation}
where $\Delta A \equiv \beta(A) - A$.

\subsubsection{Example: time translation symmetry breaking}

Now we apply the above criteria to time translation symmetry breaking. First we consider the continuous time translation symmetry group denoted by $U(t), t \in \mathbb{R}$. Then $\beta(A) = U(t) A U^{\dag}(t)$. And the continuous time translation symmetry group is denoted by $U(a), a \in \mathbb{R}^s$. It is easy to see that the long-range order does not exist:
\begin{align}
\lim_{n \rightarrow \infty} \omega(T^n(\Delta A)B) & = \lim_{n \rightarrow \infty} \omega(T^n( U(t) A U^{\dag}(t) - A) B) \nonumber \\
& = \lim_{n \rightarrow \infty} \omega(T^n(U(t) A U^{\dag}(t) - A) \omega(B) \nonumber \\
& = \lim_{n \rightarrow \infty} \omega(U(a_n) \cdots U(a_1) U(t) A U^{\dag}(t) U^{\dag}(a_1) \cdots U^{\dag}(a_n) - A) \omega(B) \nonumber \\
& = 0
\end{align}
The limit goes to 0 as we can always choose infinite runs of space translations to mimic a time evolution such that after time translations and space translation the operator goes back to itself under the average of the ground state or the thermal state. 

For discrete time translation symmetry, we consider a one-dimensional Floquet many-body localised spin chain again. 
Recall that the Floquet evolution is given by $U_f = U_2 U_1$, where $U_1 = \exp[i t_1 \sum_i \sigma_i^x]$ and $U_2 = \exp[-i H_{MBL} t_2]$ where $H_{MBL} = \sum_i J_i \sigma_i^z \sigma_{i+1}^z + h_i^z \sigma_i^z$ for simplicity here. For $t_1 \approx \pi/2$, $U_1 = \prod_i i\sigma_i^x$.
Take $A = \sigma_i^z$ and $B = \sigma_j^z$. 
After a period of $T$, $\beta_1(A) = U_f \sigma_i^z U^{\dag}_f = U_2 U_1 \sigma_i^z U_1^{\dag} U_2^{\dag} = - \sigma_i^z$, then
\begin{equation}
\lim_{n \rightarrow \infty} \omega(T^n(\Delta A)B)  = \lim_{n \rightarrow \infty} \omega(T^n( -2\sigma_i^z) \sigma_j^z)  = 0,
\end{equation}
where $n$ and $|i-j|$ go to infinity. The long-range order does not exist for a single period. 
After two periods of $T$, $\beta_2(A) = U_f U_f \sigma_i^z U^{\dag}_f U^{\dag}_f = \sigma_i^z$
\begin{equation}
\lim_{n \rightarrow \infty} \omega(T^n(\Delta A)B)  = \lim_{n \rightarrow \infty} \omega(T^n( \sigma_i^z - \sigma_i^z) \sigma_j^z)  = \lim_{n \rightarrow \infty} \omega(T^n(0) \sigma_j^z)  \neq 0,
\end{equation}
here $T^n(0)$ gives a constant when $n$ goes to infinity. This suggests the existence of long-range order after $2T$, showing the $2T$-periodicity of discrete time crystals. 
For a general case, we may choose an arbitrary rotational invariant $H_{MBL}$ but with the same $U_1$. Since $\beta(A) = U_f \sigma_i^z U^{\dag}_f = U_2 U_1 \sigma_i^z U_1^{\dag} U_2^{\dag} = - U_2 \sigma_i^z U_2^{\dag}$, any rotational invariant $U_2$ will give the expectation values under $T^n$ for $T \in \mathcal{T}$. This result is similar as in Ref.~\cite{heissenberg2019generalized}.

\subsection{Temporal correlations}

In the previous sections the long-range order in time has different representations in terms of the mixture of spatial and temporal correlations. 
For time translation symmetry breaking, these representations do not make much difference in the algebraic language as algebraic symmetries are already assumed to commute with both of space translations and time translations. 
Here we discuss the possibility of a measure of temporal correlations based on operator algebra in which the study of spatial correlations is nicely formulated in terms of the hierarchy in the Tsirelson's problem. 

One possibility is to use generalised non-local games in the time domain. As we already discussed in the last chapter, quantum-classical signalling games give temporal correlations formulated by pseudo-density matrices. We may consider other variants of signalling games as an analogue of finite input-output games and synchronous games. What is more, even for a particular kind of signalling games, we may discuss different possibilities for strategies. 

Another possibility is to generalise the pseudo-density matrix beyond the tensor product structure and projective measurements. It is known that, for spatial correlations, the hierarchy from the smallest set to the largest set is classical correlations, correlations of tensor product structures in finite dimensional Hilbert spaces, correlations of tensor product structures in infinite dimensional Hilbert spaces, the closure of correlations of tensor product structures in infinite dimensional Hilbert spaces, correlations of commutative structures in arbitrary dimensional Hilbert spaces. We may have generalised temporal correlations in terms of commutative structures. Indefinite causal structures may be involved in such representation. 

We will leave the formal establishment as a future work. It is interesting to ask what kind of temporal correlations current time crystals hold in the hierarchy and whether generalised temporal correlations may lead to different understanding for time crystals. And it might be possible to generalise all these discussion to the relativistic setting via algebraic quantum field theory.

\chapter{\label{ch:7-co}Conclusion and outlook}

\minitoc

\clearpage

We conclude this thesis with the summary of results in the main chapters and provide the outlook for future possible work. 

The results are summarised as follows. 
\begin{itemize}

\item In Chapter 3, we generalise the pseudo-density matrix formalism to continuous variables and general measurement processes. First we define spacetime Gaussian states from the first two statistical moments which fully characterise Gaussian states, and compare temporal Gaussian states with spatial Gaussian state to show a similar correlation relationship as the qubit case. Via the Wigner function representation, we define spacetime density matrices in continuous variables in general, and show that spacetime Wigner functions hold the similar properties which uniquely determine spatial Wigner functions. We further discuss the possibilities of defining spacetime states via position measurements and weak measurements, and generalise the pseudo-density matrix formulation to more general measurement processes. An experimental tomography based on quantum optics is proposed to verify the operational meaning for the generalised pseudo-density matrix formalism from measurement correlations. 

\item In Chapter 4, we use quantum correlation in time to compare the pseudo-density matrix formalism with indefinite causal structures, consistent histories, generalised non-local games, and out-of-time-order correlation functions, and path integrals. We aim to argue that spacetime formulations in non-relativistic quantum mechanics are remarkably similar. In the section of indefinite causal structures, we use the process matrix formalism in particular, compare it with the pseudo-density matrix formalism via correlations, formulate causal inequalities, and discuss the role of post-selection in indefinite causal structures. In consistent histories, the consistency conditions give the generalised pseudo-density matrix a better argument for its existence. Pseudo-density matrices can be formulated in terms of quantum-classical signalling games as well. We also provide a simple calculation for out-of-time-order correlation functions and apply it to black hole information paradox. Nevertheless, the path integral formalism has a different representation of quantum correlations from the pseudo-density matrix approach, suggesting interesting directions for quantum measure and relativistic quantum theory. 

\item In Chapter 5, we apply the temporal correlations in the pseudo-density matrix formalism to time crystals. We define time crystals as long-range order in time, a particular kind of temporal correlations which do not vanish after a long time. Then we analyse continuous time translation symmetry in terms of general decoherent processes and a generalised version of Mermin-Wagner theorem. We also discuss discrete time translation symmetry via a stabilisation protocol of quantum computation, phase flip codes of quantum error correction and Floquet many-body localisation. Finally we explore the possibility of time crystals from the algebraic point of view. 

\end{itemize}

Some of the possible future directions for work are listed as below.
\begin{itemize}

\item \emph{Mutual information in time.} Mutual information of two random variables $X$ and $Y$ measures how much information $X$ and $Y$ have in common~\cite{nielsen2002quantum}. It is also a measure of the total correlations between two subsystems of a bipartite quantum system~\cite{vedral2006introduction}. Classically, the mutual information for two systems at different times is defined as the same as two systems at different positions. However, quantifying the mutual information for two quantum systems evolving in time is still a difficult open problem. Note that a basic fact of quantum mutual information between two entangled systems of a pure state is that, it is equal to twice the von Neumann entropy of a reduced subsystem, while it can only be at most the same as the Shannon entropy of a single subsystem in the classical case. On the one hand, it shows that quantum correlations are stronger than classical ones; on the other hand, the quantum mutual information in time is supposed to show its quantum advantages over the classical mutual information. So far, we investigate different proposals for quantum mutual information in time but none of them could be called quantum. Difficulties of defining mutual information in time in the pseudo-density matrix formalism come from the negativity of temporal pseudo-density matrices. One possible solution is to purify pseudo-density matrices and make them to be positive semi-definite, then we may define the mutual information in time for subsystems at different times. 

\item \emph{Tripartite correlations in spacetime.} Bipartite quantum correlations in space-time are well-studied in the pseudo-density matrix formalism~\cite{zhao2018geometry}. A symmetric structure has been shown in two-point quantum correlations in space and time. Specifically, two-point spatial correlations in arbitrary bipartite quantum states and two-point temporal correlations for a single qubit evolving under a unitary quantum channel are mapped to each other under the operation of partial transposition. This suggests an interesting relationship between spatial and temporal correlations in the bipartite case. We further analyse tripartite correlations. One question remaining unknown is that given a tripartite correlation, how can we distinguish it from a qubit at three times, one qubit at one time and another at two time, or three qubits at a single time? Another interesting question may be the spatial-temporal analogue of monogamy of entanglement. As we know that the subsystem of a maximally entangled state cannot be entangled with a third system, a maximally temporally correlated system, that is a system under the identity evolution, may still be maximally temporally correlated with the system under the identity evolution at a later time. What will be the temporal analogue of monogamy of entanglement? Or is it a fundamental difference between spatial and temporal correlations? 

\item \emph{Spacetime from spatial-temporal correlations rather than entanglement.} It is claimed among AdS/CFT community that it will be possible to build up spacetime with quantum entanglement~\cite{raamsdonk2010building}. However, quantum entanglement is only a particular kind of spatial correlation. A better argument may be to build spacetime from spatial-temporal correlations rather than entanglement. We are discussing the possibilities of deriving the Einstein field equation from a spacetime area law of quantum correlations. The Einstein field equation can be derived through the area law for entanglement entropy~\cite{jacobson1995thermodynamics} as well as the quantum geometrical limit for the energy density of clocks and signals~\cite{lloyd2012quantum}. We want to argue that quantum correlations are much more than entanglement, and temporal correlations in quantum mechanics may provide better insights for understanding spacetime or gravity in the quantum sense. 

\item \emph{Application in black hole information paradox.} In Chapter 4, we already applied the pseudo-density matrix formalism to the out-of-time-order correlation functions in the black hole final state proposal. We are looking for further applications in black hole information paradox. One possibility still lies in the out-of-time-order correlation functions. We may use the out-of-time-order correlation functions as a tool to analyse the behaviours in the black hole formation and evaporation. We may also understand the information loss via temporal correlations. Spatial correlations like entanglement have been discussed in the black hole scenarios. Will temporal correlations between early radiation and late radiation help to understand the information loss? These questions are worth exploring. 

\end{itemize}

As we have asked in the introductory chapter, ``what is time'', we briefly report on our little lessons from quantum correlations. 

The thesis is based on the assumption that space and time should be treated on an equal footing. 
The pseudo-density matrix formulation treats temporal correlations equally in form as spatial correlations. 
We are a bit concerned about this assumption under a simple argument on monogamy. As we mentioned before, monogamy of entanglement cannot find a temporal analogue. 
Entanglement is a kind of spatial correlation; nevertheless, we cannot observe the monogamy of any temporal correlation. The maximally temporal correlated states are under the identity evolution and we can make as many copies as we want. Thus temporal correlations have no monogamy constraint; this suggests intrinsic difference between spatial correlations and temporal correlations. 
Another example is from time crystals. We have continuous space translation symmetry breaking but no continuous time translation symmetry breaking. 
One deep concern from Ref.~\cite{khemani2019brief} is that ``causality distinguishes between spacelike and timelike separations''. While generators of space translations are the momenta, generators of time translations are the Hamiltonians which is much more system dependent. 
We suspect the assumption on the equal treatment of space and time to be too strong. It is a possible route to learn about temporal correlations by taking them operationally equal as spatial correlations; but we would carefully keep in mind that, space is space, time is time. 

One possible link between spatial and temporal correlations is the partial transpose. We cannot see exactly why this operation is so important in space-time inversion; a simple understanding might be that for two systems in space converting to two systems in time, one evolves forwards under normal evolution while the other evolves backwards under the transpose. 
Path integrals are important to understand spacetime. They have shown the difference in the operational meaning of quantum correlations. Further investigation in terms of quantum measure and relativistic quantum information are ongoing.  
We are also concerned about indefinite causal structures, as it might not be enough to quantising gravity as a linear superposition of causal structures. 
It is interesting to explore further on algebraic field theory in search for the relativistic version for quantum correlations in space and time.

Anyway, the long long journey towards time just started.

\startappendices
%

\chapter{\label{app:1-cj}Choi-Jamio\l{}kowski isomorphism}

\minitoc

Here we introduce the Choi-Jamio\l{}kowski isomorphism based on Ref.~\cite{chiribella2009theoretical}. 

The set of linear operators on the finite dimensional Hilbert space $\mathcal{H}$ is denoted as $\mathcal{L}(\mathcal{H})$. The set of linear operators from $\mathcal{H}_0$ to $\mathcal{H}_1$ is denoted as $\mathcal{L}(\mathcal{H}_0, \mathcal{H}_1)$. An operator $X \in \mathcal{L}(\mathcal{H}_0, \mathcal{H}_1)$ has a one-to-one correspondence with a vector $|X\rrangle \in \mathcal{H}_1 \otimes \mathcal{H}_0$ as 
\begin{equation}
|X\rrangle = (X \otimes I_{\mathcal{H}_0}) |I_{\mathcal{H}_0}\rrangle
= (I_{\mathcal{H}_1} \otimes  X^T)  |I_{\mathcal{H}_1}\rrangle
\end{equation}
where $ \mathcal{I}_{\mathcal{H}}$ is the identity operator in $\mathcal{H}$, $ |I_{\mathcal{H}}\rrangle \in \mathcal{H} \otimes \mathcal{H}$ is the maximally entangled vector $|I_{\mathcal{H}}\rrangle = \sum_n \ket{n}\ket{n}$ ($\ket{n}$ is the orthonormal basis in $\mathcal{H}$), $X^T \in \mathcal{L}(\mathcal{H}_1, \mathcal{H}_0)$ is the transpose of $X$ with respect to two given bases in $\mathcal{H}_0$ and $\mathcal{H}_1$. 

The set of linear maps from $\mathcal{L}(\mathcal{H}_0)$ to $\mathcal{L}(\mathcal{H}_1)$ is denoted by $\mathcal{L}(\mathcal{L}(\mathcal{H}_0), \mathcal{L}(\mathcal{H}_1))$. 
A linear map $\mathcal{M} \in \mathcal{L}(\mathcal{L}(\mathcal{H}_0), \mathcal{L}(\mathcal{H}_1))$ has a one-to-one correspondence with a linear operator $M \in \mathcal{L}(\mathcal{H}_1 \otimes \mathcal{H}_0)$ as 
\begin{equation}
M = \mathcal{M} \otimes \mathcal{I}_{\mathcal{L}(\mathcal{H}_0)}(|I_{\mathcal{H}_0}\rrangle \llangle I_{\mathcal{H}_0} | )
\end{equation}
where $\mathcal{I}_{\mathcal{L}(\mathcal{H}_0)}$ is the identity map on $\mathcal{L}(\mathcal{H}_0)$. 
This is called Choi-Jamio\l{}kowski isomorphism. The operator $M$ is called Choi-Jamio\l{}kowski operator of $\mathcal{M}$. 
Its inverse transforms $M \in \mathcal{L}(\mathcal{H}_1 \otimes \mathcal{H}_0)$ to a map $\mathcal{M} \in \mathcal{L}(\mathcal{L}(\mathcal{H}_0), \mathcal{L}(\mathcal{H}_1))$ that acts on an operator $X \in \mathcal{L}(\mathcal{H}_0)$ as 
\begin{equation}
\mathcal{M}(X) = \Tr_{\mathcal{H}_0} [(I_{\mathcal{H}_1} \otimes X^T) M]
\end{equation}

A linear map $\mathcal{M}$ is trace preserving if and only if its Choi-Jamio\l{}kowski operator $M$ satisfies 
\begin{equation}
\Tr_{\mathcal{H}_1} [M] = I_{\mathcal{H}_0}.
\end{equation}
A linear map $\mathcal{M}$ is Hermitian preserving if and only if its Choi-Jamio\l{}kowski operator $M$ is Hermitian. 
A linear map $\mathcal{M}$ is completely positive if and only if its Choi-Jamio\l{}kowski operator $M$ is positive semi-definite.

\chapter{\label{app:3-properties}Proofs for the properties for spacetime Wigner functions}

\minitoc
Here we provide the proof for six properties for spacetime Wigner functions. The additional one is listed before the five properties in the main text, about the expectation value of an arbitrary operator $\hat{A}$. Before that, we introduce the Wigner representation in Liouville Space~\cite{royer1989measurement}. 

\section{Wigner Representation in Liouville Space}
Ref.~\cite{royer1989measurement} gives an introduction to the Wigner representation in Liouville Space. 
In Liouville space, operators are treated as vectors in a superspace. 
For a bra-ket notation, we call $|A\}$ a L-ket and $\{A|$ a L-bra for an operator $A$, with the scalar product as
\begin{equation}
\{B|A\} = \Tr\{B^{\dag}A\}.
\end{equation}
Different from Ref.~\cite{royer1989measurement}, we take $\hbar = 1$. 
Define a Liouville basis
\begin{equation}
|qp\} = \left( \frac{2}{\pi} \right)^{1/2} |\Pi_{qp}\},
\end{equation}
where $\Pi_{qp}$ is given by
\begin{align}
\Pi_{qp} & = \frac{1}{2} \int_{-\infty}^{\infty} \textrm{d}s e^{isp} \ket{q+\frac{\hbar}{2}s}\bra{x-\frac{\hbar}{2}s} \nonumber \\
& = \frac{1}{2} \int_{-\infty}^{\infty} \textrm{d}k e^{-ikq} \ket{p+\frac{\hbar}{2}k}\bra{p-\frac{\hbar}{2}k} \nonumber \\
& =\frac{1}{4\pi} \int_{-\infty}^{\infty} \textrm{d}k \int_{-\infty}^{\infty} \textrm{d}s e^{ik(\hat{q} - q)-is(\hat{p}-p)}.
\end{align}
In fact $\Pi_{qp}$ is the parity operator about the phase point $(x, p)$:
\begin{equation}
\Pi_{qp}(\hat{q}-q)\Pi_{qp} = -(\hat{q} - q), \ \Pi_{qp}(\hat{p}-p)\Pi_{qp} = -(\hat{p} - p).
\end{equation}
It is the same as the displaced parity operator $U(\alpha)$ with the mapping $\alpha = \frac{1}{\sqrt{2}}(q+ip)$.

$|qp\}$ forms an orthogonal and complete basis: 
\begin{align}
\{q'p'|qp\} = \delta(q'-q)\delta(p'-p)\\
\int_{-\infty}^{\infty} \int_{-\infty}^{\infty} \textrm{d}q \textrm{d}p |qp\}\{qp| = \hat{\hat{1}},
\end{align}
where $\hat{\hat{1}}$ is a unit L-operator.
However, we need to remember that $|qp\}$ is not a valid quantum state because $\Pi_{qp}$ is not positive definite.

The Weyl form of an operator $\hat{A}$ is defined as 
\begin{equation}
A(q, p) \equiv (2\pi)^{1/2} \{qp | A\} = 2\Tr[\Pi_{qp} \hat{A}].
\end{equation}
Then the Wigner function of a state $\hat{\rho}$ is given by
\begin{equation}
W(q, p) \equiv (2\pi)^{-1/2} \{qp | \rho\} = (2\pi)^{-1} \int \textrm{d}s e^{-isp}\bra{q+\frac{1}{2}\hbar s}\rho \ket{q-\frac{1}{2}\hbar s},
\end{equation}
where the normalisation holds for $\iint \textrm{d}q \textrm{d}p W(q, p) = 1$.
For an operator $\hat{A}$ measured in the state $\hat{\rho}$, its expectation value is given as
\begin{equation}\label{eqn: expofa}
\langle \hat{A} \rangle_{\rho} = \{A|\rho\} = \iint \textrm{d}q\textrm{d}p\{A|qp\}\{qp|\rho\}  = \iint \textrm{d}q\textrm{d}p A^*(q, p) W(q, p).
\end{equation}

\section{Proofs for the properties}
We prove all the six properties listed as (0) to (5) in this subsection. 
Following the notation in the previous subsection, we have the bipartite spacetime Wigner function
\begin{equation}
\mathcal{W}(q_1, p_1, q_2, p_2) = (2\pi)^{-1} \{q_1p_1, q_2p_2 | R\} =  4 \Tr [(\Pi_{q_1p_1} \otimes \Pi_{q_2p_2})\hat{R}],
\end{equation}
for a bipartite spacetime density matrix in continuous variables $\hat{R}$.

~\\
(0) For bipartite case,
\begin{equation}
\langle \hat{A} \rangle_R =  \Tr[\hat{A} \hat{R}] 
=   \iiiint \textrm{d}q_1 \textrm{d}q_2 \textrm{d}p_1 \textrm{d}p_2 A^*(q_1, p_1, q_2, p_2)\mathcal{W}(q_1, p_1, q_2, p_2),\label{aexp}
\end{equation}
where 
\begin{equation}
A(q_1, p_1, q_2, p_2) = (2\pi) \{qp | A\} = 4 \Tr [(\Pi_{q_1p_1} \otimes \Pi_{q_2p_2})\hat{A}].
\end{equation}
Note that $T(\alpha) = 2 U(\alpha) = 2\Pi(q_1, p_1)$ and $T(\beta) = 2 U(\beta) = 2\Pi(q_2, p_2)$. The above statement is equivalent to 
\begin{equation}
\langle \hat{A} \rangle_R =  \Tr[\hat{A} \hat{R}]
=  \iint\textrm{d}^2\alpha \textrm{d}^2\beta A^*(\alpha, \beta)\mathcal{W}(\alpha, \beta),
\end{equation}
where
\begin{equation}
A(\alpha, \beta) = \Tr\{[T(\alpha) \otimes T(\beta)]\hat{A}\}.
\end{equation}
%

\begin{proof}

Compared to Eqn.~\eqref{eqn: expofa}, 
\begin{align}
\langle \hat{A} \rangle_R =  & \{A | R\}  \nonumber \\
= & \iiiint \textrm{d}q_1 \textrm{d}q_2 \textrm{d}p_1 \textrm{d}p_2 \{A|q_1p_1, q_2p_2\}\{q_1p_1, q_2p_2|R\} \nonumber \\
=  & \iiiint \textrm{d}q_1 \textrm{d}q_2 \textrm{d}p_1 \textrm{d}p_2 A^*(q_1, p_1, q_2, p_2)\mathcal{W}(q_1, p_1, q_2, p_2).
\end{align}
\end{proof}
Generalisation to $n$ events is straightforward.

~\\
(1) $\mathcal{W}(q_1, p_1, q_2, p_2)$ is given by $\mathcal{W}(q_1, p_1, q_2, p_2)$ $=$ $\Tr[M(q_1, p_1, q_2, p_2) R]$ for $M(q_1, p_1, q_2, p_2)$ $=$ $M^{\dag}(q_1, p_1, q_2, p_2)$. Therefore, it is real.

\begin{proof}

Compared to Eqn.~\eqref{eqn: densitytowigner}, $M(q_1, p_1, q_2, p_2) = 4\Pi_{q_1p_1} \otimes \Pi_{q_2p_2}$, thus it is obvious that $M(q_1, p_1, q_2, p_2)$ $=$ $M^{\dag}(q_1, p_1, q_2, p_2)$.

Because a spacetime density matrix is Hermitian, the spacetime Wigner function is real. 
\end{proof}
Note that we prove the Hermicity of a spacetime density matrix from the property that spacetime Wigner function is real.

~\\
(2)
\begin{align}
\iint \textrm{d}p_1 \textrm{d}p_2  \mathcal{W}(q_1, p_1, q_2, p_2) = \bra{q_1, q_2}\hat{R}\ket{q_1, q_2}, \nonumber \\
\iint \textrm{d}q_1 \textrm{d}q_2  \mathcal{W}(q_1, p_1, q_2, p_2) = \bra{p_1, p_2}\hat{R}\ket{p_1, p_2}, \nonumber \\
\iiiint \textrm{d}q_1 \textrm{d}q_2 \textrm{d}p_1 \textrm{d}p_2 \mathcal{W}(q_1, p_1, q_2, p_2) = \Tr \hat{R} = 1.
\end{align}

\begin{proof}

Taking $\hat{A}$ in the property (0) to be
\begin{equation} 
\hat{A} = \delta(\hat{q}_1-q_1)\delta(\hat{q}_2-q_2),
\end{equation}
then 
\begin{equation} 
A^*(q_1, p_1, q_2, p_2) = \delta(\hat{q}_1-q_1)\delta(\hat{q}_2-q_2).
\end{equation}
Thus
\begin{equation} 
\Tr[\hat{A} \hat{R}] = \bra{q_1, q_2}\hat{R}\ket{q_1, q_2},
\end{equation} 
and
\begin{equation} 
\iiiint \textrm{d}q_1 \textrm{d}q_2  \textrm{d}p_1 \textrm{d}p_2 A^*(q_1, p_1, q_2, p_2)\mathcal{W}(q_1, p_1, q_2, p_2) 
=   \iint \textrm{d}p_1 \textrm{d}p_2  \mathcal{W}(q_1, p_1, q_2, p_2).
\end{equation} 
Via Eqn.~(\ref{aexp}), the first equality holds.

Similar for the second equality. The normalisation property is already proven before.
\end{proof}

~\\
(3) $\mathcal{W}(q_1, p_1, q_2, p_2)$ is Galilei covariant, that is,
if $\bra{q_1, q_2}R\ket{q'_1, q'_2}$ $\rightarrow$ $\bra{q_1+a, q_2+b}R\ket{q'_1+a, q'_2+b}$, then $\mathcal{W}(q_1, p_1, q_2, p_2)$ $\rightarrow$ $\mathcal{W}(q_1+a, p_1, q_2+b, p_2)$ and if $\bra{q_1, q_2}R\ket{q'_1, q'_2}$ $\rightarrow$ $\exp\{[ip'_1(-q_1+q'_1)+ip'_2(-q_2+q'_2)]/\hbar\}\bra{q_1, q_2}R\ket{q'_1, q'_2}$, then $\mathcal{W}(q_1, p_1, q_2, p_2)$ $\rightarrow$ $\mathcal{W}(q_1, p_1-p'_1, q_2, p_2-p'_2)$.

\begin{proof}

If $$\bra{q_1, q_2}\hat{R}\ket{q'_1, q'_2} \rightarrow \bra{q_1+a, q_2+b}\hat{R}\ket{q'_1+a, q'_2+b},$$ that is, 
$$
\hat{R} \rightarrow D^{\dag}_{a0} \otimes D^{\dag}_{b0} \hat{R} D_{a0} \otimes D_{b0},
$$
then
\begin{align*}
\mathcal{W}(q_1, p_1, &q_2, p_2) = 4\Tr[(\Pi_{q_1p_1} \otimes \Pi_{q_2p_2})\hat{R}] \rightarrow\\
&4\Tr[(\Pi_{q_1p_1} \otimes \Pi_{q_2p_2})(D^{\dag}_{a0} \otimes D^{\dag}_{b0} \hat{R} D_{a0} \otimes D_{b0})] = \mathcal{W}(q_1+a, p_1, q_2+b, p_2).
\end{align*}
If 
$$\bra{q_1, q_2}\hat{R}\ket{q'_1, q'_2} \rightarrow \exp\{[ip'_1(-q_1+q'_1)+ip'_2(-q_2+q'_2)]/\hbar\}\bra{q_1, q_2}\hat{R}\ket{q'_1, q'_2},$$
that is, 
$$
\hat{R} \rightarrow D^{\dag}_{0, -p'_1} \otimes D^{\dag}_{0, -p'_2} \hat{R} D_{0, -p'_1} \otimes D_{0, -p'_2},
$$
then
\begin{align*}
\mathcal{W}&(q_1, p_1, q_2, p_2) = 4\Tr[(\Pi_{q_1p_1} \otimes \Pi_{q_2p_2})\hat{R}] \rightarrow\\
& 4 \Tr \big[  (\Pi_{q_1p_1} \otimes \Pi_{q_2p_2}) (D^{\dag}_{0, -p'_1} \otimes D^{\dag}_{0, -p'_2} \hat{R} D_{0, -p'_1} \otimes D_{0, -p'_2} ) \big] =  \mathcal{W}(q_1, p_1-p'_1, q_2, p_2-p'_2).
\end{align*}
\end{proof}

~\\
(4) $\mathcal{W}(q_1, p_1, q_2, p_2)$ has the following property under space and time reflections: if $\bra{q_1, q_2}\hat{R}\ket{q'_1, q'_2}$ $\rightarrow$ $\bra{-q_1, -q_2}\hat{R}\ket{-q'_1, -q'_2}$, then $\mathcal{W}(q_1, p_1, q_2, p_2)$ $\rightarrow$ $\mathcal{W}(-q_1, -p_1, -q_2, -p_2)$ and if $\bra{q_1, q_2}\hat{R}\ket{q'_1, q'_2}$ $\rightarrow$ $\bra{q'_1, q'_2}\hat{R}\ket{q_1, q_2}$, then $\mathcal{W}(q_1, p_1, q_2, p_2)$ $\rightarrow$ $\mathcal{W}(q_1, -p_1, q_2, -p_2)$.

\begin{proof}

If $\bra{q_1, q_2}\hat{R}\ket{q'_1, q'_2} \rightarrow \bra{-q_1, -q_2}\hat{R}\ket{-q'_1, -q'_2}$, that is, 
$$\hat{R} \rightarrow \Pi_{00}\hat{R}\Pi_{00},$$ 
then
\begin{align*}
\mathcal{W}(q_1, p_1, &q_2, p_2) = 4\Tr[(\Pi_{q_1p_1} \otimes \Pi_{q_2p_2})\hat{R}] \rightarrow\\
&4\Tr[(\Pi_{q_1p_1} \otimes \Pi_{q_2p_2})(\Pi_{00}\hat{R}\Pi_{00})] \mathcal{W}(-q_1, -p_1, -q_2, -p_2).
\end{align*}
For $\bra{q_1, q_2}\hat{R}\ket{q'_1, q'_2} \rightarrow \bra{q'_1, q'_2}\hat{R}\ket{q_1, q_2}$, it is similar to transpose. Consider
$\hat{q}^T = q$ and $\hat{p}^T = -p$,
\begin{align*}
\mathcal{W}(q_1, p_1, &q_2, p_2) \rightarrow \mathcal{W}(q_1, -p_1, q_2, -p_2).
\end{align*}
\end{proof}

~\\
(5) Take $\hbar=1$.
\begin{equation}
\Tr(\hat{R}_1\hat{R}_2) = (2\pi) \iint \textrm{d}q \textrm{d}p \mathcal{W}_{R_1}(q, p) \mathcal{W}_{R_2}(q, p),
\end{equation}
for $\mathcal{W}_{R_1}(q, p)$ and $\mathcal{W}_{R_2}(q, p)$ are pseudo-Wigner functions for pseudo-density matrices $\hat{R}_1$ and $\hat{R}_2$ respectively.

\begin{proof}

\begin{equation}
\Tr(\hat{R}_1\hat{R}_2) = \{ R_1 | R_2 \} =\iint \textrm{d}q \textrm{d}p \{ R_1 | qp\}\{qp | R_2 \} = (2\pi) \iint \textrm{d}q \textrm{d}p \mathcal{W}_{R_1}(q, p) \mathcal{W}_{R_2}(q, p).
\end{equation}
\end{proof}

\chapter{\label{app:4-tc}Proof for continuous time translation symmetry in 1+1 dimensions}


Now we prove that there is no continuous time translation symmetry breaking in the Heisenberg model at finite temperature in 1+1 dimensions. As the original Mermin-Wagner theorem, we use the Bogoliubov inequality:
\begin{equation}
\frac{1}{2} \beta \langle [ A , A^{\dag}]_+ \rangle \langle [ [C, H]_-, C^{\dag} ]_- \rangle \geq |\langle [C, A]_-\rangle|^2
\end{equation}
where $\beta = 1 / k_B T$ is the inverse temperature, $A$ and $C$ are arbitrary operators and $H$ is the Hamiltonian of the system. $\langle \cdots \rangle$ gives the expectation value in the thermal state. 
In the one-dimensional Heisenberg model, the Hamiltonian is given as 
\begin{equation}
H = - \sum_{ij} J_{ij} S_i^z S_j^z - b \sum_i S_i^z
\end{equation}
where $S_i^z$ is the spin  $i$ along the $z$-direction and $b$ is the parameter for a small perturbation. We assume that $Q = \frac{1}{N} \sum_{i, j} |R_i - R_j|^2 |J_{ij}|$ remains finite where $R_i$ denotes the position of spin $i$. 
We assign $A$ and $C$ to be
\begin{align}
A & = e^{iHt} S^-(-k) e^{-iHt} \\
C & = e^{iHt} S^+(k) e^{-iHt}
\end{align}
where $S^{\alpha}(k) = \sum_i S_i^{\alpha} e^{-ikR_i}$ and $S^{\pm} = S_x \pm i S_y$. 
Then 
\begin{equation}
\langle [C, A]_-\rangle = 2\hbar\sum_i \langle e^{iHt} S_i^z e^{-iHt} \rangle = 2\hbar N Z(t),
\end{equation}
where $Z(t) = \langle e^{iHt} S_i^z e^{-iHt} \rangle$ is the temporal correlation in the model. 
\begin{equation}
\sum_k \langle [ A , A^{\dag}]_+ \rangle \leq 2\hbar^2 N^2 S(S+1),
\end{equation}
and 
\begin{equation}
\langle [ [C, H]_-, C^{\dag} ]_- \rangle \leq 4\hbar^2bNZ(t) + 4N\hbar^2k^2QS(S+1)
\end{equation}
Substituting the above inequalities into the Bogoliubov inequality and summing over all the wavevectors, we have 
\begin{equation}
S(S+1) \geq \frac{C(t)^2 v}{2\pi \beta \hbar^2} \int_0^{k_0} \frac{\textrm{d}k}{bZ(t) + k^2QS(S+1)} =  \frac{Z(t)^2 v}{2\pi \beta \hbar^2}\frac{\arctan (k_0 \sqrt{\frac{QS(S+1)}{bZ(t)}})}{\sqrt{QS(S+1)bZ(t)}}.
\end{equation}
Thus 
\begin{equation}
Z(t) \leq \text{const} \cdot \frac{b^{1/3}}{T^{2/3}} \quad \text{as} \ b \rightarrow 0.
\end{equation}
The temporal correlation vanishes as the perturbation parameter goes to 0 under finite temperature; thus, there is no spontaneous continuous time translation symmetry breaking in this case. 

\setlength{\baselineskip}{0pt} 

{\renewcommand*\MakeUppercase[1]{#1}%
\printbibliography[heading=bibintoc,title={\bibtitle}]}

@article{ji2020mip,
    author = "Ji, Zhengfeng and Natarajan, Anand and Vidick, Thomas and Wright, John and Yuen, Henry",
    title = "{MIP*=RE}",
    eprint = "2001.04383",
    archivePrefix = "arXiv",
    primaryClass = "quant-ph",
    month = "1",
    year = "2020"
}

@article{zhang2020different,
	doi = {10.1088/1367-2630/ab6b9f},
	url = {https://doi.org/10.1088%2F1367-2630%2Fab6b9f},
	year = 2020,
	month = {feb},
	publisher = {{IOP} Publishing},
	volume = {22},
	number = {2},
	pages = {023029},
	author = {Tian Zhang and Oscar Dahlsten and Vlatko Vedral},
	title = {Different instances of time as different quantum modes: quantum states across space-time for continuous variables},
	journal = {New Journal of Physics}
}

@article{zhang2020quantum,
    author = "Zhang, Tian and Dahlsten, Oscar and Vedral, Vlatko",
    title = "{Quantum correlations in time}",
    eprint = "2002.10448",
    archivePrefix = "arXiv",
    primaryClass = "quant-ph",
    month = "2",
    year = "2020"
}

@article{khemani2019brief,
    author = "Khemani, Vedika and Moessner, Roderich and Sondhi, S.L.",
    title = "{A Brief History of Time Crystals}",
    eprint = "1910.10745",
    archivePrefix = "arXiv",
    primaryClass = "cond-mat.str-el",
    month = "10",
    year = "2019"
}

@article{anderson2010problem,
    author = "Anderson, Edward",
    title = "{The Problem of Time in Quantum Gravity}",
    eprint = "1009.2157",
    archivePrefix = "arXiv",
    primaryClass = "gr-qc",
    month = "9",
    year = "2010"
}

@article{dowker1996consistent,
  title={On the consistent histories approach to quantum mechanics},
  author={Dowker, Fay and Kent, Adrian},
  journal={Journal of Statistical Physics},
  volume={82},
  number={5-6},
  pages={1575--1646},
  year={1996},
  publisher={Springer},
  url={https://link.springer.com/article/10.1007/BF02183396}
}

@article{dowker1992quantum,
  title = {Quantum mechanics of history: The decoherence functional in quantum mechanics},
  author = {Dowker, H. F. and Halliwell, J. J.},
  journal = {Phys. Rev. D},
  volume = {46},
  issue = {4},
  pages = {1580--1609},
  numpages = {0},
  year = {1992},
  month = {Aug},
  publisher = {American Physical Society},
  doi = {10.1103/PhysRevD.46.1580},
  url = {https://link.aps.org/doi/10.1103/PhysRevD.46.1580}
}

@article{branciard2015simplest,
	doi = {10.1088/1367-2630/18/1/013008},
	url = {https://doi.org/10.1088%2F1367-2630%2F18%2F1%2F013008},
	year = 2015,
	month = {dec},
	publisher = {{IOP} Publishing},
	volume = {18},
	number = {1},
	pages = {013008},
	author = {Cyril Branciard and Mateus Ara{\'{u}}jo and Adrien Feix and Fabio Costa and {\v{C}}aslav Brukner},
	title = {The simplest causal inequalities and their violation},
	journal = {New Journal of Physics}
}

@article{costa2018unifying,
  title = {Unifying framework for spatial and temporal quantum correlations},
  author = {Costa, Fabio and Ringbauer, Martin and Goggin, Michael E. and White, Andrew G. and Fedrizzi, Alessandro},
  journal = {Phys. Rev. A},
  volume = {98},
  issue = {1},
  pages = {012328},
  numpages = {7},
  year = {2018},
  month = {Jul},
  publisher = {American Physical Society},
  doi = {10.1103/PhysRevA.98.012328},
  url = {https://link.aps.org/doi/10.1103/PhysRevA.98.012328}
}

@article{maldacena2015bound,
    author = "Maldacena, Juan and Shenker, Stephen H. and Stanford, Douglas",
    title = "{A bound on chaos}",
    eprint = "1503.01409",
    archivePrefix = "arXiv",
    primaryClass = "hep-th",
    doi = "10.1007/JHEP08(2016)106",
    journal = "JHEP",
    volume = "08",
    pages = "106",
    year = "2016"
}

@article{roberts2016chaos,
    author = "Roberts, Daniel A. and Yoshida, Beni",
    title = "{Chaos and complexity by design}",
    eprint = "1610.04903",
    archivePrefix = "arXiv",
    primaryClass = "quant-ph",
    doi = "10.1007/JHEP04(2017)121",
    journal = "JHEP",
    volume = "04",
    pages = "121",
    year = "2017"
}

@book{zinn2010path,
  title={Path integrals in quantum mechanics},
  author={Zinn-Justin, Jean},
  year={2010},
  publisher={Oxford University Press}
}

@book{feynman2010quantum,
  title={Quantum mechanics and path integrals},
  author={Feynman, Richard P and Hibbs, Albert R and Styer, Daniel F},
  year={2010},
  publisher={Courier Corporation}
}

@article{araujo2017purification,
  title={A purification postulate for quantum mechanics with indefinite causal order},
  author={Ara{\'u}jo, Mateus and Feix, Adrien and Navascu{\'e}s, Miguel and Brukner, {\v{C}}aslav},
  journal={Quantum},
  volume={1},
  pages={10},
  year={2017},
  publisher={Verein zur F{\"o}rderung des Open Access Publizierens in den Quantenwissenschaften}
}

@article{pisarczyk2019causal,
  title = {Causal Limit on Quantum Communication},
  author = {Pisarczyk, Robert and Zhao, Zhikuan and Ouyang, Yingkai and Vedral, Vlatko and Fitzsimons, Joseph F.},
  journal = {Phys. Rev. Lett.},
  volume = {123},
  issue = {15},
  pages = {150502},
  numpages = {6},
  year = {2019},
  month = {Oct},
  publisher = {American Physical Society},
  doi = {10.1103/PhysRevLett.123.150502},
  url = {https://link.aps.org/doi/10.1103/PhysRevLett.123.150502}
}

@article{page1983evolution,
  title = {Evolution without evolution: Dynamics described by stationary observables},
  author = {Page, Don N. and Wootters, William K.},
  journal = {Phys. Rev. D},
  volume = {27},
  issue = {12},
  pages = {2885--2892},
  numpages = {0},
  year = {1983},
  month = {Jun},
  publisher = {American Physical Society},
  doi = {10.1103/PhysRevD.27.2885},
  url = {https://link.aps.org/doi/10.1103/PhysRevD.27.2885}
}

@book{barbour2001end,
  title={The end of time: The next revolution in physics},
  author={Barbour, Julian},
  year={2001},
  publisher={Oxford University Press}
}

@article{arnowitt1959dynamical,
  title = {Dynamical Structure and Definition of Energy in General Relativity},
  author = {Arnowitt, R. and Deser, S. and Misner, C. W.},
  journal = {Phys. Rev.},
  volume = {116},
  issue = {5},
  pages = {1322--1330},
  numpages = {0},
  year = {1959},
  month = {Dec},
  publisher = {American Physical Society},
  doi = {10.1103/PhysRev.116.1322},
  url = {https://link.aps.org/doi/10.1103/PhysRev.116.1322}
}

@article{hartle1983wave,
  title = {Wave function of the Universe},
  author = {Hartle, J. B. and Hawking, S. W.},
  journal = {Phys. Rev. D},
  volume = {28},
  issue = {12},
  pages = {2960--2975},
  numpages = {0},
  year = {1983},
  month = {Dec},
  publisher = {American Physical Society},
  doi = {10.1103/PhysRevD.28.2960},
  url = {https://link.aps.org/doi/10.1103/PhysRevD.28.2960}
}

@article{dewitt1967quantum,
  title = {Quantum Theory of Gravity. I. The Canonical Theory},
  author = {DeWitt, Bryce S.},
  journal = {Phys. Rev.},
  volume = {160},
  issue = {5},
  pages = {1113--1148},
  numpages = {0},
  year = {1967},
  month = {Aug},
  publisher = {American Physical Society},
  doi = {10.1103/PhysRev.160.1113},
  url = {https://link.aps.org/doi/10.1103/PhysRev.160.1113}
}

@book{smolin2013time,
  title={Time reborn: From the crisis in physics to the future of the universe},
  author={Smolin, Lee},
  year={2013},
  publisher={HMH}
}

@article{smolin2015temporal,
title = "Temporal naturalism",
journal = "Studies in History and Philosophy of Science Part B: Studies in History and Philosophy of Modern Physics",
volume = "52",
pages = "86 - 102",
year = "2015",
note = "Cosmology and Time: Philosophers and Scientists in Dialogue",
issn = "1355-2198",
doi = "https://doi.org/10.1016/j.shpsb.2015.03.005",
url = "http://www.sciencedirect.com/science/article/pii/S1355219815000271",
author = "Lee Smolin"
}

@article{cortes2014quantum,
  title = {Quantum energetic causal sets},
  author = {Cort\^es, Marina and Smolin, Lee},
  journal = {Phys. Rev. D},
  volume = {90},
  issue = {4},
  pages = {044035},
  numpages = {5},
  year = {2014},
  month = {Aug},
  publisher = {American Physical Society},
  doi = {10.1103/PhysRevD.90.044035},
  url = {https://link.aps.org/doi/10.1103/PhysRevD.90.044035}
}

@article{weedbrook2012gaussian,
   title = {Gaussian quantum information},
   author = {Weedbrook, Christian and Pirandola, Stefano and Garc\'{\i}a-Patr\'on, Ra\'ul and Cerf, Nicolas J. and Ralph, Timothy C. and Shapiro, Jeffrey H. and Lloyd, Seth},
  journal = {Rev. Mod. Phys.},
  volume = {84},
  issue = {2},
   pages = {621--669},
   numpages = {0},
   year = {2012},
   month = {May},
   publisher = {American Physical Society},
 	doi = {10.1103/RevModPhys.84.621},
 	url = {https://link.aps.org/doi/10.1103/RevModPhys.84.621}
}

@article{wang2007quantum,
  title = {Quantum information with Gaussian states},
  journal = {Physics Reports}, 
  volume = {448},
  number = {1},
  pages = {1 - 111},
  year = {2007},
  issn = {0370-1573},
  doi = {https://doi.org/10.1016/j.physrep.2007.04.005},
  url = {http://www.sciencedirect.com/science/article/pii/S0370157307001822},
  author = {Xiang-Bin Wang and Tohya Hiroshima and Akihisa Tomita and Masahito Hayashi}
}

@article{adesso2014continuous,
  title={Continuous variable quantum information: Gaussian states and beyond},
  author={Adesso, Gerardo and Ragy, Sammy and Lee, Antony R},
  journal={Open Systems \& Information Dynamics},
  volume={21},
  number={01n02},
  pages={1440001},
  year={2014},
  publisher={World Scientific}
}

@article{simon1994quantum,
  title = {Quantum-noise matrix for multimode systems: U(n) invariance, squeezing, and normal forms},
  author = {Simon, R. and Mukunda, N. and Dutta, Biswadeb},
  journal = {Phys. Rev. A},
  volume = {49},
  issue = {3},
  pages = {1567--1583},
  numpages = {0},
  year = {1994},
  month = {Mar},
  publisher = {American Physical Society},
  doi = {10.1103/PhysRevA.49.1567},
  url = {https://link.aps.org/doi/10.1103/PhysRevA.49.1567}
}

@article{wigner1932quantum,
  title = {On the Quantum Correction For Thermodynamic Equilibrium},
  author = {Wigner, E.},
  journal = {Phys. Rev.},
  volume = {40},
  issue = {5},
  pages = {749--759},
  numpages = {0},
  year = {1932},
  month = {Jun},
  publisher = {American Physical Society},
  doi = {10.1103/PhysRev.40.749},
  url = {https://link.aps.org/doi/10.1103/PhysRev.40.749}
}

@article{horodecki1996information,
  title = {Information-theoretic aspects of inseparability of mixed states},
  author = {Horodecki, Ryszard and Horodecki, Michal/},
  journal = {Phys. Rev. A},
  volume = {54},
  issue = {3},
  pages = {1838--1843},
  numpages = {0},
  year = {1996},
  month = {Sep},
  publisher = {American Physical Society},
  doi = {10.1103/PhysRevA.54.1838},
  url = {https://link.aps.org/doi/10.1103/PhysRevA.54.1838}
}

@article{zhao2018geometry,
  title = {Geometry of quantum correlations in space-time},
  author = {Zhao, Zhikuan and Pisarczyk, Robert and Thompson, Jayne and Gu, Mile and Vedral, Vlatko and Fitzsimons, Joseph F.},
  journal = {Phys. Rev. A},
  volume = {98},
  issue = {5},
  pages = {052312},
  numpages = {5},
  year = {2018},
  month = {Nov},
  publisher = {American Physical Society},
  doi = {10.1103/PhysRevA.98.052312},
  url = {https://link.aps.org/doi/10.1103/PhysRevA.98.052312}
}

@book{myerson2013game,
  title={Game theory},
  author={Myerson, Roger B},
  year={2013},
  publisher={Harvard University Press}
}

@INPROCEEDINGS{cleve2004consequences, 
author={R. {Cleve} and P. {Hoyer} and B. {Toner} and J. {Watrous}}, 
booktitle={Proceedings. 19th IEEE Annual Conference on Computational Complexity, 2004.}, 
title={Consequences and limits of nonlocal strategies}, 
year={2004}, 
volume={}, 
number={}, 
pages={236-249}, 
doi={10.1109/CCC.2004.1313847}, 
ISSN={1093-0159}, 
month={June},}

@article{clauser1969proposed,
  title = {Proposed Experiment to Test Local Hidden-Variable Theories},
  author = {Clauser, John F. and Horne, Michael A. and Shimony, Abner and Holt, Richard A.},
  journal = {Phys. Rev. Lett.},
  volume = {23},
  issue = {15},
  pages = {880--884},
  numpages = {0},
  year = {1969},
  month = {Oct},
  publisher = {American Physical Society},
  doi = {10.1103/PhysRevLett.23.880},
  url = {https://link.aps.org/doi/10.1103/PhysRevLett.23.880}
}

@article{cahill1969ordered,
  title = {Ordered Expansions in Boson Amplitude Operators},
  author = {Cahill, K. E. and Glauber, R. J.},
  journal = {Phys. Rev.},
  volume = {177},
  issue = {5},
  pages = {1857--1881},
  numpages = {0},
  year = {1969},
  month = {Jan},
  publisher = {American Physical Society},
  doi = {10.1103/PhysRev.177.1857},
  url = {https://link.aps.org/doi/10.1103/PhysRev.177.1857}
}

@article{cahill1969density,
  title = {Density Operators and Quasiprobability Distributions},
  author = {Cahill, K. E. and Glauber, R. J.},
  journal = {Phys. Rev.},
  volume = {177},
  issue = {5},
  pages = {1882--1902},
  numpages = {0},
  year = {1969},
  month = {Jan},
  publisher = {American Physical Society},
  doi = {10.1103/PhysRev.177.1882},
  url = {https://link.aps.org/doi/10.1103/PhysRev.177.1882}
}

@article{royer1977wigner,
  title = {Wigner function as the expectation value of a parity operator},
  author = {Royer, Antoine},
  journal = {Phys. Rev. A},
  volume = {15},
  issue = {2},
  pages = {449--450},
  numpages = {0},
  year = {1977},
  month = {Feb},
  publisher = {American Physical Society},
  doi = {10.1103/PhysRevA.15.449},
  url = {https://link.aps.org/doi/10.1103/PhysRevA.15.449}
}

@article{banaszek1998nonlocality,
  title = {Nonlocality of the Einstein-Podolsky-Rosen state in the Wigner representation},
  author = {Banaszek, Konrad and W\'odkiewicz, Krzysztof},
  journal = {Phys. Rev. A},
  volume = {58},
  issue = {6},
  pages = {4345--4347},
  numpages = {0},
  year = {1998},
  month = {Dec},
  publisher = {American Physical Society},
  doi = {10.1103/PhysRevA.58.4345},
  url = {https://link.aps.org/doi/10.1103/PhysRevA.58.4345}
}

@article{hillery1984distribution,
title = "Distribution functions in physics: Fundamentals",
journal = "Physics Reports",
volume = "106",
number = "3",
pages = "121 - 167",
year = "1984",
issn = "0370-1573",
doi = "https://doi.org/10.1016/0370-1573(84)90160-1",
url = "http://www.sciencedirect.com/science/article/pii/0370157384901601",
author = "M. Hillery and R.F. O'Connell and M.O. Scully and E.P. Wigner"
}

@Inbook{o1981quantum,
author="O'Connell, R. F.
and Wigner, E. P.",
editor="Wightman, Arthur S.",
title="Quantum-Mechanical Distribution Functions: Conditions for Uniqueness",
bookTitle="Part I: Physical Chemistry. Part II: Solid State Physics",
year="1997",
publisher="Springer Berlin Heidelberg",
address="Berlin, Heidelberg",
pages="263--266",
isbn="978-3-642-59033-7",
doi="10.1007/978-3-642-59033-7_26",
url="https://doi.org/10.1007/978-3-642-59033-7_26"
}

@article{caves1986quantum1,
  title = {Quantum mechanics of measurements distributed in time. A path-integral formulation},
  author = {Caves, Carlton M.},
  journal = {Phys. Rev. D},
  volume = {33},
  issue = {6},
  pages = {1643--1665},
  numpages = {0},
  year = {1986},
  month = {Mar},
  publisher = {American Physical Society},
  doi = {10.1103/PhysRevD.33.1643},
  url = {https://link.aps.org/doi/10.1103/PhysRevD.33.1643}
}

@article{caves1987quantum2,
  title = {Quantum mechanics of measurements distributed in time. II. Connections among formulations},
  author = {Caves, Carlton M.},
  journal = {Phys. Rev. D},
  volume = {35},
  issue = {6},
  pages = {1815--1830},
  numpages = {0},
  year = {1987},
  month = {Mar},
  publisher = {American Physical Society},
  doi = {10.1103/PhysRevD.35.1815},
  url = {https://link.aps.org/doi/10.1103/PhysRevD.35.1815}
}

@book{kraus1983states,
  title={States, effects and operations: fundamental notions of quantum theory},
  author={Kraus, Karl},
  year={1983},
  publisher={Springer}
}

@Article{barchielli1982model,
author="Barchielli, A.
and Lanz, L.
and Prosperi, G. M.",
title="A model for the macroscopic description and continual observations in quantum mechanics",
journal="Il Nuovo Cimento B (1971-1996)",
year="1982",
month="Nov",
day="01",
volume="72",
number="1",
pages="79--121",
issn="1826-9877",
doi="10.1007/BF02894935",
url="https://doi.org/10.1007/BF02894935"
}

@article{horsman2017can,
  title={Can a quantum state over time resemble a quantum state at a single time?},
  author={Horsman, Dominic and Heunen, Chris and Pusey, Matthew F and Barrett, Jonathan and Spekkens, Robert W},
  journal={Proc. R. Soc. A},
  volume={473},
  number={2205},
  pages={20170395},
  year={2017},
  publisher={The Royal Society}
}

@article{steinlechner2013quantum,
  title={Quantum-dense metrology},
  author={Steinlechner, Sebastian and Bauchrowitz, J{\"o}ran and Meinders, Melanie and M{\"u}ller-Ebhardt, Helge and Danzmann, Karsten and Schnabel, Roman},
  journal={Nature Photonics},
  volume={7},
  number={8},
  pages={626},
  year={2013},
  publisher={Nature Publishing Group}
}

@article{zhang2019pseudo,
  title={Pseudo-density matrix: the relation with indefinite causal structures, consistent histories and generalised non-local games and out-of-time-order correlations},
  author={Zhang, Tian},
  journal={In Preparation},
  year={2019}
}

@article{zhang2019constructing,
  title={Constructing continuous-variable spacetime quantum states from measurement correlations},
  author={Zhang, Tian and Dahlsten, Oscar and Vedral, Vlatko},
  journal={arXiv preprint arXiv:1903.06312},
  year={2019}
}

@article{zhang2019long,
  title={Long-range temporal correlations in quantum error correction and time crystals},
  author={Zhang, Tian and Zhu, Zhennan and Tennie, Felix and Yang, Xiaodong and Peng, Xinhua and Vedral, Vlatko},
  journal={In Preparation},
  year={2019}
}

@Article{royer1989measurement,
author="Royer, Antoine",
title="Measurement of quantum states and the Wigner function",
journal="Foundations of Physics",
year="1989",
month="Jan",
day="01",
volume="19",
number="1",
pages="3--32",
issn="1572-9516",
doi="10.1007/BF00737764",
url="https://doi.org/10.1007/BF00737764"
}

@article{fitzsimons2015quantum,
  title={Quantum correlations which imply causation},
  author={Fitzsimons, Joseph F and Jones, Jonathan A and Vedral, Vlatko},
  journal={Scientific reports},
  volume={5},
  pages={18281},
  year={2015},
  publisher={Nature Publishing Group}
}

@Inbook{wigner1973epistemological,
author="Wigner, E. P.",
editor="Mehra, Jagdish",
title="Epistemological Perspective on Quantum Theory",
bookTitle="Philosophical Reflections and Syntheses",
year="1995",
publisher="Springer Berlin Heidelberg",
address="Berlin, Heidelberg",
pages="55--71",
isbn="978-3-642-78374-6",
doi="10.1007/978-3-642-78374-6_5",
url="https://doi.org/10.1007/978-3-642-78374-6_5"
}

@article{caves1987quantum3,
  title = {Quantum-mechanical model for continuous position measurements},
  author = {Caves, Carlton M. and Milburn, G. J.},
  journal = {Phys. Rev. A},
  volume = {36},
  issue = {12},
  pages = {5543--5555},
  numpages = {0},
  year = {1987},
  month = {Dec},
  publisher = {American Physical Society},
  doi = {10.1103/PhysRevA.36.5543},
  url = {https://link.aps.org/doi/10.1103/PhysRevA.36.5543}
}

@book{weinberg1995quantum,
      author         = "Weinberg, Steven",
      title          = "{The Quantum theory of fields. Vol. 1: Foundations}",
      publisher      = "Cambridge University Press",
      year           = "2005",
      ISBN           = "9780521670531, 9780511252044",
      SLACcitation   = "%%CITATION = INSPIRE-406190;%%"
}

@misc{nielsen2002quantum,
  title={Quantum computation and quantum information},
  author={Nielsen, Michael A and Chuang, Isaac},
  year={2002},
  publisher={AAPT}
}

@Article{griffiths1984consistent,
author="Griffiths, Robert B.",
title="Consistent histories and the interpretation of quantum mechanics",
journal="Journal of Statistical Physics",
year="1984",
month="Jul",
day="01",
volume="36",
number="1",
pages="219--272",
issn="1572-9613",
doi="10.1007/BF01015734",
url="https://doi.org/10.1007/BF01015734"
}

@book{griffiths2003consistent,
  title={Consistent quantum theory},
  author={Griffiths, Robert B},
  year={2003},
  publisher={Cambridge University Press}
}

@article{gell2018quantum,
      author         = "Gell-Mann, Murray and Hartle, James B.",
      title          = "{Quantum Mechanics in the Light of Quantum Cosmology}",
      year           = "1989",
      eprint         = "1803.04605",
      archivePrefix  = "arXiv",
      primaryClass   = "gr-qc",
      SLACcitation   = "%%CITATION = ARXIV:1803.04605;%%"
}

@article{gell1993classical,
  title = {Classical equations for quantum systems},
  author = {Gell-Mann, Murray and Hartle, James B.},
  journal = {Phys. Rev. D},
  volume = {47},
  issue = {8},
  pages = {3345--3382},
  numpages = {0},
  year = {1993},
  month = {Apr},
  publisher = {American Physical Society},
  doi = {10.1103/PhysRevD.47.3345},
  url = {https://link.aps.org/doi/10.1103/PhysRevD.47.3345}
}

@article{omnes1990hilbert,
  title={From Hilbert space to common sense: A synthesis of recent progress in the interpretation of quantum mechanics},
  author={Omn{\'e}s, Roland},
  journal={Annals of Physics},
  volume={201},
  number={2},
  pages={354--447},
  year={1990},
  publisher={Elsevier},
  url={https://www.sciencedirect.com/science/article/pii/000349169090045P}
}

@article{hardy2007towards,
	doi = {10.1088/1751-8113/40/12/s12},
	url = {https://doi.org/10.1088%2F1751-8113%2F40%2F12%2Fs12},
	year = 2007,
	month = {mar},
	publisher = {{IOP} Publishing},
	volume = {40},
	number = {12},
	pages = {3081--3099},
	author = {Lucien Hardy},
	title = {Towards quantum gravity: a framework for probabilistic theories with non-fixed causal structure},
	journal = {Journal of Physics A: Mathematical and Theoretical}
}

@Inbook{hardy2009quantum,
author="Hardy, Lucien",
title="Quantum Gravity Computers: On the Theory of Computation with Indefinite Causal Structure",
bookTitle="Quantum Reality, Relativistic Causality, and Closing the Epistemic Circle: Essays in Honour of Abner Shimony",
year="2009",
publisher="Springer Netherlands",
address="Dordrecht",
pages="379--401",
isbn="978-1-4020-9107-0",
doi="10.1007/978-1-4020-9107-0_21",
url="https://doi.org/10.1007/978-1-4020-9107-0_21"
}

@article{araujo2015witnessing,
	doi = {10.1088/1367-2630/17/10/102001},
	url = {https://doi.org/10.1088%2F1367-2630%2F17%2F10%2F102001},
	year = 2015,
	month = {oct},
	publisher = {{IOP} Publishing},
	volume = {17},
	number = {10},
	pages = {102001},
	author = {Mateus Ara{\'{u}}jo and Cyril Branciard and Fabio Costa and Adrien Feix and Christina Giarmatzi and {\v{C}}aslav Brukner},
	title = {Witnessing causal nonseparability},
	journal = {New Journal of Physics}
}

@article{oreshkov2012quantum,
  title={Quantum correlations with no causal order},
  author={Oreshkov, Ognyan and Costa, Fabio and Brukner, {\v{C}}aslav},
  journal={Nature communications},
  volume={3},
  pages={1092},
  year={2012},
  publisher={Nature Publishing Group}
}

@article{chiribella2008quantum,
  title = {Quantum Circuit Architecture},
  author = {Chiribella, G. and D'Ariano, G. M. and Perinotti, P.},
  journal = {Phys. Rev. Lett.},
  volume = {101},
  issue = {6},
  pages = {060401},
  numpages = {4},
  year = {2008},
  month = {Aug},
  publisher = {American Physical Society},
  doi = {10.1103/PhysRevLett.101.060401},
  url = {https://link.aps.org/doi/10.1103/PhysRevLett.101.060401}
}

@article{chiribella2009theoretical,
  title = {Theoretical framework for quantum networks},
  author = {Chiribella, Giulio and D'Ariano, Giacomo Mauro and Perinotti, Paolo},
  journal = {Phys. Rev. A},
  volume = {80},
  issue = {2},
  pages = {022339},
  numpages = {20},
  year = {2009},
  month = {Aug},
  publisher = {American Physical Society},
  doi = {10.1103/PhysRevA.80.022339},
  url = {https://link.aps.org/doi/10.1103/PhysRevA.80.022339}
}

@article{hardy2012operator,
  title={The operator tensor formulation of quantum theory},
  author={Hardy, Lucien},
  journal={Philosophical Transactions of the Royal Society A: Mathematical, Physical and Engineering Sciences},
  volume={370},
  number={1971},
  pages={3385--3417},
  year={2012},
  publisher={The Royal Society Publishing}
}

@article{hardy2018construction,
  title={The construction interpretation: a conceptual road to quantum gravity},
  author={Hardy, Lucien},
  journal={arXiv preprint arXiv:1807.10980},
  year={2018}
}

@article{pollock2018non,
  title = {Non-Markovian quantum processes: Complete framework and efficient characterization},
  author = {Pollock, Felix A. and Rodr\'{\i}guez-Rosario, C\'esar and Frauenheim, Thomas and Paternostro, Mauro and Modi, Kavan},
  journal = {Phys. Rev. A},
  volume = {97},
  issue = {1},
  pages = {012127},
  numpages = {13},
  year = {2018},
  month = {Jan},
  publisher = {American Physical Society},
  doi = {10.1103/PhysRevA.97.012127},
  url = {https://link.aps.org/doi/10.1103/PhysRevA.97.012127}
}

@article{milz2017introduction,
  title={An introduction to operational quantum dynamics},
  author={Milz, Simon and Pollock, Felix A and Modi, Kavan},
  journal={Open Systems \& Information Dynamics},
  volume={24},
  number={04},
  pages={1740016},
  year={2017},
  publisher={World Scientific}
}

@Article{cotler2018superdensity,
author="Cotler, Jordan
and Jian, Chao-Ming
and Qi, Xiao-Liang
and Wilczek, Frank",
title="Superdensity operators for spacetime quantum mechanics",
journal="Journal of High Energy Physics",
year="2018",
month="Sep",
day="17",
volume="2018",
number="9",
pages="93",
issn="1029-8479",
doi="10.1007/JHEP09(2018)093",
url="https://doi.org/10.1007/JHEP09(2018)093"
}

@Article{cotler2019quantum,
author="Cotler, Jordan
and Han, Xizhi
and Qi, Xiao-Liang
and Yang, Zhao",
title="Quantum causal influence",
journal="Journal of High Energy Physics",
year="2019",
month="Jul",
day="08",
volume="2019",
number="7",
pages="42",
issn="1029-8479",
doi="10.1007/JHEP07(2019)042",
url="https://doi.org/10.1007/JHEP07(2019)042"
}

@article{kretschmann2005quantum,
  title = {Quantum channels with memory},
  author = {Kretschmann, Dennis and Werner, Reinhard F.},
  journal = {Phys. Rev. A},
  volume = {72},
  issue = {6},
  pages = {062323},
  numpages = {19},
  year = {2005},
  month = {Dec},
  publisher = {American Physical Society},
  doi = {10.1103/PhysRevA.72.062323},
  url = {https://link.aps.org/doi/10.1103/PhysRevA.72.062323}
}

@inproceedings{gutoski2007toward,
  title={Toward a general theory of quantum games},
  author={Gutoski, Gus and Watrous, John},
  booktitle={Proceedings of the thirty-ninth annual ACM symposium on Theory of computing},
  pages={565--574},
  year={2007},
  organization={ACM}
}

@article{aharonov1964time,
  title = {Time Symmetry in the Quantum Process of Measurement},
  author = {Aharonov, Yakir and Bergmann, Peter G. and Lebowitz, Joel L.},
  journal = {Phys. Rev.},
  volume = {134},
  issue = {6B},
  pages = {B1410--B1416},
  numpages = {0},
  year = {1964},
  month = {Jun},
  publisher = {American Physical Society},
  doi = {10.1103/PhysRev.134.B1410},
  url = {https://link.aps.org/doi/10.1103/PhysRev.134.B1410}
}

@article{aharonov2009multiple,
  title = {Multiple-time states and multiple-time measurements in quantum mechanics},
  author = {Aharonov, Yakir and Popescu, Sandu and Tollaksen, Jeff and Vaidman, Lev},
  journal = {Phys. Rev. A},
  volume = {79},
  issue = {5},
  pages = {052110},
  numpages = {16},
  year = {2009},
  month = {May},
  publisher = {American Physical Society},
  doi = {10.1103/PhysRevA.79.052110},
  url = {https://link.aps.org/doi/10.1103/PhysRevA.79.052110}
}

@article{silva2017connecting,
	doi = {10.1088/1367-2630/aa84fe},
	url = {https://doi.org/10.1088%2F1367-2630%2Faa84fe},
	year = 2017,
	month = {oct},
	publisher = {{IOP} Publishing},
	volume = {19},
	number = {10},
	pages = {103022},
	author = {Ralph Silva and Yelena Guryanova and Anthony J Short and Paul Skrzypczyk and Nicolas Brunner and Sandu Popescu},
	title = {Connecting processes with indefinite causal order and multi-time quantum states},
	journal = {New Journal of Physics}
}

@article{oeckl2003general,
title = "A “general boundary” formulation for quantum mechanics and quantum gravity",
journal = "Physics Letters B",
volume = "575",
number = "3",
pages = "318 - 324",
year = "2003",
issn = "0370-2693",
doi = "https://doi.org/10.1016/j.physletb.2003.08.043",
url = "http://www.sciencedirect.com/science/article/pii/S0370269303013066",
author = "Robert Oeckl"
}

@article{costa2016quantum,
	doi = {10.1088/1367-2630/18/6/063032},
	url = {https://doi.org/10.1088%2F1367-2630%2F18%2F6%2F063032},
	year = 2016,
	month = {jun},
	publisher = {{IOP} Publishing},
	volume = {18},
	number = {6},
	pages = {063032},
	author = {Fabio Costa and Sally Shrapnel},
	title = {Quantum causal modelling},
	journal = {New Journal of Physics}
}

@article{allen2017quantum,
  title = {Quantum Common Causes and Quantum Causal Models},
  author = {Allen, John-Mark A. and Barrett, Jonathan and Horsman, Dominic C. and Lee, Ciar\'an M. and Spekkens, Robert W.},
  journal = {Phys. Rev. X},
  volume = {7},
  issue = {3},
  pages = {031021},
  numpages = {22},
  year = {2017},
  month = {Jul},
  publisher = {American Physical Society},
  doi = {10.1103/PhysRevX.7.031021},
  url = {https://link.aps.org/doi/10.1103/PhysRevX.7.031021}
}

@article{jamiolkowski1972linear,
title = "Linear transformations which preserve trace and positive semidefiniteness of operators",
journal = "Reports on Mathematical Physics",
volume = "3",
number = "4",
pages = "275 - 278",
year = "1972",
issn = "0034-4877",
doi = "https://doi.org/10.1016/0034-4877(72)90011-0",
url = "http://www.sciencedirect.com/science/article/pii/0034487772900110",
author = "A. Jamiołkowski"
}

@article{choi1975completely,
title = "Completely positive linear maps on complex matrices",
journal = "Linear Algebra and its Applications",
volume = "10",
number = "3",
pages = "285 - 290",
year = "1975",
issn = "0024-3795",
doi = "https://doi.org/10.1016/0024-3795(75)90075-0",
url = "http://www.sciencedirect.com/science/article/pii/0024379575900750",
author = "Man-Duen Choi"
}

@article{godel1949example,
  title = {An Example of a New Type of Cosmological Solutions of Einstein's Field Equations of Gravitation},
  author = {G\"odel, Kurt},
  journal = {Rev. Mod. Phys.},
  volume = {21},
  issue = {3},
  pages = {447--450},
  numpages = {0},
  year = {1949},
  month = {Jul},
  publisher = {American Physical Society},
  doi = {10.1103/RevModPhys.21.447},
  url = {https://link.aps.org/doi/10.1103/RevModPhys.21.447}
}

@article{deutsch1991quantum,
  title = {Quantum mechanics near closed timelike lines},
  author = {Deutsch, David},
  journal = {Phys. Rev. D},
  volume = {44},
  issue = {10},
  pages = {3197--3217},
  numpages = {0},
  year = {1991},
  month = {Nov},
  publisher = {American Physical Society},
  doi = {10.1103/PhysRevD.44.3197},
  url = {https://link.aps.org/doi/10.1103/PhysRevD.44.3197}
}

@article{bennett2005teleportation,
  title={Teleportation, Simulated Time Travel, and How to Flirt with Someone Who Has Fallen into a Black Hole},
  author={Bennett, CH and Schumacher, B},
  journal={QUPON, Wien},
  year={2005}
}

@Article{svetlichny2011time,
author="Svetlichny, George",
title="Time Travel: Deutsch vs. Teleportation",
journal="International Journal of Theoretical Physics",
year="2011",
month="Dec",
day="01",
volume="50",
number="12",
pages="3903--3914",
issn="1572-9575",
doi="10.1007/s10773-011-0973-x",
url="https://doi.org/10.1007/s10773-011-0973-x"
}

@Article{brun2012prefect,
author="Brun, Todd A.
and Wilde, Mark M.",
title="Perfect State Distinguishability and Computational Speedups with Postselected Closed Timelike Curves",
journal="Foundations of Physics",
year="2012",
month="Mar",
day="01",
volume="42",
number="3",
pages="341--361",
issn="1572-9516",
doi="10.1007/s10701-011-9601-0",
url="https://doi.org/10.1007/s10701-011-9601-0"
}

@article{lloyd2011closed,
  title = {Closed Timelike Curves via Postselection: Theory and Experimental Test of Consistency},
  author = {Lloyd, Seth and Maccone, Lorenzo and Garcia-Patron, Raul and Giovannetti, Vittorio and Shikano, Yutaka and Pirandola, Stefano and Rozema, Lee A. and Darabi, Ardavan and Soudagar, Yasaman and Shalm, Lynden K. and Steinberg, Aephraim M.},
  journal = {Phys. Rev. Lett.},
  volume = {106},
  issue = {4},
  pages = {040403},
  numpages = {4},
  year = {2011},
  month = {Jan},
  publisher = {American Physical Society},
  doi = {10.1103/PhysRevLett.106.040403},
  url = {https://link.aps.org/doi/10.1103/PhysRevLett.106.040403}
}

@article{araujo2017quantum,
  title = {Quantum computation with indefinite causal structures},
  author = {Ara\'ujo, Mateus and Gu\'erin, Philippe Allard and Baumeler, \"Amin},
  journal = {Phys. Rev. A},
  volume = {96},
  issue = {5},
  pages = {052315},
  numpages = {11},
  year = {2017},
  month = {Nov},
  publisher = {American Physical Society},
  doi = {10.1103/PhysRevA.96.052315},
  url = {https://link.aps.org/doi/10.1103/PhysRevA.96.052315}
}

@article{yang1962concept,
  title = {Concept of Off-Diagonal Long-Range Order and the Quantum Phases of Liquid He and of Superconductors},
  author = {Yang, C. N.},
  journal = {Rev. Mod. Phys.},
  volume = {34},
  issue = {4},
  pages = {694--704},
  numpages = {0},
  year = {1962},
  month = {Oct},
  publisher = {American Physical Society},
  doi = {10.1103/RevModPhys.34.694},
  url = {https://link.aps.org/doi/10.1103/RevModPhys.34.694}
}

@article{khemani2017defining,
  title = {Defining time crystals via representation theory},
  author = {Khemani, Vedika and von Keyserlingk, C. W. and Sondhi, S. L.},
  journal = {Phys. Rev. B},
  volume = {96},
  issue = {11},
  pages = {115127},
  numpages = {7},
  year = {2017},
  month = {Sep},
  publisher = {American Physical Society},
  doi = {10.1103/PhysRevB.96.115127},
  url = {https://link.aps.org/doi/10.1103/PhysRevB.96.115127}
}

@article{barenco1997stabilization,
  title={Stabilization of quantum computations by symmetrization},
  author={Barenco, Adriano and Berthiaume, Andr{\'e} and Deutsch, David and Ekert, Artur and Jozsa, Richard and Macchiavello, Chiara},
  journal={SIAM Journal on Computing},
  volume={26},
  number={5},
  pages={1541--1557},
  year={1997},
  publisher={SIAM}
}

@article{goldstone1962broken,
  title = {Broken Symmetries},
  author = {Goldstone, Jeffrey and Salam, Abdus and Weinberg, Steven},
  journal = {Phys. Rev.},
  volume = {127},
  issue = {3},
  pages = {965--970},
  numpages = {0},
  year = {1962},
  month = {Aug},
  publisher = {American Physical Society},
  doi = {10.1103/PhysRev.127.965},
  url = {https://link.aps.org/doi/10.1103/PhysRev.127.965}
}

@Article{coleman1973there,
author="Coleman, Sidney",
title="There are no Goldstone bosons in two dimensions",
journal="Communications in Mathematical Physics",
year="1973",
month="Dec",
day="01",
volume="31",
number="4",
pages="259--264",
issn="1432-0916",
doi="10.1007/BF01646487",
url="https://doi.org/10.1007/BF01646487"
}

@article{wilczek2012quantum,
  title = {Quantum Time Crystals},
  author = {Wilczek, Frank},
  journal = {Phys. Rev. Lett.},
  volume = {109},
  issue = {16},
  pages = {160401},
  numpages = {5},
  year = {2012},
  month = {Oct},
  publisher = {American Physical Society},
  doi = {10.1103/PhysRevLett.109.160401},
  url = {https://link.aps.org/doi/10.1103/PhysRevLett.109.160401}
}

@article{watanabe2015absence,
  title = {Absence of Quantum Time Crystals},
  author = {Watanabe, Haruki and Oshikawa, Masaki},
  journal = {Phys. Rev. Lett.},
  volume = {114},
  issue = {25},
  pages = {251603},
  numpages = {5},
  year = {2015},
  month = {Jun},
  publisher = {American Physical Society},
  doi = {10.1103/PhysRevLett.114.251603},
  url = {https://link.aps.org/doi/10.1103/PhysRevLett.114.251603}
}

@article{else2016floquet,
  title = {Floquet Time Crystals},
  author = {Else, Dominic V. and Bauer, Bela and Nayak, Chetan},
  journal = {Phys. Rev. Lett.},
  volume = {117},
  issue = {9},
  pages = {090402},
  numpages = {5},
  year = {2016},
  month = {Aug},
  publisher = {American Physical Society},
  doi = {10.1103/PhysRevLett.117.090402},
  url = {https://link.aps.org/doi/10.1103/PhysRevLett.117.090402}
}

@article{yao2017discrete,
  title = {Discrete Time Crystals: Rigidity, Criticality, and Realizations},
  author = {Yao, N. Y. and Potter, A. C. and Potirniche, I.-D. and Vishwanath, A.},
  journal = {Phys. Rev. Lett.},
  volume = {118},
  issue = {3},
  pages = {030401},
  numpages = {6},
  year = {2017},
  month = {Jan},
  publisher = {American Physical Society},
  doi = {10.1103/PhysRevLett.118.030401},
  url = {https://link.aps.org/doi/10.1103/PhysRevLett.118.030401}
}

@article{zhang2017observation,
  title={Observation of a discrete time crystal},
  author={Zhang, J and Hess, PW and Kyprianidis, A and Becker, P and Lee, A and Smith, J and Pagano, G and Potirniche, I-D and Potter, Andrew C and Vishwanath, A and others},
  journal={Nature},
  volume={543},
  number={7644},
  pages={217},
  year={2017},
  publisher={Nature Publishing Group}
}

@article{choi2017observation,
  title={Observation of discrete time-crystalline order in a disordered dipolar many-body system},
  author={Choi, Soonwon and Choi, Joonhee and Landig, Renate and Kucsko, Georg and Zhou, Hengyun and Isoya, Junichi and Jelezko, Fedor and Onoda, Shinobu and Sumiya, Hitoshi and Khemani, Vedika and others},
  journal={Nature},
  volume={543},
  number={7644},
  pages={221},
  year={2017},
  publisher={Nature Publishing Group}
}

@article{else2017prethermal,
  title = {Prethermal Phases of Matter Protected by Time-Translation Symmetry},
  author = {Else, Dominic V. and Bauer, Bela and Nayak, Chetan},
  journal = {Phys. Rev. X},
  volume = {7},
  issue = {1},
  pages = {011026},
  numpages = {21},
  year = {2017},
  month = {Mar},
  publisher = {American Physical Society},
  doi = {10.1103/PhysRevX.7.011026},
  url = {https://link.aps.org/doi/10.1103/PhysRevX.7.011026}
}

@article{iemini2018boundary,
  title = {Boundary Time Crystals},
  author = {Iemini, F. and Russomanno, A. and Keeling, J. and Schir\`o, M. and Dalmonte, M. and Fazio, R.},
  journal = {Phys. Rev. Lett.},
  volume = {121},
  issue = {3},
  pages = {035301},
  numpages = {6},
  year = {2018},
  month = {Jul},
  publisher = {American Physical Society},
  doi = {10.1103/PhysRevLett.121.035301},
  url = {https://link.aps.org/doi/10.1103/PhysRevLett.121.035301}
}

@article{das2018cosmological,
  title = {Cosmological time crystal: Cyclic universe with a small cosmological constant in a toy model approach},
  author = {Das, Praloy and Pan, Supriya and Ghosh, Subir and Pal, Probir},
  journal = {Phys. Rev. D},
  volume = {98},
  issue = {2},
  pages = {024004},
  numpages = {12},
  year = {2018},
  month = {Jul},
  publisher = {American Physical Society},
  doi = {10.1103/PhysRevD.98.024004},
  url = {https://link.aps.org/doi/10.1103/PhysRevD.98.024004}
}

@article{feng2018cosmological,
  title={Cosmological time crystals from Einstein-cubic gravities},
  author={Feng, Xing-Hui and Huang, Hyat and Li, Shou-Long and Lu, H and Wei, Hao},
  journal={arXiv preprint arXiv:1807.01720},
  year={2018}
}

@article{autti2018observation,
  title = {Observation of a Time Quasicrystal and Its Transition to a Superfluid Time Crystal},
  author = {Autti, S. and Eltsov, V. B. and Volovik, G. E.},
  journal = {Phys. Rev. Lett.},
  volume = {120},
  issue = {21},
  pages = {215301},
  numpages = {5},
  year = {2018},
  month = {May},
  publisher = {American Physical Society},
  doi = {10.1103/PhysRevLett.120.215301},
  url = {https://link.aps.org/doi/10.1103/PhysRevLett.120.215301}
}

@article{rovny2018observation,
  title = {Observation of Discrete-Time-Crystal Signatures in an Ordered Dipolar Many-Body System},
  author = {Rovny, Jared and Blum, Robert L. and Barrett, Sean E.},
  journal = {Phys. Rev. Lett.},
  volume = {120},
  issue = {18},
  pages = {180603},
  numpages = {5},
  year = {2018},
  month = {May},
  publisher = {American Physical Society},
  doi = {10.1103/PhysRevLett.120.180603},
  url = {https://link.aps.org/doi/10.1103/PhysRevLett.120.180603}
}

@article{rovny2018p,
  title = {$^{31}\mathrm{P}$ NMR study of discrete time-crystalline signatures in an ordered crystal of ammonium dihydrogen phosphate},
  author = {Rovny, Jared and Blum, Robert L. and Barrett, Sean E.},
  journal = {Phys. Rev. B},
  volume = {97},
  issue = {18},
  pages = {184301},
  numpages = {16},
  year = {2018},
  month = {May},
  publisher = {American Physical Society},
  doi = {10.1103/PhysRevB.97.184301},
  url = {https://link.aps.org/doi/10.1103/PhysRevB.97.184301}
}

@article{von2016phase,
  title = {Phase structure of one-dimensional interacting Floquet systems. II. Symmetry-broken phases},
  author = {von Keyserlingk, C. W. and Sondhi, S. L.},
  journal = {Phys. Rev. B},
  volume = {93},
  issue = {24},
  pages = {245146},
  numpages = {11},
  year = {2016},
  month = {Jun},
  publisher = {American Physical Society},
  doi = {10.1103/PhysRevB.93.245146},
  url = {https://link.aps.org/doi/10.1103/PhysRevB.93.245146}
}

@book{robinson2012course,
  title={A Course in the Theory of Groups},
  author={Robinson, Derek JS},
  volume={80},
  year={2012},
  publisher={Springer Science \& Business Media}
}

@book{strocchi2005symmetry,
  title={Symmetry breaking},
  author={Strocchi, Franco},
  volume={643},
  year={2005},
  publisher={Springer}
}

@article{heissenberg2019generalized,
      author         = "Heissenberg, Carlo and Strocchi, Franco",
      title          = "{Generalized criteria of symmetry breaking. A strategy
                        for quantum time crystals}",
      year           = "2019",
      eprint         = "1906.12293",
      archivePrefix  = "arXiv",
      primaryClass   = "cond-mat.stat-mech",
      SLACcitation   = "%%CITATION = ARXIV:1906.12293;%%"
}

@book{sheldon2002first,
  title={A first course in probability},
  author={Sheldon, Ross and others},
  year={2002},
  publisher={Pearson Education India}
}

@article{preskill1998lecture,
  title={Lecture notes for physics 229: Quantum information and computation},
  author={Preskill, John},
  journal={California Institute of Technology},
  volume={16},
  year={1998}
}

@article{schumacher1995quantum,
  title = {Quantum coding},
  author = {Schumacher, Benjamin},
  journal = {Phys. Rev. A},
  volume = {51},
  issue = {4},
  pages = {2738--2747},
  numpages = {0},
  year = {1995},
  month = {Apr},
  publisher = {American Physical Society},
  doi = {10.1103/PhysRevA.51.2738},
  url = {https://link.aps.org/doi/10.1103/PhysRevA.51.2738}
}

@article{slofstra2019set, 
  title={The set of quantum correlations is not closed}, 
  volume={7}, 
  DOI={10.1017/fmp.2018.3}, 
  journal={Forum of Mathematics, Pi}, 
  publisher={Cambridge University Press}, 
  author={Slofstra, William}, 
  year={2019},
  pages={e1}
}

@article{pisarczyk2018causal,
  title={Causal limit on quantum communication},
  author={Pisarczyk, Robert and Zhao, Zhikuan and Ouyang, Yingkai and Vedral, Vlatko and Fitzsimons, Joseph F},
  journal={arXiv preprint arXiv:1804.02594},
  year={2018}
}

@book{sethna2006statistical,
  title={Statistical mechanics: entropy, order parameters, and complexity},
  author={Sethna, James},
  volume={14},
  year={2006},
  publisher={Oxford University Press}
}

@book{peskin2018introduction,
  title={An introduction to quantum field theory},
  author={Peskin, Michael E},
  year={2018},
  publisher={CRC Press}
}

@book{vedral2006introduction,
  title={Introduction to quantum information science},
  author={Vedral, Vlatko},
  year={2006},
  publisher={Oxford University Press on Demand}
}

@Article{raamsdonk2010building,
author="Van Raamsdonk, Mark",
title="Building up spacetime with quantum entanglement",
journal="General Relativity and Gravitation",
year="2010",
month="Oct",
day="01",
volume="42",
number="10",
pages="2323--2329",
issn="1572-9532",
doi="10.1007/s10714-010-1034-0",
url="https://doi.org/10.1007/s10714-010-1034-0"
}

@article{jacobson1995thermodynamics,
  title = {Thermodynamics of Spacetime: The Einstein Equation of State},
  author = {Jacobson, Ted},
  journal = {Phys. Rev. Lett.},
  volume = {75},
  issue = {7},
  pages = {1260--1263},
  numpages = {0},
  year = {1995},
  month = {Aug},
  publisher = {American Physical Society},
  doi = {10.1103/PhysRevLett.75.1260},
  url = {https://link.aps.org/doi/10.1103/PhysRevLett.75.1260}
}

@article{lloyd2012quantum,
  title={The quantum geometric limit},
  author={Lloyd, Seth},
  journal={arXiv preprint arXiv:1206.6559},
  year={2012}
}

@book{ohya2004quantum,
  title={Quantum entropy and its use},
  author={Ohya, Masanori and Petz, D{\'e}nes},
  year={2004},
  publisher={Springer Science \& Business Media}
}

@book{von2018mathematical,
  title={Mathematical Foundations of Quantum Mechanics: New Edition},
  author={Von Neumann, John},
  year={2018},
  publisher={Princeton university press}
}

@article{grunbaum1967convex,
  title={Convex polytopes},
  author={Gr{\"u}nbaum, Branko and Klee, Victor and Perles, Micha A and Shephard, Geoffrey Colin},
  year={1967},
  publisher={Springer}
}

@article{uola2019quantum,
  title={Quantum Steering},
  author={Uola, Roope and Costa, Ana and Nguyen, H Chau and G{\"u}hne, Otfried},
  journal={arXiv preprint arXiv:1903.06663},
  year={2019}
}

@article{bell1964einstein,
  title={On the Einstein-Podolsky-Rosen paradox},
  author={Bell, John S},
  journal={Physics Physique Fizika},
  volume={1},
  number={3},
  pages={195},
  year={1964},
  publisher={APS}
}

@article{brunner2014bell,
  title = {Bell nonlocality},
  author = {Brunner, Nicolas and Cavalcanti, Daniel and Pironio, Stefano and Scarani, Valerio and Wehner, Stephanie},
  journal = {Rev. Mod. Phys.},
  volume = {86},
  issue = {2},
  pages = {419--478},
  numpages = {60},
  year = {2014},
  month = {Apr},
  publisher = {American Physical Society},
  doi = {10.1103/RevModPhys.86.419},
  url = {https://link.aps.org/doi/10.1103/RevModPhys.86.419}
}

@article{ollivier2001quantum,
  title = {Quantum Discord: A Measure of the Quantumness of Correlations},
  author = {Ollivier, Harold and Zurek, Wojciech H.},
  journal = {Phys. Rev. Lett.},
  volume = {88},
  issue = {1},
  pages = {017901},
  numpages = {4},
  year = {2001},
  month = {Dec},
  publisher = {American Physical Society},
  doi = {10.1103/PhysRevLett.88.017901},
  url = {https://link.aps.org/doi/10.1103/PhysRevLett.88.017901}
}

@article{henderson2001classical,
	doi = {10.1088/0305-4470/34/35/315},
	url = {https://doi.org/10.1088%2F0305-4470%2F34%2F35%2F315},
	year = 2001,
	month = {aug},
	publisher = {{IOP} Publishing},
	volume = {34},
	number = {35},
	pages = {6899--6905},
	author = {L Henderson and V Vedral},
	title = {Classical, quantum and total correlations},
	journal = {Journal of Physics A: Mathematical and General}
}

@article{modi2012classical,
  title = {The classical-quantum boundary for correlations: Discord and related measures},
  author = {Modi, Kavan and Brodutch, Aharon and Cable, Hugo and Paterek, Tomasz and Vedral, Vlatko},
  journal = {Rev. Mod. Phys.},
  volume = {84},
  issue = {4},
  pages = {1655--1707},
  numpages = {0},
  year = {2012},
  month = {Nov},
  publisher = {American Physical Society},
  doi = {10.1103/RevModPhys.84.1655},
  url = {https://link.aps.org/doi/10.1103/RevModPhys.84.1655}
}

@article{devetak2004distilling,
  title={Distilling common randomness from bipartite quantum states},
  author={Devetak, Igor and Winter, Andreas},
  journal={IEEE Transactions on Information Theory},
  volume={50},
  number={12},
  pages={3183--3196},
  year={2004},
  publisher={IEEE}
}

@article{horodecki2003local,
  title = {Local Information as a Resource in Distributed Quantum Systems},
  author = {Horodecki, Micha\l{} and Horodecki, Karol and Horodecki, Pawe\l{} and Horodecki, Ryszard and Oppenheim, Jonathan and Sen(De), Aditi and Sen, Ujjwal},
  journal = {Phys. Rev. Lett.},
  volume = {90},
  issue = {10},
  pages = {100402},
  numpages = {4},
  year = {2003},
  month = {Mar},
  publisher = {American Physical Society},
  doi = {10.1103/PhysRevLett.90.100402},
  url = {https://link.aps.org/doi/10.1103/PhysRevLett.90.100402}
}

@article{luo2008using,
  title = {Using measurement-induced disturbance to characterize correlations as classical or quantum},
  author = {Luo, Shunlong},
  journal = {Phys. Rev. A},
  volume = {77},
  issue = {2},
  pages = {022301},
  numpages = {5},
  year = {2008},
  month = {Feb},
  publisher = {American Physical Society},
  doi = {10.1103/PhysRevA.77.022301},
  url = {https://link.aps.org/doi/10.1103/PhysRevA.77.022301}
}

@article{wu2009correlations,
  title = {Correlations in local measurements on a quantum state, and complementarity as an explanation of nonclassicality},
  author = {Wu, Shengjun and Poulsen, Uffe V. and M\o{}lmer, Klaus},
  journal = {Phys. Rev. A},
  volume = {80},
  issue = {3},
  pages = {032319},
  numpages = {11},
  year = {2009},
  month = {Sep},
  publisher = {American Physical Society},
  doi = {10.1103/PhysRevA.80.032319},
  url = {https://link.aps.org/doi/10.1103/PhysRevA.80.032319}
}

@article{modi2010unified,
  title = {Unified View of Quantum and Classical Correlations},
  author = {Modi, Kavan and Paterek, Tomasz and Son, Wonmin and Vedral, Vlatko and Williamson, Mark},
  journal = {Phys. Rev. Lett.},
  volume = {104},
  issue = {8},
  pages = {080501},
  numpages = {4},
  year = {2010},
  month = {Feb},
  publisher = {American Physical Society},
  doi = {10.1103/PhysRevLett.104.080501},
  url = {https://link.aps.org/doi/10.1103/PhysRevLett.104.080501}
}

@article{scholz2008tsirelson,
  title={Tsirelson's problem},
  author={Scholz, Volkher B and Werner, Reinhard F},
  journal={arXiv preprint arXiv:0812.4305},
  year={2008}
}

@article{tsirelson2006bell,
  title={Bell inequalities and operator algebras},
  author={Tsirelson, Boris},
  year={2006},
  publisher={Citeseer}
}

@article{dykema2016synchronous,
  title={Synchronous correlation matrices and Connes’ embedding conjecture},
  author={Dykema, Kenneth J and Paulsen, Vern},
  journal={Journal of Mathematical Physics},
  volume={57},
  number={1},
  pages={015214},
  year={2016},
  publisher={AIP Publishing}
}

@article{slofstra2019tsirelson,
  title={Tsirelson’s problem and an embedding theorem for groups arising from non-local games},
  author={Slofstra, William},
  journal={Journal of the American Mathematical Society},
  year={2019}
}

@article{lu2018entanglement,
  title = {Entanglement Structure: Entanglement Partitioning in Multipartite Systems and Its Experimental Detection Using Optimizable Witnesses},
  author = {Lu, He and Zhao, Qi and Li, Zheng-Da and Yin, Xu-Fei and Yuan, Xiao and Hung, Jui-Chen and Chen, Luo-Kan and Li, Li and Liu, Nai-Le and Peng, Cheng-Zhi and Liang, Yeong-Cherng and Ma, Xiongfeng and Chen, Yu-Ao and Pan, Jian-Wei},
  journal = {Phys. Rev. X},
  volume = {8},
  issue = {2},
  pages = {021072},
  numpages = {20},
  year = {2018},
  month = {Jun},
  publisher = {American Physical Society},
  doi = {10.1103/PhysRevX.8.021072},
  url = {https://link.aps.org/doi/10.1103/PhysRevX.8.021072}
}

@article{horodecki2009quantum,
  title = {Quantum entanglement},
  author = {Horodecki, Ryszard and Horodecki, Pawe\l{} and Horodecki, Micha\l{} and Horodecki, Karol},
  journal = {Rev. Mod. Phys.},
  volume = {81},
  issue = {2},
  pages = {865--942},
  numpages = {0},
  year = {2009},
  month = {Jun},
  publisher = {American Physical Society},
  doi = {10.1103/RevModPhys.81.865},
  url = {https://link.aps.org/doi/10.1103/RevModPhys.81.865}
}

@book{park2018fundamentals,
  title={Fundamentals of Probability and Stochastic Processes with Applications to Communications},
  author={Park, Kun Il and Park},
  year={2018},
  publisher={Springer}
}

@article{ahn2013quantum,
  title = {Quantum-state cloning in the presence of a closed timelike curve},
  author = {Ahn, D. and Myers, C. R. and Ralph, T. C. and Mann, R. B.},
  journal = {Phys. Rev. A},
  volume = {88},
  issue = {2},
  pages = {022332},
  numpages = {5},
  year = {2013},
  month = {Aug},
  publisher = {American Physical Society},
  doi = {10.1103/PhysRevA.88.022332},
  url = {https://link.aps.org/doi/10.1103/PhysRevA.88.022332}
}

@article{brun2013quantum,
  title = {Quantum State Cloning Using Deutschian Closed Timelike Curves},
  author = {Brun, Todd A. and Wilde, Mark M. and Winter, Andreas},
  journal = {Phys. Rev. Lett.},
  volume = {111},
  issue = {19},
  pages = {190401},
  numpages = {5},
  year = {2013},
  month = {Nov},
  publisher = {American Physical Society},
  doi = {10.1103/PhysRevLett.111.190401},
  url = {https://link.aps.org/doi/10.1103/PhysRevLett.111.190401}
}

@article{buscemi2012all,
  title = {All Entangled Quantum States Are Nonlocal},
  author = {Buscemi, Francesco},
  journal = {Phys. Rev. Lett.},
  volume = {108},
  issue = {20},
  pages = {200401},
  numpages = {5},
  year = {2012},
  month = {May},
  publisher = {American Physical Society},
  doi = {10.1103/PhysRevLett.108.200401},
  url = {https://link.aps.org/doi/10.1103/PhysRevLett.108.200401}
}

@article{rosset2018resource,
  title = {Resource Theory of Quantum Memories and Their Faithful Verification with Minimal Assumptions},
  author = {Rosset, Denis and Buscemi, Francesco and Liang, Yeong-Cherng},
  journal = {Phys. Rev. X},
  volume = {8},
  issue = {2},
  pages = {021033},
  numpages = {15},
  year = {2018},
  month = {May},
  publisher = {American Physical Society},
  doi = {10.1103/PhysRevX.8.021033},
  url = {https://link.aps.org/doi/10.1103/PhysRevX.8.021033}
}

@Article{hawking1975particle,
author="Hawking, S. W.",
title="Particle creation by black holes",
journal="Communications in Mathematical Physics",
year="1975",
month="Aug",
day="01",
volume="43",
number="3",
pages="199--220",
issn="1432-0916",
doi="10.1007/BF02345020",
url="https://doi.org/10.1007/BF02345020"
}

@article{hawking1976breakdown,
  title = {Breakdown of predictability in gravitational collapse},
  author = {Hawking, S. W.},
  journal = {Phys. Rev. D},
  volume = {14},
  issue = {10},
  pages = {2460--2473},
  numpages = {0},
  year = {1976},
  month = {Nov},
  publisher = {American Physical Society},
  doi = {10.1103/PhysRevD.14.2460},
  url = {https://link.aps.org/doi/10.1103/PhysRevD.14.2460}
}

@inproceedings{preskill1992black,
  title={Do black holes destroy information},
  author={Preskill, John},
  booktitle={Proceedings of the International Symposium on Black Holes, Membranes, Wormholes and Superstrings, S. Kalara and DV Nanopoulos, eds.(World Scientific, Singapore, 1993) pp},
  pages={22--39},
  year={1992},
  organization={World Scientific}
}

@article{horowitz2003black,
      author         = "Horowitz, Gary T. and Maldacena, Juan Martin",
      title          = "{The Black hole final state}",
      journal        = "JHEP",
      volume         = "02",
      year           = "2004",
      pages          = "008",
      doi            = "10.1088/1126-6708/2004/02/008",
      eprint         = "hep-th/0310281",
      archivePrefix  = "arXiv",
      primaryClass   = "hep-th",
      SLACcitation   = "%%CITATION = HEP-TH/0310281;%%"
}

@article{gottesman2003comment,
      author         = "Gottesman, Daniel and Preskill, John",
      title          = "{Comment on `The Black hole final state'}",
      journal        = "JHEP",
      volume         = "03",
      year           = "2004",
      pages          = "026",
      doi            = "10.1088/1126-6708/2004/03/026",
      eprint         = "hep-th/0311269",
      archivePrefix  = "arXiv",
      primaryClass   = "hep-th",
      reportNumber   = "CALT-68-2466",
      SLACcitation   = "%%CITATION = HEP-TH/0311269;%%"
}

@article{lloyd2013unitarity,
      author         = "Lloyd, Seth and Preskill, John",
      title          = "{Unitarity of black hole evaporation in final-state
                        projection models}",
      journal        = "JHEP",
      volume         = "08",
      year           = "2014",
      pages          = "126",
      doi            = "10.1007/JHEP08(2014)126",
      eprint         = "1308.4209",
      archivePrefix  = "arXiv",
      primaryClass   = "hep-th",
      reportNumber   = "CALT-68-2856",
      SLACcitation   = "%%CITATION = ARXIV:1308.4209;%%"
}

@article{bousso2014measurements,
  title = {Measurements without probabilities in the final state proposal},
  author = {Bousso, Raphael and Stanford, Douglas},
  journal = {Phys. Rev. D},
  volume = {89},
  issue = {4},
  pages = {044038},
  numpages = {8},
  year = {2014},
  month = {Feb},
  publisher = {American Physical Society},
  doi = {10.1103/PhysRevD.89.044038},
  url = {https://link.aps.org/doi/10.1103/PhysRevD.89.044038}
}

@article{banks1992horned,
  title = {Are horned particles the end point of Hawking evaporation?},
  author = {Banks, T. and Dabholkar, A. and Douglas, M. R. and O'Loughlin, M.},
  journal = {Phys. Rev. D},
  volume = {45},
  issue = {10},
  pages = {3607--3616},
  numpages = {0},
  year = {1992},
  month = {May},
  publisher = {American Physical Society},
  doi = {10.1103/PhysRevD.45.3607},
  url = {https://link.aps.org/doi/10.1103/PhysRevD.45.3607}
}

@article{hawking1988wormholes,
  title = {Wormholes in spacetime},
  author = {Hawking, S. W.},
  journal = {Phys. Rev. D},
  volume = {37},
  issue = {4},
  pages = {904--910},
  numpages = {0},
  year = {1988},
  month = {Feb},
  publisher = {American Physical Society},
  doi = {10.1103/PhysRevD.37.904},
  url = {https://link.aps.org/doi/10.1103/PhysRevD.37.904}
}

@article{page1993average,
  title = {Average entropy of a subsystem},
  author = {Page, Don N.},
  journal = {Phys. Rev. Lett.},
  volume = {71},
  issue = {9},
  pages = {1291--1294},
  numpages = {0},
  year = {1993},
  month = {Aug},
  publisher = {American Physical Society},
  doi = {10.1103/PhysRevLett.71.1291},
  url = {https://link.aps.org/doi/10.1103/PhysRevLett.71.1291}
}

@article{page1993information,
  title = {Information in black hole radiation},
  author = {Page, Don N.},
  journal = {Phys. Rev. Lett.},
  volume = {71},
  issue = {23},
  pages = {3743--3746},
  numpages = {0},
  year = {1993},
  month = {Dec},
  publisher = {American Physical Society},
  doi = {10.1103/PhysRevLett.71.3743},
  url = {https://link.aps.org/doi/10.1103/PhysRevLett.71.3743}
}

@article{hawking1990baby,
  title={Baby Universes II},
  author={Hawking, Stephen W},
  journal={Mod. Phys. Lett. A},
  volume={5},
  pages={453--466},
  year={1990}
}

@article{nambu1960quasi,
  title = {Quasi-Particles and Gauge Invariance in the Theory of Superconductivity},
  author = {Nambu, Yoichiro},
  journal = {Phys. Rev.},
  volume = {117},
  issue = {3},
  pages = {648--663},
  numpages = {0},
  year = {1960},
  month = {Feb},
  publisher = {American Physical Society},
  doi = {10.1103/PhysRev.117.648},
  url = {https://link.aps.org/doi/10.1103/PhysRev.117.648}
}

@Article{goldstone1961field,
author="Goldstone, J.",
title="Field theories with « Superconductor » solutions",
journal="Il Nuovo Cimento (1955-1965)",
year="1961",
month="Jan",
day="01",
volume="19",
number="1",
pages="154--164",
issn="1827-6121",
doi="10.1007/BF02812722",
url="https://doi.org/10.1007/BF02812722"
}

@article{mermin1966absence,
  title = {Absence of Ferromagnetism or Antiferromagnetism in One- or Two-Dimensional Isotropic Heisenberg Models},
  author = {Mermin, N. D. and Wagner, H.},
  journal = {Phys. Rev. Lett.},
  volume = {17},
  issue = {22},
  pages = {1133--1136},
  numpages = {0},
  year = {1966},
  month = {Nov},
  publisher = {American Physical Society},
  doi = {10.1103/PhysRevLett.17.1133},
  url = {https://link.aps.org/doi/10.1103/PhysRevLett.17.1133}
}

@article{hohenberg1967existence,
  title = {Existence of Long-Range Order in One and Two Dimensions},
  author = {Hohenberg, P. C.},
  journal = {Phys. Rev.},
  volume = {158},
  issue = {2},
  pages = {383--386},
  numpages = {0},
  year = {1967},
  month = {Jun},
  publisher = {American Physical Society},
  doi = {10.1103/PhysRev.158.383},
  url = {https://link.aps.org/doi/10.1103/PhysRev.158.383}
}

@article{gelfert2001absence,
	doi = {10.1088/0953-8984/13/27/201},
	url = {https://doi.org/10.1088%2F0953-8984%2F13%2F27%2F201},
	year = 2001,
	month = {jun},
	publisher = {{IOP} Publishing},
	volume = {13},
	number = {27},
	pages = {R505--R524},
	author = {Axel Gelfert and Wolfgang Nolting},
	title = {The absence of finite-temperature phase transitions in low-dimensional many-body models: a survey and new results},
	journal = {Journal of Physics: Condensed Matter}
}

@article{bruno2013comment,
  title = {Comment on ``Quantum Time Crystals''},
  author = {Bruno, Patrick},
  journal = {Phys. Rev. Lett.},
  volume = {110},
  issue = {11},
  pages = {118901},
  numpages = {1},
  year = {2013},
  month = {Mar},
  publisher = {American Physical Society},
  doi = {10.1103/PhysRevLett.110.118901},
  url = {https://link.aps.org/doi/10.1103/PhysRevLett.110.118901}
}

@article{bruno2013impossibility,
  title = {Impossibility of Spontaneously Rotating Time Crystals: A No-Go Theorem},
  author = {Bruno, Patrick},
  journal = {Phys. Rev. Lett.},
  volume = {111},
  issue = {7},
  pages = {070402},
  numpages = {5},
  year = {2013},
  month = {Aug},
  publisher = {American Physical Society},
  doi = {10.1103/PhysRevLett.111.070402},
  url = {https://link.aps.org/doi/10.1103/PhysRevLett.111.070402}
}

\end{document}